\documentclass[aos, authoryear]{imsart}

\usepackage[utf8]{inputenc}
\usepackage[authoryear]{natbib} 
\usepackage{hyperref}
\RequirePackage{graphicx}
\RequirePackage{amsthm,amsmath,amsfonts,amssymb}
\numberwithin{equation}{section}
\startlocaldefs

\usepackage{etoolbox}

\usepackage{
	amsmath,
	amssymb,
	wrapfig,
	cases,
	mathtools,
	thmtools,
	array,
	bbm,
	bm,
	enumerate,
	makecell,
	esvect,
	mathrsfs,
	booktabs,
	upgreek,
	changepage,
	dsfont,
	fixme,
	listings,
	multirow,
	xargs,
	xstring,
	multicol,
	float,
	enumerate,
	indentfirst,
	}

\usepackage[capitalize,nameinlink]{cleveref}
\usepackage{crossreftools}
\crefname{section}{Section}{Sections}
\crefname{subsection}{Subsection}{Subsections}
\crefname{lemma}{Lemma}{Lemmas}
\crefname{corollary}{Corollary}{Corollaries}
\crefname{theorem}{Theorem}{Theorems}
\crefname{assumption}{Assumption}{Assumptions}

\newtheorem{definition}{Definition}
\newtheorem{theorem}{Theorem}
\newtheorem{lemma}{Lemma}
\newtheorem{assumption}{Assumption}
\newtheorem{corollary}{Corollary}
\newtheorem{proposition}{Proposition}
\newtheorem{remark}{Remark}

\newcommand{\EE}{\mathbb{E}}

\newcommand{\II}{\mathbb{I}}

\newcommand{\NN}{\mathbb{N}}

\newcommand{\PP}{\mathbb{P}}

\newcommand{\RR}{\mathbb{R}}

\newcommand{\ZZ}{\mathbb{Z}}

\newcommand{\Bb}{\mathcal{B}}

\newcommand{\Uu}{\mathcal{U}}

\newcommand{\zero}{\mathbf{0}}

\newcommand{\eps}{\varepsilon}

\def\[#1\]{\begin{equation}\begin{aligned}#1\end{aligned}\end{equation}}
\def\*[#1\]{\begin{equation*}\begin{aligned}#1\end{aligned}\end{equation*}}

\def\s*[#1\s]{\small\begin{align*}#1\end{align*}\normalsize}

\newcommand{\lcrx}[4][{-1}]{ 
	\IfEq{#1}{-1}{\left #2 {{{{#3}}}} \right #4}{
   	\IfEq{#1}{0}{#2 {{{{#3}}}} #4}{
	\IfEq{#1}{1}{\bigl #2 {{{{#3}}}} \bigr #4}{
	\IfEq{#1}{2}{\Bigl #2 {{{{#3}}}} \Bigr #4}{
	\IfEq{#1}{3}{\biggl #2 {{{{#3}}}} \biggr #4}{
	\IfEq{#1}{4}{\Biggl #2 {{{{#3}}}} \Biggr #4}{
    \GenericWarning{"4th argument to lcrx must be -1, 0, 1, 2, 3, or 4"}
    }}}}}}} 

\newcommand{\setdelim}{\ \vert \ }
\newcommand{\Bigsetdelim}{\ \Big\vert \ }

\newcommand{\ind}{\II} 

\def\multiset#1#2{\ensuremath{\left(\kern-.3em\left(\genfrac{}{}{0pt}{}{#1}{#2}\right)\kern-.3em\right)}}

\DeclareMathOperator*{\argmax}{\arg\max} 

\DeclareMathOperator*{\newlim}{\mathrm{lim}\vphantom{\mathrm{infsup}}}
\DeclareMathOperator*{\newmin}{\mathrm{min}\vphantom{\mathrm{infsup}}}
\DeclareMathOperator*{\newmax}{\mathrm{max}\vphantom{\mathrm{infsup}}}
\DeclareMathOperator*{\newinf}{\mathrm{inf}\vphantom{\mathrm{infsup}}}
\DeclareMathOperator*{\newsup}{\mathrm{sup}\vphantom{\mathrm{infsup}}}
\renewcommand{\lim}{\newlim}
\renewcommand{\min}{\newmin}
\renewcommand{\max}{\newmax}
\renewcommand{\inf}{\newinf}
\renewcommand{\sup}{\newsup}

\newcommand{\Tr}{^{\scriptscriptstyle\text{T}}} 
\newcommand{\inv}{^{-1}}
\newcommand{\tr}{\mathrm{Tr}} 
\newcommand{\detbar}[1]{\lvert #1 \rvert}

\newcommand{\dee}{\mathrm{d}} 

\newcommand{\bernoullidist}{\mathrm{Ber}}

\newcommand{\normaldist}{\mathrm{Gaussian}}

\newcommand{\sbra}[2][{-1}]{\lcrx[#1] [ {#2} ] }

\newcommand{\abs}[2][{-1}]{\lcrx[#1] \vert {#2} \vert }

\newcommand{\norm}[2][{-1}]{\lcrx[#1] \Vert {#2} \Vert}

\newcommand{\Nats}{\NN}

\newcommand{\Ints}{\ZZ}

\newcommand{\Reals}{\RR}

\newcommand{\PosInts}{\Ints_+}

\newcommand{\range}[2][{1}]{
	\IfEq{#1}{1}{\sbra{#2}}{\sbra{#2}_{#1}}}
\newcommand{\rangeO}[2][{0}]{
	\IfEq{#1}{0}{\sbra{#2}_0}{\sbra{#2}_{#1}}}

\newcommand{\dummydim}{p}

\newcommand{\dummytau}{b}

\newcommand{\const}{C}
\newcommand{\constt}{D}
\newcommand{\paramconst}{C_{\quadnum, \redim}}

\newcommand{\quadnumadj}{\kappa}

\newcommand{\shrinkingrad}{\gamma_i}
\newcommand{\shrinkingconst}{\gamma}

\newcommand{\tvec}{\bm{t}}

\newcommand{\smalltsetdef}[2][1]{ \tau^{(#1)}_{{ < #2 }} }
\newcommand{\equaltsetdef}[2][1]{ \tau^{(#1)}_{{ = #2 }} }
\newcommand{\bigtset}[1]{\tau^{(#1)}_{\geq2\quadnum}}
\newcommand{\bigtsetdef}[2][1]{\tau^{(#1)}_{\geq #2}}
\newcommand{\customtset}[2]{\tau^{(#1)}_{\geq#2}}

\newcommand{\tsum}{\tau(\tvec)}

\newcommand{\absbig}[1]{\Big\lvert #1 \Big\rvert}
\newcommand{\abssmall}[1]{\lvert #1 \rvert}

\newcommand{\AGHQ}{{\upshape AQ}}
\newcommand{\AQ}{{\texttt{\tiny \upshape AQ}}}

\newcommand{\glmm}{GLMM}
\newcommand{\glmms}{GLMMs}

\newcommand{\adaptiveQuadrature}{AQ}

\newcommand{\genmodelname}{mixed model}
\newcommand{\genmodelnames}{mixed models}

\newcommand{\pto}{\overset{p}{\longrightarrow}}

\newcommand{\modeaccent}[1]{\widehat{#1}}
\newcommand{\mb}[1]{\boldsymbol{#1}}

\newcommand{\ball}[2]{\Bb_{#1}(#2)}
\newcommand{\ballc}[2]{[\ball{#1}{#2}]^{c}}

\newcommand{\bigball}[2]{\Bb_{#1}(#2)} 
\newcommand{\bigballc}[2]{\Big[\Bb_{#1}(#2)\Big]^c} 

\newcommand{\universalradius}{\delta}
\newcommand{\paramradius}{\zeta_\numobs}
\newcommand{\paramradiusi}{\zeta_i}
\newcommand{\shrinkingradius}{\gamma_{m}}
\newcommand{\eigen}{\lambda}
\newcommand{\conmargin}{\beta}
\newcommand{\derivnum}{t} 
\newcommand{\derivbound}{D}
\newcommand{\llhoodmargin}{b}
\newcommand{\priorbig}{c_2}
\newcommand{\priorsmall}{c_1}

\newcommand{\derivboundmodeip}{\modeaccent{M}_{i}^{\params}}
\newcommand{\dumderivvec}{\mb{\alpha}}

\newcommand{\multitaylorexpboundip}{\modeaccent{\Lambda}_{\numpergroup}^{\params}}

\newcommand{\hessbig}{\overline{\eta}}

\newcommand{\hessbigmodeip}{{{\overline{\eta}}_{i}^{\params}}}
\newcommand{\hesssmallmode}{\underline{\eta_n}}
\newcommand{\hesssmallmodeip}{{{\underline{\eta}}_{i}^{\params}}}
\newcommand{\hessbigtrue}{\overline{\eta^*_n}}

\newcommand{\hesssmalltrueip}{\underline{\eta}^{*,\params}_i}
\newcommand{\hesssmall}{\underline{\eta}}

\newcommand{\gausscovsmall}{\underline{\xi}_1}
\newcommand{\gausscovbig}{\overline{\xi}_1}
\newcommand{\gausscovderivbig}{\overline{\xi^{\prime}_1}}

\newcommand{\recovsmall}{\underline{\xi}_2}
\newcommand{\recovbig}{\overline{\xi}_2}

\newcommand{\covbound}{M}
\newcommand{\linpredbound}{\recovbig^{*}}

\newcommand{\paramsbigball}{\bigball{\paramstrue}{\paramradius}}
\newcommand{\paramsbigballfixed}{\bigball{\paramstrue}{\universalradius}}
\newcommand{\paramssmallball}{\bigball{\paramstrue}{\shrinkingradius}}
\newcommand{\reball}{\bigball{\retruepi}{\universalradius}}

\newcommand{\weightmat}{\mb{W}}
\newcommand{\weightmatp}{\weightmat^{\params}}
\newcommand{\weightmatip}{\weightmatp_i}

\newcommand{\natparamspace}{\Xi}
\newcommand{\samplespace}{\mathcal{S}}

\newcommand{\pergrouprate}{\gamma^*(\quadnum)}

\newcommand{\dist}{\pi} 

\newcommand{\param}{\theta}
\newcommand{\paramspace}{\Theta}
\newcommand{\regparamspace}{\Theta_{\text{\upshape Mod}}}
\newcommand{\resdparamspace}{\Theta_{\text{\upshape Re}}}
\newcommand{\dispparamspace}{\Theta_{\text{\upshape Disp}}}

\newcommand{\midweight}{\tau}

\newcommand{\property}[2]{\mathscr{P}(#1,#2)}

\newcommand{\polyfunc}{P}

\newcommand{\normaldens}[3]{\phi(#1; #2, #3)}

\newcommand{\identmat}{\mb{I}}

\newcommand{\respace}{\Uu}

\newcommand{\obsi}{y}
\newcommand{\obs}{\mb{\obsi}}
\newcommand{\obsall}{\obs}
\newcommand{\responsei}{Y}
\newcommand{\responseall}{\mb{\responsei}}
\newcommand{\cov}{\mb{x}}
\newcommand{\recov}{\mb{v}}
\newcommand{\redesign}{\mb{V}}
\newcommand{\design}{\mb{X}}
\newcommand{\fulldesign}{\mb{D}}

\newcommand{\reidx}{u}
\newcommand{\re}{\mb{\reidx}}

\newcommand{\regparam}{\mb{\beta}}
\newcommand{\regparamtrue}{\mb{\beta}_{*}}
\newcommand{\linpred}{\eta}
\newcommand{\linpredp}{\linpred^{\params}}

\newcommand{\resd}{\mb{\sigma}}
\newcommand{\obssd}{\mb{\phi}}

\newcommand{\params}{\mb{\param}}

\newcommand{\paramsmle}{\widehat{\params}}
\newcommand{\intparamsmle}{\params^\circ}

\newcommand{\approxparamsmle}{\widetilde{\params}}

\newcommand{\retrue}{\re_{*}}
\newcommand{\retruep}{\retrue}
\newcommand{\retruepi}{\re_{*,i}}
\newcommand{\retrueptrue}{\re_{*,i}}
\newcommand{\paramtrue}{\param_{*}}
\newcommand{\paramstrue}{\params_{*}}

\newcommand{\trueprob}[1]{P_{#1}^*}

\newcommand{\linkfunc}{h}

\newcommand{\responsedist}{F}
\newcommand{\redist}{G}
\newcommand{\responsedens}{f}
\newcommand{\redens}{g}
\newcommand{\margdens}{\pi}
\newcommand{\approxmargdens}{\widetilde{\pi}}
\newcommand{\approxmargdensAQ}{\approxmargdens^{\AQ}}

\newcommand{\jointdens}{\pi}
\newcommand{\logjointdens}{\ell}
\newcommand{\logmargdens}{\ell}
\newcommand{\approxlogmargdens}{\widetilde{\logmargdens}}

\newcommand{\condmode}{\widehat{\re}}
\newcommand{\complementmode}{\overline{\re}}
\newcommand{\paramscomplementmode}{\overline{\params}}
\newcommand{\condmodei}{\condmode_{i}}
\newcommand{\condmodeip}{\condmodei^{\params}}
\newcommand{\condmodeipmle}{\condmodei^{\paramsmle}}
\newcommand{\condmodeiptrue}{\condmodei^{\paramstrue}}

\newcommand{\complementmodei}{\complementmode_{i}}
\newcommand{\complementmodeip}{\complementmodei^{\params}}
\newcommand{\complementmodeipmode}{\complementmodei^{\paramscomplementmode}}

\newcommand{\requadmid}{\mb{V}_i^{\cholip \, \quadpointvec + \condmodeip}}

\newcommand{\hess}{\mb{H}}
\newcommand{\hessi}{\hess_{i}}
\newcommand{\hessip}{\hessi^{\params}}
\newcommand{\hessipinv}{(\hessi^{\params})\inv}

\newcommand{\hessre}{\mb{G}}
\newcommand{\hessrei}{\hessre_{i}}
\newcommand{\hessreip}{\hessrei^{\params}}

\newcommand{\chol}{\mb{L}}
\newcommand{\choli}{\chol_{i}}
\newcommand{\cholip}{\choli^{\params}}
\newcommand{\cholipTr}{(\choli^{\params})\Tr}

\newcommand{\llhoodi}{\ell_{i}}
\newcommand{\llhoodip}{\llhoodi^{\params}}

\newcommand{\covmat}{\boldsymbol{\Sigma}}

\newcommand{\indsim}{\overset{\text{\upshape \tiny ind.}}{\sim}}

\newcommand{\dto}{\overset{d}{\to}}

\newcommand{\quadnum}{k}

\newcommand{\quadpointvec}{\mb{z}}

\newcommand{\quadpointbound}{\overline{\mb{z}}}
\newcommand{\quadpointset}{\mathcal{Q}}

\newcommand{\quadrule}[2]{\mathfrak{R}(#1, #2)}

\newcommand{\weight}{\omega}
\newcommand{\weightmin}{\underline{\weight}}
\newcommand{\weightmax}{\overline{\weight}}

\newcommand{\numgroups}{m}
\newcommand{\numpergroup}{n}
\newcommand{\minpergroup}{n_{\min}}
\newcommand{\redim}{d}
\newcommand{\regparamdim}{q}
\newcommand{\dispparamdim}{l}
\newcommand{\paramdim}{p}
\newcommand{\covdim}{\regparamdim}

\newcommand{\resddim}{s}
\newcommand{\numobs}{N}
\newcommand{\mlerate}{r^*}

\newcommand{\ed}{\overset{d}{=}}
\newcommand{\normalrv}{\boldsymbol{Z}}

\newcommand{\manualendproof}{\hfill\qedsymbol\\[2mm]}

\theoremstyle{remark}
\newtheorem{example}{Example}

\endlocaldefs

\begin{document}

\begin{frontmatter}
\title{Asymptotics of numerical integration for two-level mixed models}
\runtitle{Asymptotics of numerical integtration for two-level mixed models}
\thankstext{T1}{Equal contribution authorship.}
\begin{aug}
\author[A]{\fnms{Blair}~\snm{Bilodeau}\ead[label=e1]{blair.bilodeau@mail.utoronto.ca}}
\author[B]{\fnms{Alex}~\snm{Stringer}\ead[label=e2]{alex.stringer@uwaterloo.ca}}
\author[C]{\fnms{Yanbo}~\snm{Tang*}\ead[label=e3]{yanbo.tang@imperial.ac.uk}}
 \address[A]{University of Toronto \printead[presep={,\ }]{e1}}
 \address[B]{University of Waterloo \printead[presep={,\ }]{e2}}
\address[C]{Imperial College London \printead[presep={,\ }]{e3}}
\end{aug}

\begin{abstract}
We study \genmodelnames{} with a single grouping factor,
where inference about unknown parameters 
requires optimizing a marginal likelihood defined by an intractable integral.
Low-dimensional numerical integration techniques are regularly used to approximate these integrals,
with inferences about parameters based on the resulting approximate marginal likelihood.
For a generic class of \genmodelnames{} that satisfy explicit regularity conditions, 
we derive the stochastic relative error rate incurred for both the likelihood and maximum likelihood estimator when adaptive numerical integration is used to approximate the marginal likelihood.
We then specialize the analysis to well-specified generalized linear mixed models having exponential family response and multivariate Gaussian random effects, 
verifying that the regularity conditions hold, and hence that the convergence rates apply.
We also prove that for models with likelihoods satisfying very weak concentration conditions
that the maximum likelihood estimators from non-adaptive numerical integration approximations of the marginal likelihood are not consistent,
further motivating adaptive numerical integration as the preferred tool for inference in mixed models.
Code to reproduce the simulations in this paper is provided at \url{https://github.com/awstringer1/aq-theory-paper-code}.
\end{abstract}


\end{frontmatter}


\section{Introduction}
\label{sec:intro}

\subsection{Approximate integration in statistical problems}\label{subsec:introapproximateintegration}

\noindent Approximate integration is often required when fitting
statistical models by maximum likelihood.
The concentration properties of likelihood functions
as the number of data increase render such statistical problems
distinct from deterministic numerical integration problems, and a distinct 
perspective on convergence is required.
In this paper we provide comprehensive convergence theory 
of maximum likelihood estimation in two-level linear and nonlinear mixed models in which the integral defining the marginal likelihood is approximated using numerical integration.
We show that if the maximum likelihood estimate based on the exact likelihood converges in distribution to a random variable, then so does the maximum likelihood estimate based on an approximate likelihood computed using adaptive numerical integration, as long as the order of the accuracy of the integration rule is chosen high enough.
We further show the negative result that non-adaptive integration rules cannot yield consistent maximum likelihood estimators in any statistical model in which the normalized likelihood function concentrates around its mode.
The conclusion is that non-adaptive numerical integration should not be used for approximate likelihood inference in any statistical model, while in mixed models adaptive integration
can always be made to provide inferences asymptotically indistinguishable from inferences based on the exact likelihood by choosing an accurate enough integration rule.

In contrast to previous work \citep{laplacemaxlikglmm,jiang21glmm,aghqmle}, rather than focussing on a particular class of response and random effect distributions, in the present work we devise a set of explicit and verifiable regularity conditions about these distributions under which our convergence results hold. We then verify that these conditions hold for well-specified generalized linear models, i.e. those having exponential family response distributions and multivariate Gaussian random effects.
This makes our convergence theory general and also generalizable: in order to apply our theorems to a particular model not explicitly treated in the present work, all that must be done is to verify that the conditions hold.
By explicitly verifying our conditions for a particular class of models we ensure that the conditions are reasonable, and not so strong as to rule out potentially interesting models.

\subsection{Adaptive numerical integration}
\noindent In ``regular'' statistical models, 
suitably normalized likelihood functions concentrate
around their modes as more data are obtained.
This concentration behaviour is what enables 
consistent inference and quantification of uncertainty using classical asymptotic normality results, e.g. \citet[Ch. 5, 7]{vandervaart}.
Failure to account for this concentration behaviour
can lead to failure of standard numerical integration 
techniques when applied to statistical problems. 
This phenomenon has long been
observed to occur when using Gaussian quadrature to fit mixed models, e.g. \citet{numpoints},
and is discussed by others \citep{aghqmle}.
However, the failure of numerically accurate integration
methods in simple statistical problems has not been explained
theoretically. 
In \cref{fact:nonconvergence} in \cref{subsec:nonconvergence} we prove that any likelihood that concentrates around its mode (at any rate) as more data are obtained cannot lead to consistent maximum likelihood estimates when approximated using a fixed quadrature rule.

A standard solution to this problem is
to adapt the quadrature rule to the location and curvature
of the likelihood by shifting and scaling the points and weights in a data-dependent manner.
In the statistical literature this is referred to as \emph{adaptive quadrature} (AQ)
at least as far back as \citet{nayloradaptive}, although this term conflicts somewhat with the numerical analysis literature which ascribes a different meaning to ``adaptive''. 
Adaptive quadrature rules have enjoyed
common use in fitting generalized linear and nonlinear mixed models \citep{approximationsnlme} and are widely
available in common software including \texttt{R} (package \texttt{lme4}, \citealt{lme4}; package \texttt{GLMMadaptive}, \citealt{glmmadaptive}),
\texttt{SAS} (\texttt{PROC NLMIXED}),
\texttt{STATA} (function \texttt{gllamm}). 
However, convergence theory for these methods is limited. 
In \cref{sec:mainresults} we prove that
the approximate maximum likelihood estimator obtained
by maximizing an adaptive quadrature approximation to 
the marginal likelihood in a mixed model attains higher-order asymptotic accuracy (\cref{fact:likelihood}), leading to a consistent approximate maximum likelihood estimator (\cref{lem:consistency}) 
having the same large-sample statistical properties as the
maximum of the exact likelihood if enough quadrature points
are used (\cref{fact:consistency}). 
We prove that these results hold under very general yet precise and verifiable regularity conditions (\cref{assn:kderiv,assn:hessian,assn:limsup,assn:limsup-out,assn:consistency,assn:prior} in \cref{sec:uniform-assumptions}) which
we formally verify for generalized linear mixed models with
exponential family responses and multivariate Gaussian random effects (\cref{prop:glmmregularity} in \cref{sec:examples}).
Simulations in \cref{subsec:empirical} illustrate the 
practical implications of the theory. 

\subsection{Technical approach and related work}\label{subsec:relatedwork}

\noindent Asymptotic convergence results for exact maximum likelihood estimators (disregarding integration error) have recently been studied for mixed models; see \citet{mleconvergence,jiang21glmm,bhaskaran2023dispersion,maestrini2024second}.
Our \cref{fact:consistency} complements these novel results by including the error incurred by the approximate integration required to implement maximum likelihood estimation in these models, rendering them directly applicable to practice; see \cref{cor:glmmerrorrate} in \cref{subsec:mlerates}.

The Laplace approximation (quadrature with a single point) is the default estimation method in the popular \texttt{lme4} software package \citep{lme4} and is closely related to the penalized quasi-likelihood method \citep{pql}.
\citet{laplacemaxlikglmm} studies the rate of convergence of Laplace-approximate maximum marginal likelihood estimators in Gaussian nonlinear mixed effects models. 
Our \cref{fact:consistency}: (a) recovers the rate of consistency of \citet{laplacemaxlikglmm} up to a small constant; (b) applies to a much broader class of models and likelihood approximations; and (c)
extends it in combination with \citet{jiang21glmm} to include
asymptotic normality in addition to consistency.
\citet{approximatelikelihood} studies the much more general
problem of obtaining the rate of consistency and the asymptotic normality of an approximate maximum likelihood 
estimator obtained by maximizing \emph{any} approximate
likelihood. 
Their theory is proved under the very strong conditions that two derivatives of the approximate log-likelihood converge uniformly in probability to the corresponding likelihood derivatives at a specific rate.
\citet{approximatelikelihood} applies this theory to Bernoulli
generalized linear mixed models fit by Laplace-approximate marginal likelihood.
However, the uniform convergence of derivatives is not
formally established, and this appears highly nontrivial.
We provide results for mixed models fit by adaptive Gaussian quadrature with any number of quadrature points,
under conditions on the model
and data-generating distribution which we formally verify for
generalized linear mixed models with any exponential family distribution.
Our rates recover those of \citet{approximatelikelihood} again up to a small constant factor.

Beyond a single quadrature point, a deterministic rate of convergence for \adaptiveQuadrature{} 
was first derived by \citet{adaptive_GH_1994} and later
corrected by \citet{adaptive}.
\citet{aghqmle} studies stochastic convergence of the approximate maximum marginal likelihood estimator using
\adaptiveQuadrature{} in a class of latent variable models
with exponential family response that are included in the class of mixed models we consider in the present paper;
see \cref{sec:mixedmodelsprelims} for details.
Like \citet{approximatelikelihood}, they assume uniform convergence in probability of two derivatives of the likelihood approximation.
They further assume the highly non-trivial result of uniform convergence in probability of the likelihood approximation over an unspecified region in the parameter space.
We formally prove a precise version of this assumption in \cref{fact:likelihood}, 
and this yields a rate of convergence which is fundamental to the subsequent rate of consistency we derive in \cref{fact:consistency}.
Therein, we clarify the important detail that such uniform convergence only occurs in a suitably
shrinking region in the parameter space.
Again, our rates recover those of \citet{aghqmle} up to a small constant factor.

The small additional constant factor mentioned in the previous paragraphs is due to our uniform regularity conditions.
This term facilitates verifying that the conditions actually hold for well-specified exponential family models, specifically through \cref{lem:exponentialfamilyderivatives,lemma:uniform-markov} in \cref{subsec:proof-expfam}. Previous authors do not include this term in their conditions and hence it does not show up in their convergence rates, but they also do not verify that their conditions hold. It is not clear how to verify these conditions without this small additional factor.
We note that there is no apparent practical difference between our slightly looser rates and those reported by previous authors.
A similar $\eps$ factor appears in the error rate of Quasi-Monte Carlo integration \citep{owen2019monte}. 
This factor is not expected to be practically impactful for quadrature since the number of quadrature points can be increased in practice to obtain a faster rate if desired.

Our results depend directly on the recent work of \citet{bilodeau2021stochastic},
who gave the first 
stochastic rate of convergence for the related task of Bayesian posterior normalization under model-based regularity conditions.
Their result may be applied to our present situation pointwise, for any fixed parameter value.
A primary technical contribution of the present work is to upgrade their result to hold uniformly across 
parameter values and numbers of groups, as is required to assess the convergence of the maximum likelihood estimator in mixed models.
This requires: (a) suitably upgraded regularity conditions, which we formally
verify for generalized linear mixed models with multivariate Gaussian random effects; and (b)
an appropriate modification of the entire proof of the main result of \citet{bilodeau2021stochastic} to hold
under these new conditions at a slightly looser rate.

A further technical challenge addressed in the present paper is that uniform convergence of the approximate likelihood cannot occur in a \emph{fixed} neighbourhood of the true parameter value.
This is because the adaptation of the quadrature points is centred around the mode of the latent variables, which itself must converge in probability to some point,
and this convergence can only be made to occur in a \emph{shrinking} neighbourhood of the true parameter value. This subtlety has not been treated by previous authors, who often implicitly assume a form of uniform convergence over too large of a space. 
This is too strong of an assumption to verify, and indeed has not been previously verified.
The primary technical challenge that we solve in order to achieve the required uniform convergence is to identify the precise region in the parameter space in
which this convergence holds, 
and the rate at which it should be made to shrink as the number of data increase. 
This requires a balanced approach, as follows.
We first apply the technical lemmas of \citet{bilodeau2021stochastic} to provide a formal proof of the rate of uniform convergence in probability of the likelihood approximation (\cref{fact:likelihood}) at a rate depending on the order of the quadrature rule, on an appropriately shrinking neighbourhood.
We then provide a proof of consistency of
the approximate maximum likelihood estimator (\cref{lem:consistency}) which
guarantees that it is eventually in this shrinking neighbourhood.
The main difficulty of the proof involves balancing these rates, where uniform consistency is easier to show for a faster shrinking ball, but it is harder to show that the approximate maximum likelihood estimator remains in a faster shrinking neighbourhood.
We then combine these two results and our regularity conditions to 
prove that the distance between the exact and approximate maximum likelihood estimators converges faster than does the exact estimator to its limit, on the shrinking neighbourhood (\cref{fact:consistency}).
The result is a general and self-contained convergence theory for approximate maximum likelihood estimation for two-level mixed models.


\section{Preliminaries}

\subsection{Two-level mixed models}\label{sec:mixedmodelsprelims}

\noindent We study two-level mixed models for repeated measurements data consisting of a total number of observations $\numobs \in \Nats$, 
number of groups $\numgroups \in [\numobs] = \{1,\ldots,\numobs\}$, and number of observations per group $(\numpergroup_i)_{i\in[\numgroups]}$ satisfying $\numobs = \sum_{i=1}^{\numgroups}\numpergroup_{i}$.
For $i\in[\numgroups]$, let $\obsall_{i} = (\obsi_{ij})_{j\in[\numpergroup_{i}]}$ denote the group's observations, where each $\obsi_{ij} \in \Reals$. 
Let $\trueprob{\numobs}$ denote the true, unknown joint distribution of $\obsall = (\obsall_{1}\Tr,\ldots,\obsall_{\numgroups}\Tr)\Tr$.
In addition, for each $i\in[\numgroups]$ and $j\in[\numpergroup_i]$, there are observed covariates 
$\cov_{ij} \in \Reals^{\covdim}$ 
which may be fixed or random; they are implicitly conditioned upon for the remainder of the paper.

A \genmodelname{} is determined by
a random effect distribution $\redist$ with known mean,
parametrized by variance parameters $\resd\in\resdparamspace \subseteq \Reals^\resddim$;
a conditional response distribution $\responsedist$ 
parametrized by regression coefficients $\regparam\in\regparamspace \subseteq \Reals^\regparamdim$;
and possibly additional dispersion parameters $\obssd\in\dispparamspace\subseteq\Reals^\dispparamdim$
such that:
\[\label{eqn:glmmdefinition}
	\obs_{ij} \setdelim \cov_{ij}, \re_i &\indsim \responsedist(\cov_{ij},\re_i; \regparam,\obssd), \\
	\re_i &\indsim \redist(\resd).
\]
The distinction between regression and variance parameters is somewhat arbitrary in the definition, and is made only for convenience.
In practice, different parameters will have different convergence rates on a model-specific basis; see \cref{cor:glmmerrorrate} for an example from generalized linear mixed models.

Model (\ref{eqn:glmmdefinition}) is quite general, and restrictions yield more familiar models. 
For example, a generalized linear mixed model (\glmm{}; \citealt{pql}) is obtained by constraining $\responsedist$ to be exponential family with canonical parameter linear in $\re_i$ and $\regparam$ and constraining $\redist$ to be Gaussian; see \cref{sec:examples}.
The latent variable models analyzed by \citet{aghqmle}
are obtained by having the canonical parameter depend linearly on $\regparam\Tr\re$
and $\redist$ to be a \emph{spherical} Gaussian, depending
on no further variance parameters. 
A nonlinear mixed model (NLMM; \citealt{approximationsnlme}) is obtained by constraining $\responsedist$ to be Gaussian with mean given by a known nonlinear function of $\cov$, $\regparam$, and $\re$,
and constraining $\redist$ to be Gaussian.
The regularity conditions for these distributions under which the convergence of approximate likelihoods holds are stated in \cref{sec:uniform-assumptions}.
\cref{prop:glmmregularity} in \cref{sec:examples} demonstrates that these regularity conditions are weaker than requiring the response to belong to the exponential family or the random effects to be Gaussian, by demonstrating
that they are satisfied by exponential family models with correlated multivariate Gaussian random effects.



We denote the model for the joint density of the group's responses and random effects by $\jointdens(\obsall_{i}, \re_i;\params) = \prod_{j=1}^{\numpergroup_{i}}\responsedens(\obsi_{ij} \setdelim \re_{i};\regparam, \obssd)\redens(\re_{i};\resd)$,
where $\responsedens$ and $\redens$ are the densities of $\responsedist$ and $\redist$.
Inferences about the unknown parameters $\params=(\regparam,\obssd,\resd)$ are based on the marginal likelihood,
\[\label{eqn:marglik}
	\margdens(\obsall; \params)
	= \prod_{i=1}^{\numgroups}\margdens(\obsall_{i}; \params)
	= \prod_{i=1}^{\numgroups} \int\jointdens(\obsall_{i}, \re_i;\params)\dee\re_i,
\]
where the first equality follows from the independence of $\re_1,\dots,\re_\numgroups$.
With respect to the latent variables, the joint distribution of data and random effects may be regarded as \emph{high-dimensional} because $\text{dim}(\re) = \redim\numgroups$ and $\numgroups\to\infty$ as more data are obtained.
Numerical integration typically incurs a computational cost that is exponential in the dimension of the integrand, and hence general high-dimensional
numerical integration is infeasible unless some structure of the integral is exploited to develop a more efficient approximation.
In two-level mixed models, the $\redim\numgroups$-dimensional integral defining the marginal likelihood factors into a product of $\numgroups$ integrals of dimension $\redim$, where $\redim = \text{dim}(\re_i)$ is typically quite small, enabling the use of accurate low-dimensional quadrature techniques within what is nominally a high-dimensional problem.
Inferences are therefore based on an approximation to $\margdens(\obsall; \params) $ obtained by approximating each integral $\margdens(\obsall_{i}; \params)$ and then taking the product of the approximations.
As the number of data increase, so does the number of approximate integrals being multiplied, and the error in the approximate likelihood will grow.
The accuracy of the approximation will therefore affect the statistical properties of inferences based on the resulting approximate marginal
likelihood, with larger data requiring integration rules that attain higher accuracy.
A reviewer points out that this phenomenon also occurs when using Monte Carlo integration methods, which may require a larger number of samples to attain
the same accuracy on a larger set of data.
This paper focusses on the statistical properties of inferences based on quadrature/cubature approximations to this marginal likelihood.


\subsection{Fixed quadrature}\label{ssec:quadrature}

\noindent A \emph{quadrature rule} $\quadrule{\quadpointset}{\weight}$ to approximate integral of the form in \cref{eqn:marglik} is a collection of \emph{points} $\quadpointset \subseteq \Reals^\redim$ and a \emph{weight function} $\weight: \quadpointset \to \Reals$, and is denoted by $\quadrule{\quadpointset}{\weight}$. Given such a rule, an approximate marginal likelihood is: 
\begin{equation}\label{eqn:nonadaptiveml}
	\approxmargdens(\obsall_{i}; \params) = \sum_{\quadpointvec\in\quadpointset}\jointdens\left(\obsall_{i},\quadpointvec;\params\right)\weight(\quadpointvec).
\end{equation}
We call any such approximation $\quadrule{\quadpointset}{\weight}$ for which the points $\quadpointvec\in\quadpointset$ and weights $\weight$ do not
depend on the data to be a \emph{non-adaptive} or \emph{fixed} quadrature rule.
We focus on quadrature rules with the following exact integration property.
\begin{definition}[Definition~1 of \citet{bilodeau2021stochastic}]\label{def:exact-integrate-property}
For any $\quadnum,\redim\in\Nats$, a quadrature rule $\quadrule{\quadpointset}{\weight}$ satisfies $\property{\quadnum}{\redim}$ if for all $\redim$-dimensional real-valued polynomials $\polyfunc$ of total order $2\quadnum-1$ or less,
\[\label{eqn:exact-integration}
	\int_{\Reals^{\redim}} \normaldens{\re}{0}{\identmat_\redim} \polyfunc(\re) \dee \re = \sum_{\quadpointvec \in \quadpointset} \normaldens{\quadpointvec}{0}{\identmat_\redim} \polyfunc(\quadpointvec) \weight(\quadpointvec),
\]
where $\normaldens{\re}{0}{\identmat_\redim}$ is the standard $\redim$-dimensional Gaussian density.
\end{definition}
The Gauss--Hermite quadrature rule satisfies $\property{\quadnum}{1}$
and the multi-dimensional product rule extension $\quadrule{\quadpointset^\redim}{\weight_1\cdots\weight_\redim}$ satisfies $\property{\quadnum}{\redim}$ for $\redim>1$; see \citet{bilodeau2021stochastic} for discussion of other rules and extensions to multiple dimensions.

Classical convergence analysis (e.g. \citealt{numint}) predicts small error when specific quadrature rules are applied to \emph{deterministic} functions satisfying specific properties.
Likelihood functions are random, and such analysis predicts that accurate results may be obtained if a particular data set leads to a realized likelihood function which is well-behaved in this sense.
However, statistical convergence theory requires a probabilistic analysis of the error incurred under assumptions on the data-generating distribution and model.
In this paper we show that no fixed quadrature rule---no matter how accurate---can yield an approximate marginal likelihood with the correct statistical properties; see \cref{fact:nonconvergence} in \cref{sec:mainresults}.
When used in statistical problems, quadrature rules must be adapted to the data.

\subsection{Adaptive quadrature}\label{prelims:aghq}


\noindent Bernstein-von Mises theory \citep[Section 10.2]{vandervaart} states that when appropriately normalized, likelihood functions satisfying weak conditions 
concentrate around their modes as more data are obtained.
Approximating integrals involving likelihood functions
(such as \cref{eqn:marglik})
using fixed quadrature rules ignores this physical behaviour
and results in a quadrature rule that misses most of the mass of
the integrand as more data are obtained; see \cref{fact:nonconvergence}
for the formal statement.
This phenomenon is the source of the empirical lack of accuracy that has frequently been observed when using GQ to fit
mixed models (e.g. \citealt{numpoints}).
Consequently, \emph{adaptive quadrature} (\AGHQ{}) is popular for fitting mixed models. 
In this paper we show that this method does have desirable convergence properties for mixed models; see \cref{fact:consistency} in \cref{subsec:consistency}.

For each $i\in[\numgroups]$ and $\params\in\paramspace$,
let $\llhoodip(\re) = \log\jointdens(\obsall_{i}, \re;\params)$, $\condmodeip = \text{argmax}_{\re\in\Reals^{\redim}}\llhoodip(\re)$, and $\hessip = -\partial^{2}_{\re}\llhoodip(\condmodeip)$. Define $\cholip$ to be the lower Cholesky triangle satisfying $\hessipinv = \cholip\cholipTr$.
For a given quadrature rule $\quadrule{\quadpointset}{\weight}$ that satisfies $\property{\quadnum}{\redim}$, the \AGHQ{} approximation to $\margdens(\obsall_{i}; \params)$ is:
\[\label{eqn:aqapprox}
	\approxmargdensAQ(\obsall_{i}; \params) = \detbar{\cholip}\sum_{\quadpointvec\in\quadpointset}\jointdens\left(\obsall_{i}, \cholip\quadpointvec + \condmodeip;\params\right)\weight(\quadpointvec).
\]
The corresponding likelihood approximation is $\approxmargdensAQ(\obsall; \params) = \prod_{i=1}^{\numgroups}\approxmargdensAQ(\obsall_{i}; \params)$. When $\quadnum=1$ and $\quadrule{\quadpointset}{\weight}$ is Gauss-Hermite quadrature, $\approxmargdensAQ(\obsall; \params)$ is called a Laplace approximation.
Adaptive quadrature incurs the computational cost of fixed quadrature along with the cost of obtaining the mode and Hessian of the log-integrand; the former cost usually dominates the latter in practice.


\section{Theoretical Guarantees}\label{sec:mainresults}

\subsection{Non-convergence of fixed quadrature approximations to likelihood functions}\label{subsec:nonconvergence}

\noindent We provide the first theoretical explanation for the empirical lack of accuracy of non-adaptive quadrature rules in statistical problems
noted by previous authors \citep{numpoints,aghqmle}.
\cref{fact:nonconvergence} states that any non-adaptive quadrature rule cannot yield an asymptotically
convergent likelihood approximation, as long as the likelihood admits weak concentration properties.
\cref{cor:mixedmodelsnonconverge} then applies \cref{fact:nonconvergence} to mixed models (\cref{eqn:glmmdefinition}).

Consider a latent variable model which assumes that
the density of random variable $\responseall$ is given by $\jointdens(\obsall) = \int\jointdens(\obsall,\re)\dee\re$ for some joint density $\jointdens(\obsall,\re)$.
This is identical to the Bayesian setup within which \citet{bilodeau2021stochastic} argued that adapting a quadrature rule as described in \cref{prelims:aghq} is \emph{sufficient} to achieve fast asymptotic convergence of the approximation 
as $\numobs\to\infty$.
\cref{fact:nonconvergence} provides a type of converse to \citet[Theorem 1]{bilodeau2021stochastic}: adapting the quadrature rule to the data---somehow---is \emph{necessary} to achieve asymptotic convergence.

\begin{theorem}\label{fact:nonconvergence}
Fix any quadrature rule $\quadrule{\quadpointset}{\weight}$ where $\quadpointset \subseteq \Reals^\redim$ and $\weight: \quadpointset \to \Reals$.
Let the random variable $\responseall\in\Reals^\numobs$ have distribution $\trueprob{\numobs}$.
Let the density of $\responseall$ under the model be $\margdens(\obsall) = \int\jointdens(\obsall,\re)\dee\re$,
and let 
$$
\approxmargdens(\obsall) = \sum_{\quadpointvec\in\quadpointset}\weight(\quadpointvec)\jointdens(\obsall, \quadpointvec).
$$
Denote by $\pto$ convergence in probability with respect to $\trueprob{\numobs}$.
If there exists $\retrue$ such that $\jointdens(\retrue | \obsall)\pto\infty$ and $\jointdens(\re | \obsall)\pto0$ for all $\re\neq\retrue$,
then there exist $\kappa > 0$ and $\gamma \in (0,1)$ such that
$$
\lim_{\numobs\to\infty}\trueprob{\numobs}\left\{ \abs{\frac{\approxmargdens(\obsall)}{\margdens(\obsall)} - 1} > \kappa\right\} > \gamma.
$$
\end{theorem}

\begin{proof}
Observe that
$$
\frac{\approxmargdens(\obsall)}{\margdens(\obsall)} = \sum_{\quadpointvec\in\quadpointset}\weight(\quadpointvec)\jointdens(\quadpointvec | \obsall),
$$
where $\jointdens(\quadpointvec | \obsall)$ is the posterior density of $\re$ evaluated at $\quadpointvec$.
We can therefore write
\begin{equation}\label{eqn:fracbound}
\weightmin\times\max_{\quadpointvec\in\quadpointset}\jointdens(\quadpointvec | \obsall)
\leq \frac{\approxmargdens(\obsall)}{\margdens(\obsall)}
\leq \abs{\quadpointset}\weightmax\times\max_{\quadpointvec\in\quadpointset}\jointdens(\quadpointvec | \obsall),
\end{equation}
where $\weightmin = \min_{\quadpointvec\in\quadpointset}\weight(\quadpointvec)$ and 
$\weightmax = \max_{\quadpointvec\in\quadpointset}\weight(\quadpointvec)$.

There are two cases to consider. 
Suppose first that $\retrue\in\quadpointset$. 
Then by \cref{eqn:fracbound} and the assumption of the theorem, $\approxmargdens(\obsall) / \margdens(\obsall) \pto\infty$.
We therefore may choose $\epsilon>0, \gamma\in(0,1)$ such that there must exist $n\in\Nats$ such that for every $\numobs>n$,
$$
\trueprob{\numobs}\left\{ \frac{\approxmargdens(\obsall)}{\margdens(\obsall)} > \eps + 1\right\} > \gamma.
$$
Set $\kappa = \epsilon > 0$ and note that $\approxmargdens(\obsall) > \margdens(\obsall)$ eventually to yield the result.
Suppose next that $\retrue\notin\quadpointset$. 
Then by \cref{eqn:fracbound}, $\approxmargdens(\obsall) / \margdens(\obsall) \pto0$.
Choose $\epsilon, \gamma\in(0,1)$ such that there must exist $n\in\Nats$ such that for every $\numobs>n$,
\*[
\gamma < \trueprob{\numobs}\left\{ \frac{\approxmargdens(\obsall)}{\margdens(\obsall)} < \eps\right\}
= \trueprob{\numobs}\left\{ 1 - \frac{\approxmargdens(\obsall)}{\margdens(\obsall)} > 1 - \eps\right\}
= \trueprob{\numobs}\left\{ \abs{\frac{\approxmargdens(\obsall)}{\margdens(\obsall)} - 1 } > 1 - \eps\right\},
\]
where the last step uses that $\approxmargdens(\obsall) < \margdens(\obsall)$ eventually.
Set $\kappa = 1-\epsilon > 0$ to yield the result.
\end{proof}

The most obvious way to guarantee the conditions of \cref{fact:nonconvergence} is for the model to satisfy a Bernstein-von Mises theorem \citep[Section 10.2]{vandervaart}.
For a very broad class of misspecified models, \citet{misspec} show that a Bernstein-von Mises-type result holds, suggesting that \cref{fact:nonconvergence}
is widely applicable and that its conclusions apply to many models used in practice.

Returning focus to the mixed models which are the subject of the present paper, \cref{cor:mixedmodelsnonconverge} specializes \cref{fact:nonconvergence} to mixed models of the form given in \cref{eqn:glmmdefinition}, with ``true'' parameter value $\paramstrue$ (see \cref{sec:uniform-assumptions} for the precise definition).

\begin{corollary}\label{cor:mixedmodelsnonconverge}
Consider the model given by \cref{eqn:glmmdefinition}.
Under \cref{assn:kderiv,assn:hessian,assn:limsup,assn:limsup-out,assn:consistency,assn:prior} in \cref{sec:uniform-assumptions} and
for $\paramstrue$ defined therein, there exists $\kappa > 0$ and $\gamma\in(0,1)$ such that
$$
\lim_{\numobs\to\infty}\trueprob{\numobs}\left\{ \abs{\frac{\approxmargdens(\obsall;\paramstrue)}{\margdens(\obsall;\paramstrue)} - 1} > \kappa\right\} > \gamma.
$$
\end{corollary}
\begin{proof}
Restricting attention to $\params = \paramstrue$ reduces the problem to exactly that considered by \citet{bilodeau2021stochastic},
and our \cref{assn:kderiv,assn:hessian,assn:limsup,assn:consistency,assn:prior} reduce to their Assumptions 1 -- 5.
In their Remark 5 they show that these assumptions imply that the Bernstein-von Mises theorem holds for $\jointdens(\obsall,\re;\paramstrue)$;
this in turn implies the conditions of \cref{fact:nonconvergence}.
\end{proof}
While \cref{cor:mixedmodelsnonconverge} only applies to the single parameter value $\paramstrue$ and only states that the error cannot reach zero
(as opposed to, say, diverging to $\infty$), it is nonetheless sufficient to rule out inferences based on $\approxmargdens(\obsall;\params)$ for most mixed models used in practice.
In most cases the error of the approximation will depend on $\params$, therefore it is not guaranteed that the approximated integrated likelihood maintains its shape locally around the mode and consequently confidence intervals constructed using the local curvature or the likelihood drop may be unreliable.


\subsection{Approximation Error for Adaptive Quadrature}
\label{subsec:approx-error}

\noindent Likelihood approximations based on \emph{adaptive}
quadrature do converge. \cref{fact:likelihood} quantifies
the rate of convergence for adaptive quadrature approximations to the marginal likelihood in mixed models.
This intermediate technical result is required to prove convergence of the approximate maximum likelihood estimator (\cref{fact:consistency}).
A similar result is assumed by \citet{aghqmle} although they do not specify the region of $\paramspace$ in which the uniform convergence occurs, 
and a stronger result about uniform convergence of derivatives of the approximate log-likelihood 
is required by \citet{approximatelikelihood}.
Our proof is self-contained, and makes use of suitably upgraded technical lemmas recently provided by \citet{bilodeau2021stochastic}.
A slight loosening of the usual error rates compared to results obtained in \citet{aghqmle} and \citet{approximatelikelihood} is required for uniformity of the approximation error to hold. 
Given that the uniformity is assumed and not show in these previous works, it is possible that their rates are too optimistic for the mixed models considered at present.

We define $\paramradiusi = \numpergroup_i^{-\alpha}$ for $0 < \alpha < 1/4$, and $\paramradius = (\min_{i=1,\ldots,\numgroups}\numpergroup_i)^\alpha$.
We let the number of groups $\numgroups = \minpergroup^q$ for some $q > 0$, so that as $\numobs \rightarrow \infty$, $\numgroups \rightarrow \infty$ as well.
The radius $\paramradius$ will define the shrinking region in the parameter space in which all our statements about uniform convergence hold; the precise rate of shrinkage $\alpha$ is
chosen to balance the concentration of the likelihood with the convergence to zero of the integration error.
We also define a fixed neighbourhood of arbitrary radius $\universalradius>0$, and a point $\paramstrue\in\paramspace$ around which the likelihood concentrates.
This may be intuitively thought of as a ``true'' value of $\params$, and under weak conditions will be the point that maximizes the expected log-likelihood; we emphasize that at no point do we assume the model is correctly specified in the sense that $\trueprob{\numobs}$
is recovered by $\dist(\params;\obsall)$ for any $\params\in\paramspace$.

\cref{fact:likelihood} upgrades the main result of \citet{bilodeau2021stochastic} to hold uniformly at a slightly loosened rate.

\begin{lemma}\label{fact:likelihood}
Fix $\quadnum\in\Nats$, let 
$\approxmargdensAQ(\obsall; \params)$ be the approximation of \cref{eqn:aqapprox} with a quadrature rule satisfying $\property{\quadnum}{\redim}$ and $\numgroups = \minpergroup^q$ for any $q> 0$.
Then, under \cref{assn:prior,assn:hessian,assn:consistency,assn:kderiv,assn:limsup,assn:limsup-out} in \cref{sec:uniform-assumptions} 
there exists $\const>0$ not depending on $\numgroups$ such that
\*[
	\lim_{\minpergroup\to\infty}\trueprob{\numobs}\left(\sup_{\params\in\paramsbigball} \abs{\log\approxmargdensAQ(\obsall; \params) - \log\margdens(\obsall; \params)} < \const \, \sum_{i=1}^\numgroups \numpergroup_i^{-\lfloor(\quadnum+2)/3\rfloor + \eps}\right) = 1,
\]
for every $\eps > 0$.
\end{lemma}
\begin{proof}
	See \cref{sec:proofs-likelihood}
\end{proof}

In general, the constant $\const$ depends on the size of the higher-order derivatives of the likelihood function in a neighbourhood around the data-generating parameter and an error term introduced by a truncation argument used in \cite{bilodeau2021stochastic}.
However, with \cref{assn:prior,assn:hessian,assn:consistency,assn:kderiv,assn:limsup,assn:limsup-out} we assume that the behaviour of the likelihood function is uniform across all groups, even as $m$ increases, which results in an bound which is independent of $i$ in \cref{fact:likelihood}.
This uniformity does worsen the error bound by some sub-polynomial factor, i.e. it grows slower than any polynomial function in $\numpergroup_i$, which results in the inclusion of the $\eps$ factor in \cref{fact:likelihood}.




\subsection{Convergence of the Approximate Maximum Marginal Likelihood Estimator}\label{subsec:consistency}

\noindent Let $\paramsmle = \argmax_{\params\in\paramspace} \log\margdens(\obsall; \params)$ and $\approxparamsmle = \argmax_{\params\in\paramspace} \log\approxmargdens(\obsall; \params)$.
The ``exact'' MLE $\paramsmle$ exists in theory and \citet{jiang21glmm} provide rigorous asymptotic convergence theory for it as an estimator of $\params$.
If $\paramsmle$ could be computed in practice, then it would be used for
inference about $\params$. However, $\paramsmle$ is intractable because $\margdens(\obsall; \params)$ is. In practice, inferences are based
on $\approxparamsmle$, and it is the statistical properties of this
approximate maximum likelihood estimator that are of direct interest to practice.
In \cref{fact:consistency} we quantify the additional error incurred by 
maximizing an approximate likelihood instead of the true likelihood.
The proof technique is to use \cref{fact:likelihood} to bound the 
distance between $\paramsmle$ and $\approxparamsmle$, and then
use this to express the asymptotic behaviour of $\approxparamsmle$
in terms of that of both $\paramsmle$ and $\approxmargdens(\obsall; \params)$
as provided by \citet{jiang21glmm} and \cref{fact:likelihood}, respectively.

Computing $\approxparamsmle$ must generally be done using numerical optimization.
While this is not the focus of the present analysis, we do require the mild assumption that the optimization algorithm return a finite answer
with probability tending to $1$ as more data are obtained.
We formalize this in \cref{assn:mle-estimate}:

\begin{assumption}\label{assn:mle-estimate}
$\lim_{\minpergroup\to\infty}\trueprob{\numobs}[\approxparamsmle \in \paramsbigballfixed] = 1$ for the $\delta$ chosen in \cref{assn:consistency,assn:hessian,assn:kderiv,assn:limsup,assn:limsup-out,assn:prior} and for $\numgroups = \minpergroup^q$ for some $q > 0$.
\end{assumption}
We note that \cref{assn:mle-estimate} is so mild as to have been made implicitly in all previous
work on this topic: it is required in order to invoke any regularity
conditions on the likelihood evaluated at $\params = \approxparamsmle$.
It is the only assumption that we do not formally verify in \cref{sec:examples} for exponential family models. 
If it does not hold,
then there is some non-zero probability that no inferences can be made at all as more data are obtained, and this would preclude further convergence analysis.

\cref{assn:consistency,assn:hessian,assn:kderiv,assn:mle-estimate,assn:prior} are sufficient to show the much
stronger result that $\approxparamsmle$ is consistent for $\params$. This is formalized in \cref{lem:consistency}:
\begin{lemma}\label{lem:consistency}
	Under \cref*{assn:consistency,assn:hessian,assn:kderiv,assn:mle-estimate,assn:prior}, 
	$$
	\lim_{\minpergroup\to\infty}\trueprob{\numobs}[\approxparamsmle \in \paramsbigball] = 1,
	$$
	where $\zeta_N = \minpergroup^{-\alpha}$ for $0<\alpha <1/4$ and $\numgroups = \minpergroup^q$ for some $q > 0$.
\end{lemma}

\begin{proof}
	Note:
	\*[
		\approxmargdensAQ(\obsall_{i}; \params) = \detbar{\cholip}\sum_{\quadpointvec\in\quadpointset}\jointdens\left(\obsall_{i}, \cholip\quadpointvec + \condmodeip;\params\right)\weight(\quadpointvec),
	\] 
	and by definition of maxima and the positivity of the likelihood function:
	\*[
		\detbar{\cholip}\sum_{\quadpointvec\in\quadpointset}\jointdens\left(\obsall_{i}, \cholip\quadpointvec + \condmodeip;\params\right)\weight(\quadpointvec) &\leq \max_{\quadpointvec\in\quadpointset} \weight(\quadpointvec) \detbar{\cholip} \detbar{\quadpointset} \jointdens\left(\obsall_{i}, \condmodeip;\params\right), \\
		\detbar{\bm{L}^{\paramstrue}_i } \sum_{\quadpointvec\in\quadpointset}\jointdens\left(\obsall_{i}, \bm{L}^{\paramstrue}_i \quadpointvec + \hat\re_i^{\paramstrue} ;\params\right)\weight(\quadpointvec) &\geq \weight(0) \detbar{\bm{L}^{\paramstrue}_i } \jointdens\left(\obsall_{i}, \hat\re_i^{\paramstrue} ;\params\right).
	\]
 	Using the above inequalities, we show the difference of the logarithms of the approximate marginal likelihood for $\params \in \paramsbigballfixed \cap \paramsbigball^C$ is negative, implying the maxima lies in $\paramsbigball$.
	Fix $\params \in \paramsbigballfixed \cap \paramsbigball^C$.
	Consider the following upper bound:
	\*[
		&\log(\approxmargdensAQ(\obsall_{i}; \params)) - \log(\approxmargdensAQ(\obsall_{i}; \paramstrue)) \\
		&\leq  \underbrace{ \left\{ \log(\detbar{\cholip}) - \log(\detbar{\bm{L}^{\paramstrue}_i })\right\}}_{\mathcal{A}_1(\params)} +  \underbrace{ \left\{\log( \detbar{\quadpointset}  \max_{\quadpointvec\in\quadpointset} \weight(\quadpointvec)) - \log(\weight(0)) \right\}}_{\mathcal{A}_2(\params)} \\
		&+  \underbrace{ \left\{\log\left(\jointdens\left(\obsall_{i}, \condmodeip;\params\right)\right) - \log\left(\jointdens\left(\obsall_{i}, \hat\re_i^{\paramstrue} ;\paramstrue\right) \right)\right\}}_{\mathcal{A}_3(\params)}.
	\]
	By \cref*{assn:limsup-out} the pair $(\params,\condmodeip ) \in \paramsbigballfixed \times \reball$. 
	We now show $\mathcal{A}_3(\params) < 0$ and that $\mathcal{A}_1(\params)$ and $\mathcal{A}_2(\params)$ are comparatively negligible in the limit.
	Observe that $\mathcal{A}_2(\params)$ does not depend on $\numobs$ nor $i$. For $\mathcal{A}_1(\params)$ we have for all $\eps > 0$
	\*[
		\mathcal{A}_1(\params) \leq  \hessbig \paramdim \log(\numpergroup_i)\numpergroup_i^\eps /\hesssmall
	\]
	by \cref{assn:hessian}. For $\mathcal{A}_3(\params)$:
	\*[
		\mathcal{A}_3(\params) = \left\{ \log\left(\jointdens\left(\obsall_{i}, \condmodeip;\params\right)\right)  - \log\left(\jointdens\left(\obsall_{i}, \retruepi ;\paramstrue\right)\right)  + \log\left(\jointdens\left(\obsall_{i}, \retruepi ;\paramstrue\right)\right)  - \log\left(\jointdens\left(\obsall_{i}, \hat\re_i^{\paramstrue} ;\paramstrue\right)\right) \right\},
	\]
	where we added and subtracted the term $\log\left(\jointdens\left(\obsall_{i}, \retruepi ;\paramstrue\right)\right)$.	
	We have
	\*[
		& \frac{1}{\numpergroup_i} \left\{\log\left(\jointdens\left(\obsall_{i}, \retruepi ;\paramstrue\right)\right)  - \log\left(\jointdens\left(\obsall_{i}, \hat\re_i^{\paramstrue} ;\paramstrue\right)\right)\right\} \\
		&= -\frac{1}{\numpergroup_i}(\retruepi - \hat\re_i^{\paramstrue} )^\top \partial^2_{\re} \logjointdens_i^{\params}(\re^\circ)(\retruepi - \hat\re_i^{\paramstrue} )\\
		&\leq \hessbig \numpergroup_i^\eps \norm{\retruepi - \hat\re_i^{\paramstrue}}_2^2 \leq \beta^\prime \hessbig \numpergroup_i^\eps \paramradius^2, 
	\] 
	where $\re^\circ = (1 - t_1)\retruepi + t_1 \hat\re_i^{\paramstrue} $ for some $t_1 \in [0,1]$, and the inequalities hold with probability tending to 1 by \cref{assn:hessian,assn:consistency}. Further,
	\*[
		& \left\{\log\left(\jointdens\left(\obsall_{i}, \condmodeip;\params\right)\right)  - \log\left(\jointdens\left(\obsall_{i}, \retruepi ;\paramstrue\right)\right) \right\} \\
		&= (\params - \paramstrue, \condmodeip -\retruepi)^\top \partial_{(\params, \re)}\log\left(\jointdens\left(\obsall_{i}, \retruepi ;\paramstrue\right)\right) \\
		&+ (\params - \paramstrue, \condmodeip -\retruepi)^\top \partial^2_{(\params, \re)}\log\left(\jointdens\left(\obsall_{i}, \re^\circ ;\params^{\circ}\right)\right)(\params - \paramstrue, \condmodeip -\retruepi),
	\]
	where $\params^{\circ} = (1-t_2)\paramstrue + t_2\params$ for some $t_2\in[0,1]$. For the first term, observe:
	\*[
		&\left| (\params - \paramstrue, \condmodeip -\retruepi)^\top \partial_{(\params, \re)}\log\left(\jointdens\left(\obsall_{i}, \retruepi ;\paramstrue\right)\right) \right| \\
		&\leq \norm{(\params - \paramstrue, \condmodeip -\retruepi)}_2 \norm{\partial_{(\params, \re)}\log\left(\jointdens\left(\obsall_{i}, \retruepi ;\paramstrue\right)\right)}_2\\
		&\leq \universalradius (\paramdim\derivbound \numpergroup_i)^{1/2 + \eps},
	\]
	from \cref{assn:kderiv}. 
	For the second term by \cref{assn:hessian} for any $\eps > 0$
	\*[
		 (\params - \paramstrue, \condmodeip -\retruepi)^\top \partial^2_{(\params, \re)}\log\left(\jointdens\left(\obsall_{i}, \re^\circ ;\params^\circ\right)\right)(\params - \paramstrue, \condmodeip -\retruepi)	
		&\leq - \hesssmall \norm{\params - \paramstrue}_2^2 \leq -\hesssmall\numpergroup_i^{1 - \eps}\paramradius^2,
	\]
	where we recall that we have fixed $\params \in \paramsbigballfixed \cap \paramsbigball^C$
	and hence $\norm{\params - \paramstrue}_2 \geq \paramradius$.

	Finally, recall that for every $i=1,\ldots,\numgroups$, 
	$\paramradius \geq \numpergroup_i^{-\alpha}$ for some $0 < \alpha < 1/4$.
	In this case as $\eps$ is arbitrary we have $\numpergroup_i^{1 - \eps}\paramradius^2 > \numpergroup_i^{1/2 + \eps^\prime}$ for some $\eps^\prime > \eps$, and hence for $\numpergroup_i$ large enough, 
	$$
	\mathcal{A}_1(\params) + \mathcal{A}_2(\params) + \mathcal{A}_3(\params) < \frac{\hessbig \paramdim \numpergroup_i^{\eps} \log\numpergroup_i}{\hesssmall}+ |\mathcal{A}_2(\params)| + \universalradius (\paramdim\derivbound \numpergroup_i)^{1/2 + \eps}-\hesssmall\numpergroup_i^{1/2 + \eps^\prime} < 0,
	$$
	where $\mathcal{A}_2$ is constant in $\numpergroup_i$ and $\params$.
	Summing over $i=1,\ldots,\numgroups$, and noting that by our assumptions these bounds holds uniformly, we have, for $\minpergroup$ large enough,
	$$
	\log(\approxmargdensAQ(\obsall; \params)) - \log(\approxmargdensAQ(\obsall; \paramstrue)) < 0
	$$
	for arbitrary $\params \in \paramsbigballfixed \cap \paramsbigball^C$.
	But by definition of $\approxparamsmle$, we have $\log(\approxmargdensAQ(\obsall; \approxparamsmle)) \geq \log(\approxmargdensAQ(\obsall; \paramstrue))$,
	and hence $\approxparamsmle\in\paramsbigball\cup\paramsbigballfixed^C$.
	However by \cref{assn:mle-estimate}, $\approxparamsmle\in\paramsbigballfixed$, so we conclude that $\approxparamsmle\in\paramsbigball$.
	This completes the proof.
\end{proof}

With \cref{fact:likelihood} and \cref{lem:consistency} available,
we are in a position to fully characterize the statistical properties
of $\approxparamsmle$ in \cref{fact:consistency}.

\begin{theorem}\label{fact:consistency}
	Fix $\quadnum\in\Nats$, let $\minpergroup = \min\{\numpergroup_{1},\dots,\numpergroup_\numgroups\}$ and $\numgroups = \minpergroup^q$ for any $q > 0$.
	Then, under \cref{assn:prior,assn:hessian,assn:consistency,assn:kderiv,assn:limsup,assn:limsup-out,assn:mle-estimate}, if there exists a sequence of vectors  $\mlerate_\numobs \rightarrow \infty$ such that as $\numpergroup_{\min} \rightarrow \infty$,
	\*[
		\mlerate_\numobs\cdot(\paramstrue - \paramsmle) \ed \normalrv + o_p(1),
	\]
	for a random variable $\normalrv$ (where the multiplication is component-wise), then 
	\*[
	\mlerate_\numobs\cdot(\paramstrue - \approxparamsmle) \ed \normalrv + o_p(1) + O_p\left\{\mlerate_\numobs \minpergroup^{-(\lfloor(\quadnum+2)/3 \rfloor + 1)/2 + \eps} \right\},
	\]
	for every $\eps > 0$.
\end{theorem}
\begin{proof}
Define $\pergrouprate = \minpergroup^{-\lfloor(\quadnum+2)/3 \rfloor + \eps}$. For notational convenience, define $\logmargdens(\obsall;\params) = \log\margdens(\obsall;\params)$ and $\approxlogmargdens(\obsall; \params) = \log\approxmargdens(\obsall;\params)$.
By definition of $\approxparamsmle$,
\*[
	\logmargdens(\obsall; \paramsmle)
	- \logmargdens(\obsall; \approxparamsmle)
	&= \logmargdens(\obsall; \paramsmle)
	- \approxlogmargdens(\obsall; \paramsmle)
	+ \approxlogmargdens(\obsall; \paramsmle)
	- \approxlogmargdens(\obsall; \approxparamsmle)
	+ \approxlogmargdens(\obsall; \approxparamsmle)
	- \logmargdens(\obsall; \approxparamsmle) \\
	&\leq \logmargdens(\obsall; \paramsmle)
	- \approxlogmargdens(\obsall; \paramsmle)
	+ \approxlogmargdens(\obsall; \approxparamsmle)
	- \logmargdens(\obsall; \approxparamsmle) \\
	&\leq 2 \sup_{\params\in\paramsbigball} \abs{\logmargdens(\obsall; \params)
	- \approxlogmargdens(\obsall; \params)},
\]
where the last step follows by the assumption of the theorem that $\paramsmle\pto\paramstrue$, and by \cref{lem:consistency}.
Thus, \cref{fact:likelihood} implies that there exists $\const>0$ for which
\[\label{eqn:mle-1}
	\lim_{\minpergroup \to\infty}\trueprob{\numobs}
	\Big( \logmargdens(\obsall; \paramsmle)
	- \logmargdens(\obsall; \approxparamsmle) < \const \numgroups \pergrouprate\Big) = 1.
\]

Further, by a first-order Taylor expansion, there exists $\alpha\in[0,1]$ such that if $\intparamsmle = \alpha \paramsmle + (1-\alpha) \approxparamsmle$,
\[\label{eqn:mle-taylor}
	\logmargdens(\obsall; \paramsmle)
	- \logmargdens(\obsall; \approxparamsmle)
	= -\frac{1}{2}(\paramsmle - \approxparamsmle)\Tr \partial^2_{\params} \logmargdens(\obsall; \intparamsmle) (\paramsmle - \approxparamsmle).
\]
Apply \cref{lem:consistency} and again use the assumption that $\paramsmle\pto\paramstrue$ to conclude
that
$$
\lim_{\minpergroup\to\infty}\trueprob{\numobs}\left\{\intparamsmle \in \paramsbigball\right\} = 1.
$$
Therefore, by \cref{assn:hessian,eqn:mle-taylor},
\*[
	\logmargdens(\obsall; \paramsmle)
	- \logmargdens(\obsall; \approxparamsmle)
	\geq \frac{1}{2}\norm[0]{\paramsmle - \approxparamsmle}^2 \inf_{\params\in\paramsbigball} \eigen_{\paramdim}(-\partial^2_{\params} \logmargdens(\obsall; \params)),
\]
so by \cref{lemma:hessian-mle} in \cref{sec:proofs-consistency},
\[\label{eqn:mle-2}
	\lim_{\minpergroup\to\infty}\trueprob{\numobs}
	\Big( \ \logmargdens(\obsall; \paramsmle)
	- \logmargdens(\obsall; \approxparamsmle)
	\geq \frac{1}{2}\norm[0]{\paramsmle - \approxparamsmle}^2 \cdot \hesssmall^\prime \sum_{i=1}^\numgroups n_i^{1- \eps} \Big) = 1,
\]
for some $\hesssmall^\prime>0$.
Since
\*[
	\frac{\numgroups \pergrouprate}{\sum_{i=1}^\numgroups n_i^{1- \eps}}
	= \frac{\pergrouprate}{\frac{1}{\numgroups}\sum_{i=1}^\numgroups n_i^{1- \eps}}
	\leq \frac{\pergrouprate}{\minpergroup^{1- \eps}}
	= \minpergroup^{-\lfloor(\quadnum+2)/3 \rfloor - 1 + 2\eps},
\]
combining \cref{eqn:mle-1,eqn:mle-2} implies that there exists $\const'>0$ such that
\[\label{eqn:mle-3}
	\lim_{\minpergroup\to\infty}\trueprob{\numobs}
	\left\{ \norm[1]{\paramsmle - \approxparamsmle}_2
	\leq \const'\, \minpergroup^{(-\lfloor(\quadnum+2)/3 \rfloor - 1)/2 + \eps }\right\} = 1.
\]
By the triangle inequality 
and \cref{eqn:mle-3} we therefore have
\*[
	\mlerate_\numobs\cdot(\approxparamsmle - \paramstrue) 
	\ed 
	\normalrv + o_p(1) + O_p\left\{\mlerate_\numobs \minpergroup^{(-\lfloor(\quadnum+2)/3 \rfloor - 1)/2 + \eps}\right\},
\]
for any $\eps > 0$.
\end{proof}
The convergence rate of the true MLE, $\mlerate_\numobs$, can be found in \citet{jiang21glmm} for \glmms{}, with $r^{*}_N$ a vector with elements equal to 
$\numgroups^{1/2}$ or $(\numgroups\minpergroup)^{1/2}$, depending on the exact structure of the linear predictor,
under the conditions that $\numgroups,\minpergroup\to\infty$ with $\minpergroup/\numgroups\to0$.
The limiting random variable $\normalrv$ is Gaussian with a tractable variance matrix, and in practice Wald confidence intervals are formed in the usual manner with marginal variances obtained from a studentized pivot based on this limiting Gaussian distribution.
We elaborate on the application of \cref{fact:consistency} to exponential family generalized linear mixed models in \cref{sec:examples}.

We emphasize that $\approxparamsmle$ is a different estimator for $\params$ for each different $\quadnum$.
\cref{fact:consistency} gives the relationship that is needed between  $\numgroups$, $\minpergroup$ and $\quadnum$ to yield asymptotically valid confidence intervals for $\params$ based on $\approxparamsmle$, and hence provides guidance on which $\quadnum$---and hence which estimator of $\params$---should be chosen.
Specifically, $\quadnum$ should be chosen large enough to ensure that the additional error term decreases as $\numgroups$ and $\numpergroup$ increase.
Since (a) these are asymptotic upper bounds depending on unknown constants and (b) in practice $\numgroups$ and $\numpergroup$ are fixed for a particular set of data, 
the usual practical advice is to choose $\quadnum$ large enough such that inferences stop changing when $\quadnum$ is further increased, and \cref{fact:consistency} supports this strategy.
We also give empirical evidence in support of this conclusion in \cref{subsec:empirical}.


\section{Generalized Linear Mixed Models}\label{sec:examples}

\subsection{Exponential Family Models}

\noindent A generalized linear mixed model (\glmm{}) is:
\[\label{eqn:glmmexample}
	\obsi_{ij} \setdelim \cov_{ij},\recov_{ij}, \re_i &\indsim \responsedist(\linpred_{ij}), \\
	\linpred_{ij} = \linkfunc(\mu_{ij}) &= \beta_0 + \cov_{ij}\Tr\regparam + \recov_{ij}\Tr\re_i, \\
	\re_i &\indsim \normaldist\left\{\zero,\covmat(\resd)\right\}.
\]
Here
$\linkfunc:\Reals\to\Reals$ is a known link function,
$\covmat(\resd)$ is a covariance matrix depending on unknown parameters $\resd$.
The distribution $\responsedist$ belongs to a natural exponential family with mean $\mu_{ij}$ and density
\[\label{eqn:exponentialfamily}
	\responsedens(\obsi_{ij} \setdelim \linpred_{ij}) = \exp\left\{ \frac{\obsi_{ij}\linpred_{ij} - b(\linpred_{ij})}{a(\obssd)} + c(\obsi;\obssd)\right\},
\]
for functions $a(\cdot),b(\cdot),c(\cdot)$ and any $\obsi\in\samplespace$ where $\samplespace\subseteq\Reals$ is the sample space. For simplicity we treat the dispersion parameter $\obssd$ as fixed and known, and assume $0<a(\obssd)<\infty$. Observe that \cref{eqn:glmmexample} is obtained from \cref{eqn:glmmdefinition}, where $\responsedist$ is chosen to have density given by \cref{eqn:exponentialfamily} and $\redist=\normaldist\left\{\zero,\covmat(\resd)\right\}$.

\cref{prop:glmmregularity} establishes that \cref{fact:nonconvergence,fact:likelihood,fact:consistency} apply to any \glmm{} that is 
non-degenerate and well specified; see
\cref{assn:non-degen,assn:well-spec} in \cref{sec:proofs-exponentialfamily} for the precise conditions required.
These conditions are very mild, and all the ``common'' exponential family distributions are permitted, including Gaussian, Binomial, Poisson, Gamma, and Negative Binomial.

\begin{proposition}\label{prop:glmmregularity}
If the generalized linear mixed model defined by \cref{eqn:glmmexample}
satisfies \cref{assn:well-spec,assn:non-degen} in \cref{sec:proofs-exponentialfamily}, then it satisfies \cref{assn:prior,assn:hessian,assn:consistency,assn:kderiv,assn:limsup,assn:limsup-out} in \cref{sec:uniform-assumptions}.
\end{proposition}
\begin{proof}
See \cref{sec:proofs-exponentialfamily}.
\end{proof}

\subsection{Exact MLE Rates}\label{subsec:mlerates}

\noindent \cref{prop:glmmregularity} states that \cref{fact:consistency} applies to the GLMM defined by \cref{eqn:glmmexample}. 
In order to make quantitative use of \cref{fact:consistency}, the convergence rates $\mlerate_\numobs$ of the components of the exact MLE must be specified.
\citet{jiang21glmm} provide a thorough analysis of this topic for GLMMs.
Our \cref{fact:consistency} complements their analysis to account for integration error, making explicit the
error rates enjoyed by the approximate MLEs in GLMMs, which are the quantities upon which inferences are based in practice.

The convergence rates for elements of $\paramsmle$ in a GLMM 
are different for variance components and for regression coefficients
that share a covariate with a random effect compared to ones that do
not. 
Let $\resd = \text{vech}\{\covmat(\resd)\}$,
and consider \cref{eqn:glmmexample}.
For simplicity, let $\redim = 1$ and take $\recov_{ij} = 1$, the random intercepts model.
In the notation of the present paper, Theorem 1 of \citet{jiang21glmm} implies that, for fixed matrix $\boldsymbol{V}\in\Reals^{\paramdim}$,
$$
\sqrt{\numgroups}
\begin{pmatrix}
\widehat{\beta}_0 - \beta_0^{*} \\
\sqrt{\numpergroup}(\widehat{\regparam} - \regparam^{*}) \\
\widehat{\resd} - \resd^{*} \\
\end{pmatrix}
\dto
\normaldist(\zero,\boldsymbol{V}),
$$
where $\paramsmle = (\widehat{\beta}_0,\widehat{\regparam}\Tr,\widehat{\resd})\Tr$ is the exact MLE and $\paramstrue$ the corresponding ``true value'' of $\params$ from \cref{assn:well-spec}.
\cref{cor:glmmerrorrate}
states the error rate of $\approxparamsmle$, the quantity upon which inferences about $\params$ are based in practice.
\begin{corollary}\label{cor:glmmerrorrate}
Assume that the covariates $x_{ij}$ and $v_{ij}$ are independent and identically distributed, and satisfy \cref{assn:well-spec} iv) with probability tending to 1 in limit as both $m$ and $\minpergroup \rightarrow \infty$. 
Further assume that \cref{assn:non-degen}: i--iii and v, \cref{assn:well-spec}, Assumption 1--3 from \cite{jiang21glmm} hold, then the model given by \cref{eqn:glmmexample} satisfies:

$$
\sqrt{\numgroups}
\begin{pmatrix}
\widetilde{\beta}_0 - \beta_0^{*} \\
\sqrt{\numpergroup}(\widetilde{\regparam} - \regparam^{*}) \\
\widetilde{\resd} - \resd^{*} \\
\end{pmatrix}
\ed
\normalrv + o_p(1) + 
\begin{pmatrix}
O_p(\numgroups^{1/2} \minpergroup^{(-\lfloor(\quadnum+2)/3 \rfloor - 1)/2+ \eps} ) \\
O_p(\numgroups^{1/2} \minpergroup^{(-\lfloor(\quadnum+2)/3 \rfloor)/2+ \eps} ) \\
O_p(\numgroups^{1/2} \minpergroup^{(-\lfloor(\quadnum+2)/3 \rfloor - 1)/2 + \eps} )
\end{pmatrix},
$$
for all $\eps > 0$, 
where $Z \sim N(0, V)$ for a positive definite matrix $V$, for the exact expression see Theorem 1 of \cite{jiang21glmm}.
\end{corollary}	
\begin{proof}
By \cref{prop:glmmregularity}, \cref{assn:non-degen,assn:well-spec}
imply \cref{assn:prior,assn:hessian,assn:consistency,assn:kderiv,assn:limsup,assn:limsup-out}.
Further, \citet[Theorem 1]{jiang21glmm} provides the asymptotics of $\paramsmle$ that 
are required by \cref{fact:consistency}.
\end{proof}
\begin{remark}
\cite{jiang21glmm} assume a random design for the covariates, and we have reflected this in the statement of the \cref{cor:glmmerrorrate}. 
This change from fixed design to random design requires a reformulation of the eigenvalue condition in \cref{assn:non-degen} iv), as a probabilistic statement in \cref{cor:glmmerrorrate}.
\end{remark}

\begin{remark}
	For certain GLMs models such as the logistic model, the cost of uniformity in the integration error can be sharpened to to logarithmic factor instead of a sub-polynomial factor for general GLMs. Specifically with more detailed accounting for the logistic model one can obtain a statement of the kind:
$$
\sqrt{\numgroups}
\begin{pmatrix}
\widetilde{\beta}_0 - \beta_0^{*} \\
\sqrt{\numpergroup}(\widetilde{\regparam} - \regparam^{*}) \\
\widetilde{\resd} - \resd^{*} \\
\end{pmatrix}
\ed
\normalrv + o_p(1) + 
\begin{pmatrix}
O_p(\numgroups^{1/2} \log(\minpergroup)^a \minpergroup^{(-\lfloor(\quadnum+2)/3 \rfloor - 1)/2} ) \\
O_p(\numgroups^{1/2}\log(\minpergroup)^a \minpergroup^{(-\lfloor(\quadnum+2)/3 \rfloor)/2} ) \\
O_p(\numgroups^{1/2}\log(\minpergroup)^a \minpergroup^{(-\lfloor(\quadnum+2)/3 \rfloor - 1)/2} )
\end{pmatrix},
$$
for some $a \in \Reals$.
\end{remark}

\citet[Assumption 1 and 2]{jiang21glmm} requires $\minpergroup/\numgroups\to 0$ as
$\numgroups,\minpergroup\to\infty$, meaning that $\numgroups$ should grow faster than $\minpergroup$. 
This is satisfied by $\numgroups = \minpergroup^{q}$ for \emph{any} $q > 1$, meaning that $\numgroups$ can grow \emph{arbitrarily} faster than $\numpergroup$ and convergence of the exact MLE is still attained.
\cref{cor:glmmerrorrate} reveals that there is a limit to how fast
$\numgroups$ can grow compared to $\numpergroup$ when approximate integration is required to compute the MLE:
$q < r(\quadnum)$ or $q < r(\quadnum) + 1$, where $r(\quadnum) = \lfloor(\quadnum+2)/3\rfloor$, is required for the integration error to be 
negligible for $\widetilde{\regparam}$ and $(\widetilde{\beta}_0, \widetilde{\resd})$, respectively.
If $\numgroups$ grows too fast compared to $\minpergroup$, the integration
error will degrade the quality of the inferences about $\params$
based on $\approxparamsmle$.
The solution is to increase $\quadnum$ and hence $r(\quadnum)$,
that is, to use a more accurate integral approximation in the case
that $\numgroups$ is too large relative to $\minpergroup$.
We illustrate this empirically in \cref{subsec:empirical}.
We reiterate that the Laplace approximation is AQ with $\quadnum = 1$, so this discussion also applies to the question of when to use or not use the Laplace approximation to fit generalized linear mixed models.

\subsection{Empirical Error Analysis}\label{subsec:empirical}

\noindent A practitioner will be faced with a fixed $\numgroups$ and $\numpergroup$,
but has control over $\quadnum$.
The practical recommendation based on the theory presented in this
paper is to choose $\quadnum$ high enough that
$\approxparamsmle$ and approximate confidence intervals based on it
do not change when $\quadnum$ is increased or decreased.
Here we present an empirical analysis which demonstrates the
impact of \cref{fact:consistency} and its \cref{cor:glmmerrorrate}.
The core idea is that if $\quadnum$ is chosen large enough, inferences
based on $\approxparamsmle$ should be indistinguishable from those based on the exact MLE.
Specifically, Wald confidence intervals should attain close to nominal coverage, on average.
However, if $\quadnum$ is chosen too low for a given $\numgroups$ and
$\numpergroup$, then the quality of the inferences should degrade as $\numgroups$, and hence $\mlerate_\numobs$, is increased.
Since the theory for the exact MLE predicts that the quality of inferences based on $\paramsmle$ should improve on average
as $\numgroups$ is increased, this contradictory behaviour can be attributed to the increasing integral approximation error
incurred as $\numgroups\to\infty$.

A reviewer points out that \cref{fact:consistency} does not make a statement about the quality of the Hessian of the approximate
log-marginal likelihood as an approximation to the Hessian of the exact log-likelihood, and that this would be required to make a formal
statement about the coverage of Wald intervals based on the former. Indeed, \citet{approximatelikelihood} shows that an accurate Hessian
approximation is sufficient for accurate Wald intervals in this context. However, the following simulations show empirical evidence of a
setup in which any error in the Hessian is not large in comparison to that in the approximate likelihood, and the behvaiour predicted by
\cref{fact:consistency} is recovered for Wald confidence intervals based on the Hessian of the approximate log-marginal likelihood.

We construct simulations to investigate how this expected behaviour
depends on $\numgroups$, $\numpergroup$, and $\quadnum$.
We simulate $1000$ sets of data from the model
\begin{align}\label{eqn:bernoullimodel}
	\obsi_{ij} \setdelim x_{ij}, \reidx_i &\indsim \bernoullidist\left\{\log\left(\frac{\linpred_{ij}}{1-\linpred_{ij}}\right)\right\}, \\
	\linpred_{ij} &= (\beta_0 + \reidx_i) + x_{ij}\beta_1, \\
	\reidx_i &\indsim \normaldist\left(0,\sigma^2\right),
\end{align}
with $i = 1,\ldots,\numgroups$ groups of size $j = 1,\ldots,\numpergroup$.
The covariate $x_{ij}$ was generated from $\text{N}(0,1)$ and hence varied within and between groups.
The parameters were $(\beta_0,\beta_1,\sigma) = (-4,2,2)$, leading to very imbalanced
binary outcomes with $P(Y = 1)$ ranging from $0.18\%$ for $x = 0$ and $u$ at it's $1\%$ percentile to $58\%$ for $x = 1$
and $u$ at its $99\%$ percentile; $P(Y=1) = 1.8\%$ and $12\%$ for $x = 0$ and $1$ and $u = 0$, its mean.
The numbers of groups were very large at $\numgroups = 1000\times (2^0,\ldots,2^4)$, and the group sizes small at $\numpergroup = 2,4,6,8,10$.
These were chosen so that (a) $\numgroups$ grows faster than $\numpergroup$ as required by \citet{jiang21glmm}, and (b)
the integration error should dominate the sampling error for lower $\numpergroup$ and higher $\numgroups$, if $\quadnum$ is chosen too small. 
Overall, this is a simulation setup in which it should
be challenging to make accurate approximate inferences.

\begin{figure}[ht]
\centering
\includegraphics[width=5in]{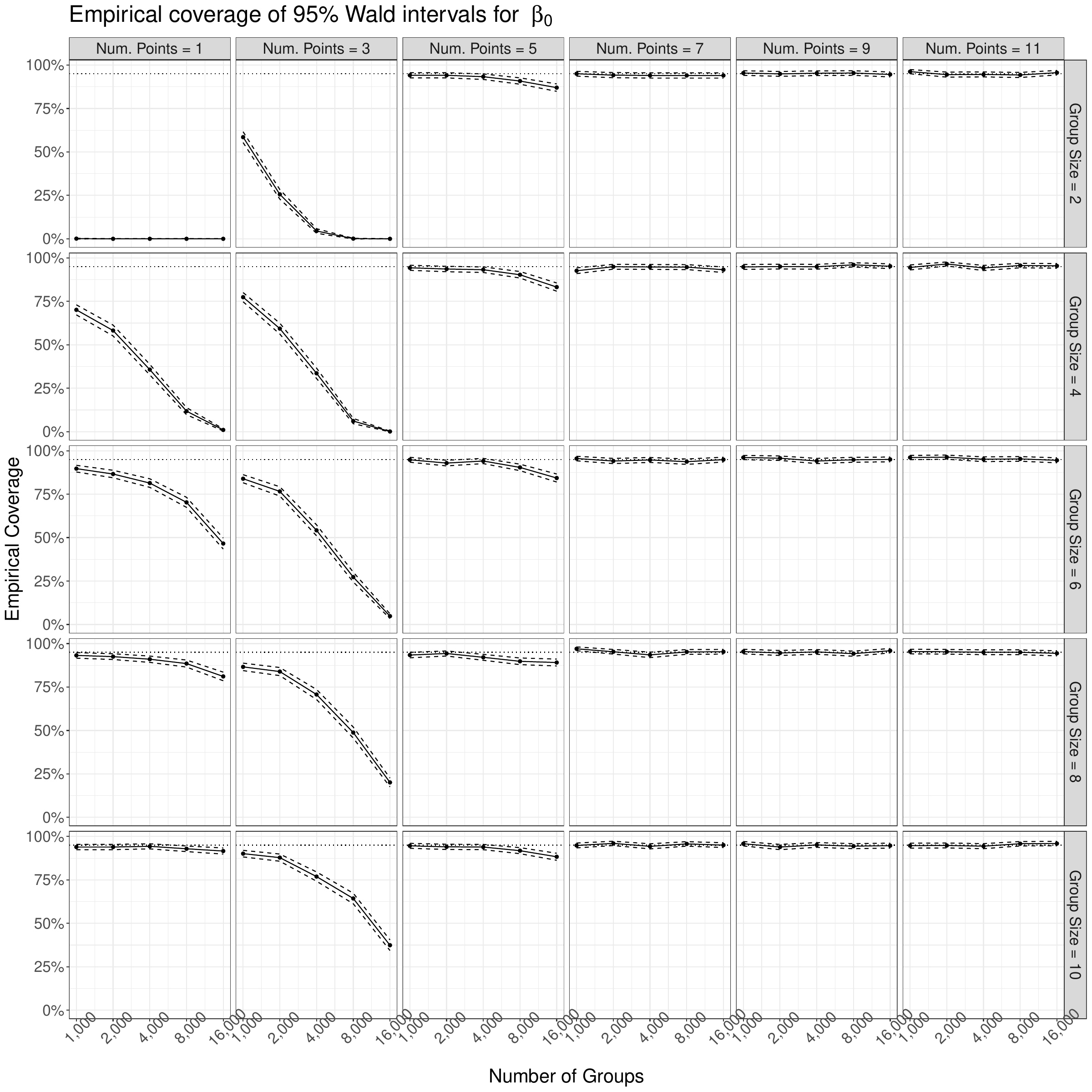}
\caption{Empirical coverage of $\beta_0$ for the simulation of \cref{subsec:empirical}. A Bernoulli random intercept model (\cref{eqn:bernoullimodel})
was fit 1000 times to each combination of numbers of groups ($\numgroups$, x-axis), group size ($\numpergroup$, rows) and numbers of quadrature points ($\quadnum$, columns). The y-axis shows empirical coverages of $95\%$ Wald confidence intervals centred at the approximate
MLE, $\approxparamsmle$, with standard errors computed using the diagonal of the inverse Hessian of the approximate marginal likelihood.
Shown are empirical coverage proportions across $1000$ simulations ($\bullet$) along with Monte Carlo confidence intervals ($- - -$).}
\label{fig:simresults}
\end{figure}

\cref{fig:simresults} shows the results.
When $\quadnum$ is too low, increasing $\numpergroup$ leads to approximate Wald confidence intervals that have worse coverage.
The effect is less dramatic for larger $\numpergroup$. Once $\quadnum$ is increased large enough, however, the coverages remain
nominal as $\numgroups$ is increased, indicating that the integration error is of a lower order than the sampling error.
This supports the practical recommendation of simply increasing $\quadnum$ until inferences stop changing; this is likely to be
the point at which numerical error is less than the sampling error. 
Additional simulation results are shown in \cref{supp:additionalsims}.

\section{Discussion and extensions}\label{sec:discussion}

\noindent Take $\numgroups = 1$ and consider the following general latent variable
model as a special case of \cref{eqn:glmmdefinition}:
\[\label{eqn:lgm}
	\obs_{i} \setdelim \cov_{i}, \re &\indsim \responsedist(\cov_{i},\re; \regparam,\obssd), i = 1,\ldots,n, \\
	\re &\sim \redist(\resd).
\]
This covers a wide range of interesting models 
including spline smoothing \citep{smoothestimation} and spatio-temporal models \citep{geostataggdata}.
\cref{fact:consistency} applies with $\numgroups=1$ and $r^{*}_N$
depending only on $\numpergroup$, to give the asymptotic 
behaviour of $\approxparamsmle$ when that for $\paramsmle$ is available.
However, in these interesting problems, the asymptotics for $\paramsmle$
require the dimension of $\re$ to increase with $\numpergroup$,
a situation which is not covered by our present analysis.
Extension of \cref{fact:consistency} and the other results in this paper
for Laplace-approximate marginal likelihood inference in the presence of
such high-dimensional latent variables $\re$ appears feasible and the subject
of ongoing work. 
The main technical challenge in proving \cref{fact:consistency} was
to obtain the rate of local uniform convergence in probability of the likelihood
approximation. \citet{tang2021laplace} provide local uniform rates of convergence
for Laplace approximate marginal likelihood in high dimensions, showing that if $\text{dim}(\re) = q$ then the approximation error is
$O(q^{3}\log(n)/n)$, implying that the likelihood approximation converges 
whenever $q = O(n^{\alpha})$ for any $\alpha < 1/3$.
\citet{marginallikelihoodpenalizedspline} shows that for spline models, $q = 2/9 < 1/3$ is sufficient
for asymptotic convergence of the exact maximum likelihood estimator.
\citet{sanz2022finite} discusses convergence rates and scaling for spatial Gaussian processes
fit using finite element approximations, which is considerably more
complicated.
Extension of \cref{fact:consistency} to the case where $\text{dim}(\re)$
increases with $\numpergroup$ via these novel results could lead to convergence results for Laplace-approximate maximum marginal likelihood estimators in these
and related problems.

\begin{acks}[Acknowledgments]
	We are grateful for helpful comments from Helen Ogden, Glen McGee, Jeffrey Negrea, Art Owen, and the Editor and two reviewers.
\end{acks}
\begin{funding}
	Blair Bilodeau was supported by an NSERC Canada Graduate Scholarship and the Vector Institute. Alex Stringer is supported by NSERC Discovery Grant RGPIN-2023-03331.
\end{funding}

\begin{appendix}
\section{Assumptions}\label{sec:uniform-assumptions}
\noindent First, we inherit some standard notation from \citet{bilodeau2021stochastic}. For a positive-definite $\dummydim\times\dummydim$ matrix $A$, let $\eigen_{1}(A) \geq \cdots \geq \eigen_{\dummydim}(A) > 0$ denote its ordered eigenvalues. For a generic $m \times n$ matrix $B$, we use $\norm{B}_{op}$ to denote it's maximal singular value, i.e. its operator norm. For any $f: \Reals^\dummydim \rightarrow \Reals$, $\dumderivvec \subseteq \Nats^\dummydim$, and $\mb{x} \in \Reals^\dummydim$, we define
\*[
	\abssmall{\dumderivvec} = \sum_{j = 1}^{\dummydim} \alpha_i, \quad \dumderivvec! = \prod_{j =1}^\dummydim \alpha_j!, \quad \mb{x}^{\dumderivvec} =\mb{x}_{\dumderivvec} =\prod_{j = 1}^\dummydim x_j^{\alpha_j}, \text{ and }\\
	\partial^{\dumderivvec} f(\mb{x}) = \partial x_1^{\alpha_1} \partial x_2^{\alpha_2} \cdots \partial x_\dummydim^{\alpha_\dummydim} f( x) = \frac{\partial^{\abssmall{\dumderivvec}} f(x)} {\partial x_1^{\alpha_1} \partial x_2^{\alpha_2}\cdots \partial x_\dummydim^{\alpha_\dummydim}}. 
\]
For any vector, $\bm{z}$, and radius, $\delta>0$, let $\bigball{\bm{z}}{\delta} = \{\bm{z'}: \norm{\bm{z}-\bm{z'}}_2 < \delta\}$, where $\norm{\cdot}_2$ denotes Euclidean norm. 


In the Assumptions that follows, let $\numgroups = \numpergroup_{\min}^q$ for some $q > 0$.
For any sequence of data-generating distributions $\trueprob{\numobs}$, we say \cref{assn:kderiv,assn:hessian,assn:limsup,assn:limsup-out,assn:consistency,assn:prior} hold if there exists
$\universalradius>0$,
$0 < \alpha < 1/4$,
$\paramstrue\in\paramspace$, and $\retruepi \in \Reals^\redim, i = 1,\ldots,\numgroups$
such that for $\paramradius = (\min_{i=1,\ldots,\numgroups}\numpergroup_i)^{-\alpha}$ as defined in \cref{sec:mainresults},
each of the following six statements holds for all values of $\epsilon > 0$.

\begin{assumption}\label{assn:kderiv}
There exists $\derivnum,\derivbound>0$ such that for all $\dumderivvec \subseteq \Nats^\paramdim$ with $0 \leq \abs{\dumderivvec} \leq \derivnum$,
\*[
	\lim_{\minpergroup \rightarrow \infty} \trueprob{\numobs}\Big[\forall_{i = 1, \dots, \numgroups} \ \sup_{\params\in\paramsbigballfixed} \sup_{\re\in\reball} \abs{\partial_{\re}^{\dumderivvec}\llhoodip(\re)} < \numpergroup_i^{1+\eps} \cdot \derivbound\Big] = 1.
\]
\end{assumption}

\begin{assumption} \label{assn:hessian}
There exist $0 < \hesssmall \leq \hessbig < \infty$ such that 
for all $i\in[\numgroups]$ 
\*[
	\lim_{\minpergroup \rightarrow \infty}  &\trueprob{\numobs}\Big[ \forall_{i = 1, \dots, \numgroups} \ \numpergroup_i^{1- \eps} \cdot \hesssmall \leq 
	\inf_{\params\in\paramsbigballfixed} \inf_{\re\in\reball}
	\eigen_{\redim}(-\partial^{2}_{(\params,\re)}\llhoodip(\re)) \\
	&\leq \sup_{\params\in\paramsbigballfixed} \sup_{\re\in\reball} \eigen_{1}(-\partial^{2}_{(\params,\re)}\llhoodip(\re))  \leq \numpergroup_i^{1 +\eps} \cdot \hessbig\Big] = 1
\]
\end{assumption}

\begin{assumption}\label{assn:limsup}
There exists $\llhoodmargin >0$ such that
\*[
	\lim_{\minpergroup \rightarrow \infty}\trueprob{\numobs}\Big[ \forall_{i = 1, \dots, \numgroups} \ \sup_{\params\in\paramsbigball} \sup_{\re\in[\reball]^{c}} \log\jointdens_{i}(\obsall_{i} \setdelim \re;\params) - \log\jointdens_{i}(\obsall_{i} \setdelim \retruepi;\params) \leq - \numpergroup_i^{1-\eps} \llhoodmargin\Big] = 1.
\]
\end{assumption}

\begin{assumption}\label{assn:limsup-out}
	There exists $\llhoodmargin^\prime >0$ such that
	\*[
		\lim_{\minpergroup \rightarrow \infty}\trueprob{\numobs}\Big[ \forall_{i = 1, \dots, \numgroups}  \sup_{\params\in[\paramsbigballfixed]^{c}} \sup_{\re\in[\reball]^{c}} \log\jointdens_{i}(\obsall_{i} \setdelim \re;\params) - \log\jointdens_{i}(\obsall_{i} \setdelim \retruepi;\params) \leq - \numpergroup_i^{1 - \eps}\llhoodmargin^\prime \Big] =1.
	\]	
\end{assumption}

\begin{assumption}\label{assn:consistency}
There exists a $\conmargin > 0$ 
such that for every $\eps > 0$
\*[
	\lim_{\minpergroup \rightarrow \infty}\trueprob{\numobs} \left[ \forall_{i = 1, \dots, \numgroups}\ \sup_{\params\in\paramsbigball} \paramradius^{-1} \norm{\condmodeip - \retruepi}_2 \leq \conmargin \numpergroup_i^\eps \right] =1.
\]
Furthermore for every $\conmargin^\prime > 0$ and for every function $G^\prime(n_i)$ such that $\lim_{n \rightarrow \infty} G^\prime(n_i) = \infty$:
\*[
	\lim_{\minpergroup \rightarrow \infty} \trueprob{\numobs} \left[\forall_{i = 1, \dots, \numgroups} \ \numpergroup_i^{1/2} G^\prime(n_i)^{-1} \norm{\hat\re_i^{\paramstrue} - \retruepi}_2 > \conmargin^\prime \numpergroup_i^\eps \right] = 0.
\]
\end{assumption}

\begin{assumption}\label{assn:prior}

There exist  $0 < \priorsmall < \priorbig < \infty$ such that 
\*[
   \priorsmall \leq \inf_{\params\in\paramsbigball} \inf_{\re\in\reball} \redens(\re;\resd) \leq \sup_{\params\in\paramsbigball} \sup_{\re\in\reball} \redens(\re;\resd) \leq \priorbig.
\]
Furthermore, uniformly for all $i = 1, \dots, \numgroups$, for some $a\in \Reals$
\*[
	\int_{\Reals^\redim} \norm{ \partial_{\params} \log\margdens (\obsall_i;\re_i, \params)}_2 \redens(\re;\resd)  \dee \re = O(\numpergroup_i^{a} ) , \\
	\int_{\Reals^\redim} \norm{ \partial^{2}_{\params} \log\margdens (\obsall_i;\re_i, \params) }_{op} \redens(\re;\resd) \dee \re = O(\numpergroup_i^{a} ),
\]
for all $\params \in \paramsbigball$ and for $\zeta_N$ as defined in \cref{assn:consistency}. 
\end{assumption}

\begin{remark}
We state our assumptions as limits in probability.
In the proofs, for the sake of clarity and brevity, algebra which invokes these assumptions is often performed outside of a probability statement.
This is understood to mean that the given statement holds for any event
upon which the relevant assumption holds;
the assumptions then state that the measure of the sets of events for which each statement holds tends to $1$ as $\minpergroup\to\infty$.
\end{remark}

\bibliographystyle{imsart-nameyear} 
\bibliography{numquadpoints}   

\section{Proof of Lemma \ref{fact:likelihood}}\label{sec:proofs-likelihood}

\noindent Our approach is to show that the strengthening of the pointwise assumptions in \citet{bilodeau2021stochastic} to the uniform assumptions in \cref{sec:uniform-assumptions} is sufficient to control the approximation error of the likelihood in a ball around the true parameter,
with rates slightly changed to reflect the additional $\epsilon$ factor in our uniform assumptions compared to their pointwise versions in \citet{bilodeau2021stochastic}.
For completeness, we carefully verify that each step from \citet{bilodeau2021stochastic} can be appropriately upgraded to hold uniformly.
In \cref{prop:glmmregularity}, we show that the stronger uniform assumptions holds for generalized linear mixed models.

To state intermediate results, we require the following notation from \citet{bilodeau2021stochastic}.
For $\dummytau > 3$, let
\*[
	\smalltsetdef[j]{\dummytau} &= \left\{(t_3,\dots,t_{2\quadnum}) \in \PosInts^{2\quadnum-3} \Bigsetdelim \sum_{s=3}^{2\quadnum} t_s = j \text{ and } \sum_{s=3}^{2\quadnum} s t_s \leq\dummytau - 1 \right\},\\
	\equaltsetdef[j]{\dummytau} &= \left\{(t_3,\dots,t_{2\quadnum}) \in \PosInts^{2\quadnum -3} \Bigsetdelim \sum_{s=3}^{2\quadnum} t_s = j \text{ and } \sum_{s=3}^{2\quadnum} s t_s = \dummytau \right\},\\
	\bigtsetdef[j]{\dummytau} &= \left\{(t_3,\dots,t_{2\quadnum}) \in \PosInts^{2\quadnum -3} \Bigsetdelim \sum_{s=3}^{2\quadnum} t_s = j \text{ and } \sum_{s=3}^{2\quadnum} s t_s \geq \dummytau \right\}.
\]
Further, let $\tsum = \sum_{s=3}^{2\quadnum} s t_s$ for any $\tvec = (t_3,\dots,t_{2\quadnum}) \in \PosInts^{2\quadnum -3}$.
Fix $\numgroups$ and $i \in [\numgroups]$.

For all $\params\in\paramspace$ and $\numobs\in\Nats$, the following holds $\trueprob{\numobs}$-a.s.
\[\label{eqn:pergroup-big-as-bound}
	\abs{\frac{\margdens_{i}(\obsall_{i};\params)}{\approxmargdensAQ_{i}(\obsall_{i};\params)} - 1}
	&= 
	\frac{\abs{\margdens_{i}(\obsall_{i};\params) - \approxmargdensAQ_{i}(\obsall_{i};\params)}}{\approxmargdensAQ_{i}(\obsall_{i};\params)}
	= 
	\frac{\abs{\margdens_{i}(\obsall_{i};\params) - \approxmargdensAQ_{i}(\obsall_{i};\params)}}{\jointdens_{i}(\obsall_{i} \setdelim \condmodeip;\params)}
	\frac{\jointdens_{i}(\obsall_{i} \setdelim \condmodeip;\params)}{\approxmargdensAQ_{i}(\obsall_{i};\params)}.
\]

We prove the following intermediate results.

\begin{lemma}\label{fact:normalizing_constant_approx}
Under assumptions \cref{assn:kderiv,assn:hessian,assn:limsup,assn:consistency,assn:prior},
if $\quadrule{\quadpointset}{\weight}$ is a quadrature rule satisfying $\property{\quadnum}{\redim}$ then for all $\eps > 0$ and for $\numgroups = \minpergroup^q$ for any $q> 0$
\*[
	\lim_{\minpergroup \to \infty} \trueprob{\numobs}
	\Bigg( \forall_{i = 1, \dots, \numgroups} \
	\sup_{\params\in\paramsbigball}
	\frac{\abs{\margdens_{i}(\obsall_{i};\params) - \approxmargdensAQ_{i}(\obsall_{i};\params)}}{\jointdens_{i}(\obsall_{i} \setdelim \condmodeip;\params)}
	\leq
	\constt \frac{1}{n_i^{\redim/2 + \lfloor (\quadnum +2)/3 \rfloor - \eps}}
	\Bigg)
	= 1.
\]
\end{lemma}

\begin{lemma}\label{fact:lowerbound_quad}
Under assumptions \cref{assn:kderiv,assn:hessian,assn:consistency,assn:prior},
if $\quadrule{\quadpointset}{\weight}$ is a quadrature rule satisfying $\property{\quadnum}{\redim}$ then for all $\eps > 0$ and for $\numgroups = \minpergroup^q$ for any $q> 0$
\*[
	\lim_{\minpergroup \to \infty} \trueprob{\numobs}
	\Bigg( \forall_{i = 1, \dots, \numgroups} \
	\sup_{\params\in\paramsbigball}
	\frac{\jointdens_{i}(\obsall_{i} \setdelim \condmodeip;\params)}{\approxmargdensAQ_{i}(\obsall_{i};\params)}
	\leq \constt
	n_i^{\redim/2 + \eps}
	\Bigg)
	= 1.
\]
\end{lemma}

Consequently, for each $i \in [\numgroups]$, under \cref{assn:kderiv,assn:hessian,assn:limsup,assn:consistency,assn:prior} there exists a $\const > 0$ such that
\*[
	\lim_{\minpergroup\to\infty}\trueprob{\numobs}\left(\forall_{i = 1, \dots, \numgroups} \ \sup_{\params\in\paramsbigball} \abs{\frac{\approxmargdensAQ_{i}(\obsall_{i};\params)}{\margdens_{i}(\obsall_{i};\params)} - 1} < \const \, \numpergroup_i^{-\lfloor(\quadnum+2)/3\rfloor + \eps}\right) = 1.
\]
for all $\eps > 0$.
Next, using $\log(1+x) \leq x$ for all $x>-1$ and $\log(1-x) \geq -2x$ for all $x \in [0,3/4]$ gives that
for each $i \in [\numgroups]$,
\*[
	\lim_{\minpergroup\to\infty}\trueprob{\numobs}\left(\forall_{i = 1, \dots, \numgroups} \ \sup_{\params\in\paramsbigball} \abs{\log \approxmargdensAQ_{i}(\obsall_i; \params) - \log \margdens_i(\obsall_i; \params)} < 2\const \, \numpergroup_i^{-\lfloor(\quadnum+2)/3\rfloor + \eps}\right) = 1.
\]

Therefore,
\*[
	\lim_{\minpergroup \to\infty}\trueprob{\numobs}\left(\sup_{\params\in\paramsbigball} \abs{\log \approxmargdensAQ(\obsall; \params) - \log\margdens(\obsall; \params)} < 2\const \, \sum_{i=1}^\numgroups \numpergroup_i^{-\lfloor(\quadnum+2)/3\rfloor+\eps}\right) = 1,
\]
for all $\eps > 0$.

\subsection{Proof of \cref{fact:normalizing_constant_approx}}
\noindent Fix arbitrary $\shrinkingconst >0$ (to be tuned at the end as a function of $\redim$ and $\quadnum$) and let $\shrinkingrad = \shrinkingconst \sqrt{(\log \numpergroup_i)/\numpergroup_i}$ for each $\numobs \in \Nats$. 

First, expand the fraction of interest, giving
\*[
	&\frac{\abs{\margdens_{i}(\obsall_{i};\params) - \approxmargdensAQ_{i}(\obsall_{i};\params)}}{\jointdens_{i}(\obsall_{i} \setdelim \condmodeip;\params)} = 
	\frac{
	\abs{
		\int\jointdens_{i}(\obsall_{i}, \re;\params)\dee\re
		- 
		\detbar{\cholip}\sum_{\quadpointvec\in\quadpointset(\redim,\quadnum)}\jointdens_{i}\left(\obsall_{i}, \cholip\quadpointvec + \condmodeip;\params\right)\weight_{\quadnum}(\quadpointvec)
		}
	}
	{\jointdens_{i}(\obsall_{i} \setdelim \condmodeip;\params)} \\
	&= \redens_{i}(\condmodeip;\resd)
	\Big|
		\int  \exp\left\{\llhoodip(\re) - \llhoodip(\condmodeip)\right\} \dee\re
		- 
		\detbar{\cholip}\sum_{\quadpointvec\in\quadpointset(\redim,\quadnum)}\weight_{\quadnum}(\quadpointvec)\exp\left\{\llhoodip(\cholip\quadpointvec + \condmodeip) - \llhoodip(\condmodeip)\right\}
	\Big|.
\]
After splitting the region of integration and applying the triangle inequality, this is upper bounded by
\[\label{eqn:initial-expansion}
	&\leq 
	\redens_{i}(\condmodeip;\resd)
	\abs{
		\int_{\bigball{\condmodeip}{\shrinkingrad}}
		\exp\left\{\llhoodip(\re) - \llhoodip(\condmodeip)\right\} \dee\re
		- 
		\detbar{\cholip}\sum_{\quadpointvec\in\quadpointset(\redim,\quadnum)}\weight_{\quadnum}(\quadpointvec)\exp\left\{\llhoodip(\cholip\quadpointvec + \condmodeip) - \llhoodip(\condmodeip)\right\}
	} \\
	&\qquad +
	\redens_{i}(\condmodeip;\resd) 
	\int_{\bigballc{\condmodeip}{\shrinkingrad}}
		\exp\left\{\llhoodip(\re) - \llhoodip(\condmodeip)\right\} \dee\re.
\]
Now, define
\*[
	\derivboundmodeip = 
	\sup_{\dumderivvec: \abssmall{\dumderivvec}\leq \derivnum} 
	\sup_{\re\in\bigball{\condmodeip}{\quadpointbound\lvert\cholip\rvert \lor \shrinkingrad}} 
	\absbig{\partial^{\dumderivvec} \llhoodip(\re)} ,
	\quad 
	\hessbigmodeip = \frac{\eigen_{1}(\hessip(\condmodeip))}{n} ,
	\quad \text{and} \quad
	\hesssmallmodeip = \frac{\eigen_{\redim}(\hessip(\condmodeip))}{n},
\]
where $\quadpointbound = \sup_{\quadpointvec \in \quadpointset} \norm{\quadpointvec}_2 < \infty$.

The main result we need to inherit from \citet{bilodeau2021stochastic} is the following.

\begin{lemma} \label{fact:normalizing_constant_as}
For all $1 \leq \quadnum \leq \lfloor \derivnum/2 \rfloor$ and $\params\in\paramsbigball$,
if $\quadrule{\quadpointset}{\weight}$ is a quadrature rule satisfying $\property{\quadnum}{\redim}$ then
there exists a constant $\paramconst>0$ depending only on $\redim$ and $\quadnum$ such that for all $\numobs \in \Nats$ it holds $\trueprob{\numobs}$-a.s. that
\[\label{eqn:almost-sure-lemma}
	&\abs{
		\int_{\bigball{\condmodeip}{\shrinkingrad}}
		\exp\left\{\llhoodip(\re) - \llhoodip(\condmodeip)\right\} \dee\re
		- 
		\detbar{\cholip}\sum_{\quadpointvec\in\quadpointset(\redim,\quadnum)}\weight_{\quadnum}(\quadpointvec)\exp\left\{\llhoodip(\cholip\quadpointvec + \condmodeip) - \llhoodip(\condmodeip)\right\}
	} \\
	&\leq
	\paramconst
	\left(
	\left(\frac{\hessbigmodeip}{\hesssmallmodeip}\right)^{\redim/2}
	+
	1
	\right) \numpergroup_i^{-\redim/2} \\
	&\qquad\times
	\Bigg[
	\max_{j \in [\quadnumadj]}
	\max_{\tvec \in \bigtset{j}}
	(\derivboundmodeip)^{j}  \,
	(\hesssmallmodeip \, \numpergroup_i)^{-\tsum/2}
	+
	\max_{\tvec \in \customtset{\quadnumadj+1}{3(\quadnumadj +1) + \ind\{2\quadnum = 2 \!\!\!\! \pmod 3\}} } (\derivboundmodeip)^{\quadnumadj + 1} 
	(\hesssmallmodeip \, \numpergroup_i)^{-\tsum/2}
	\\
	&\qquad\qquad +
	(\hesssmallmodeip)^{-1/2}
	\max_{j \in \{0\} \cup [\quadnumadj]} 
	(\derivboundmodeip)^{j} 
	\numpergroup_i^{-\frac{\shrinkingconst^2 \hesssmallmodeip + 2}{4}}
	+
	\multitaylorexpboundip
	(\derivboundmodeip)^{\quadnumadj+2} 
	\max_{\tvec \in \customtset{\quadnumadj+2}{3(\quadnumadj+2)}}
	(\hesssmallmodeip \, \numpergroup_i)^{-\tsum/2}
	\Bigg],
\]
where
\*[
	\multitaylorexpboundip
	&= \Bigg[
	\left(\frac{\hessbigmodeip}{\hesssmallmodeip}\right)^{\redim/2} 
	\exp\left\{\derivboundmodeip \Big(\frac{\log(\numpergroup_i)}{\numpergroup_i} \Big)^{3/2} \right\} 
	+
	\exp\left\{\derivboundmodeip \max\left\{(\hesssmallmodeip \, \numpergroup_i)^{-3/2}, (\hesssmallmodeip \, \numpergroup_i)^{-\quadnum}\right\}\right\}
	\Bigg]
\]
and $\quadnumadj$ is the smallest integer such that $3(\quadnumadj+1)\geq 2\quadnum$.
\end{lemma}
\begin{proof}[Proof of \cref{fact:normalizing_constant_as}]
For each $\params\in\paramsbigball$, this follows from Lemma~4 of \citet{bilodeau2021stochastic}. All that remains to be checked is that the constant $\paramconst$ does not depend on $\params$, which follows immediately from inspection of Appendix~S.3 of \citet{bilodeau2021stochastic}.
\end{proof}

We then want to make use of the following, which provides the necessary convergence for each of the quantities used in \cref{fact:normalizing_constant_as}.
\begin{lemma}\label{fact:constant_convergence}
Under \cref{assn:kderiv,assn:consistency,assn:hessian,assn:prior}, and for $\numgroups = \minpergroup^q$ for any $q> 0$, the following hold for every $\eps > 0$:
\begin{itemize}
\item[\upshape i)]
$\underset{\minpergroup\to\infty}{\text{\upshape lim}} \trueprob{\numobs}\!\bigg( \forall_{i = 1, \dots, \numgroups} \ \hesssmall \numpergroup_{i}^{-\eps} \leq \inf_{\params\in\paramsbigball} \hesssmallmodeip \leq \sup_{\params\in\paramsbigball} \hessbigmodeip \leq \hessbig \numpergroup_{i}^{\eps} \bigg) = 1.$
\item[\upshape ii)] 
$\underset{\minpergroup\to\infty}{\text{\upshape lim}} \trueprob{\numobs}\!\bigg(\forall_{i = 1, \dots, \numgroups} \ \sup_{\params\in\paramsbigball} \derivboundmodeip \leq \numpergroup_i^{1 + \eps} \cdot \derivbound \bigg) = 1.$
\item[\upshape iii)]
$\underset{\minpergroup\to\infty}{\text{\upshape lim}} \trueprob{\numobs}\!\bigg(\forall_{i = 1, \dots, \numgroups} \ \sup_{\params\in\paramsbigball} \redens_{i}(\condmodeip;\resd) \leq \priorbig \bigg) = 1.$
\end{itemize}
\end{lemma}
\begin{proof}[Proof of \cref{fact:constant_convergence}]
i) 
By \cref{assn:consistency} for every $\eps > 0$,
\*[
  \lim_{\minpergroup \to \infty} \trueprob{\numobs} \left[\forall_{i = 1, \dots, \numgroups} \sup_{\params\in\paramsbigball} \paramradius\inv\norm{\condmodeip - \retruep}_2 \leq \conmargin \numpergroup_i^\eps \right] = 1.
\]
Choose $\eps < \alpha$, then $\lim_{\minpergroup\to\infty} \trueprob{\numobs}\Big(\forall_{i = 1, \dots, \numgroups} \ \forall \params\in\paramsbigball \quad \condmodeip \in \reball \Big) = 1$, so
\*[ 
	&\lim_{\minpergroup\to\infty} \trueprob{\numobs}\left(\forall_{i = 1, \dots, \numgroups} \sup_{\params\in\paramsbigball} \hessbigmodeip \leq \hessbig \numpergroup_{i}^{\eps} \right) \\
	=& \lim_{\minpergroup\to\infty} \trueprob{\numobs}\left(\forall_{i = 1, \dots, \numgroups} \ \sup_{\params\in\paramsbigball} \frac{\eigen_{1}(-\partial^{2}_{\re}\llhoodip(\condmodeip))}{n} \leq \hessbig \numpergroup_{i}^{\eps} \right) \\
	\geq& \lim_{\minpergroup\to\infty} \trueprob{\numobs}\left(\forall_{i = 1, \dots, \numgroups} \ \forall \params\in\paramsbigball \quad \condmodeip \in \reball, \sup_{\params\in\paramsbigball} \sup_{\re\in\reball} \frac{\eigen_{1}(-\partial^{2}_{\re}\llhoodip(\re))}{\numpergroup_i} \leq \hessbig\numpergroup_i^\eps \right) \\
	=& 1,
\]
where the last step uses \cref{assn:hessian}. The inequality for $\hesssmallmode$ is proven in the same fashion. 

\noindent
ii) 
First, recall that $\abssmall{\cholip} \leq (\hesssmallmodeip \, \numpergroup_i)^{-\redim/2 + \eps}$ for all $\params\in\paramsbigball$ and $\eps > 0$. By i), we have
\*[
	\lim_{\minpergroup\to\infty} \trueprob{\numobs} \left(\forall_{i = 1, \dots, \numgroups} \ \sup_{\params\in\paramsbigball} (\hesssmallmodeip \, \numpergroup_i)^{-\redim/2 + \eps } \leq (\hesssmall \, \numpergroup_i)^{-\redim/2 +\eps} \right) = 1.
\]
That is, for all $\eps^\prime > 0$, 
\*[
	\lim_{\minpergroup\to\infty} \trueprob{\numobs} \left( \forall_{i = 1, \dots, \numgroups} \ \sup_{\params\in\paramsbigball} \quadpointbound\abssmall{\cholip} \lor \shrinkingrad < \eps^\prime \right) = 1,
\]
so by \cref{assn:consistency} again, 
\*[
	\lim_{\minpergroup\to\infty} \trueprob{\numobs} \left( \forall_{i = 1, \dots, \numgroups} \ \forall \params\in\paramsbigball \quad \bigball{\condmodeip}{\quadpointbound\abssmall{\cholip} \lor \shrinkingrad} \subseteq \reball \right) = 1.
\]

Thus,
\*[ 
	&\hspace{-2em}\lim_{\minpergroup\to\infty} \trueprob{\numobs}\left( \forall_{i = 1, \dots, \numgroups} \ \sup_{\params\in\paramsbigball} \derivboundmodeip \leq \numpergroup_i^{1 + \eps} \cdot \derivbound \right) \\
	&= \lim_{\minpergroup\to\infty} \trueprob{\numobs}\left(\forall_{i = 1, \dots, \numgroups} \ \sup_{\params\in\paramsbigball} \sup_{\dumderivvec: \abssmall{\dumderivvec}\leq \derivnum}
	\sup_{\re\in\bigball{\condmodeip}{\quadpointbound\lvert\cholip\rvert \lor \shrinkingrad}} 
	\absbig{\partial^{\dumderivvec} \llhoodip(\re)} \leq \numpergroup_i^{1 + \eps} \cdot \derivbound \right)  \\
	&\geq \lim_{\minpergroup\to\infty} \trueprob{\numobs}\Big(\forall_{i = 1, \dots, \numgroups} \  \forall \params\in\paramsbigball \quad \bigball{\condmodeip}{\quadpointbound\abssmall{\cholip} \lor \shrinkingrad} \subseteq \reball, \\
	&\sup_{\params\in\paramsbigball}
	\sup_{\dumderivvec: \abssmall{\dumderivvec}\leq \derivnum} 
	\sup_{\re\in\reball} 
	\absbig{\partial^{\dumderivvec} \llhoodip(\re)} \leq \numpergroup_i^{1 + \eps} \cdot \derivbound \Big) 
	= 1,
\]
for all $\eps > 0$, where the last step uses \cref{assn:kderiv}.

\noindent
iii) This follows directly from \cref{assn:consistency,assn:prior}.
\end{proof}

We now control the contribution of the integral of the normalized likelihood outside of a fixed ball.

\begin{lemma}\label{fact:fixed_numer}
Under \cref{assn:limsup,assn:prior}, for every $\eps > 0$ and for $\numgroups = \minpergroup^q$ for any $q> 0$
\*[
	\lim_{\minpergroup\to\infty}
	\trueprob{\numobs}
	\left[ \forall_{i = 1, \dots, \numgroups} \
	\sup_{\params\in\paramsbigball}
	\int_{\ballc{\retruep}{\universalradius}} \exp\{ \llhoodip(\re) - \llhoodip(\condmodeip) \} \dee\re
	\leq
	\frac{e^{-\numpergroup_i^{1 - \eps} \llhoodmargin}}{\priorsmall}  
	\right] = 1, 	
\]
\end{lemma}
\begin{proof}[Proof of \cref{fact:fixed_numer}]
Fix $\params\in\paramsbigball$.
By the proof of Lemma~5 in \citet{bilodeau2021stochastic},
\*[
	&\int_{\ballc{\retruep}{\universalradius}} 
	\exp\{ { \llhoodip(\re) - \llhoodip(\condmodeip)} \} \dee\re \\
	&\leq \frac{1}{\redens(\retruep;\resd)} \sup_{\re \in \ballc{\retruep}{\universalradius}}\exp\{\log\jointdens_{i}(\obsall_{i} \setdelim \re;\params) - \log\jointdens_{i}(\obsall_{i} \setdelim \retruep;\params) \}.
\]
The result then follows by applying \cref{assn:limsup,assn:prior}.
\end{proof}

\begin{lemma}\label{lem:shrinking_numer}
Under \cref{assn:hessian,assn:limsup,assn:prior,assn:consistency}, there exists a constant $\constt>0$ such that for every $\eps > 0$ and for $\numgroups = \minpergroup^q$ for any $q> 0$
\*[
  \lim_{\minpergroup\to\infty} 
  \trueprob{\numobs}\left[
	\forall_{i = 1, \dots, \numgroups} \
  \sup_{\params\in\paramsbigball}
  \int_{\ballc{\condmodeip}{\shrinkingrad} \cap \ball{\retruep}{\universalradius} } \exp\{ \llhoodip(\re) - \llhoodip(\condmodeip) \}\dee\re
  \leq \constt \frac{1}{\numpergroup_i^{ \shrinkingconst^2\hesssmall/4 + \redim/2- \eps} } \right] = 1.
\]
\end{lemma}
\begin{proof}[Proof of \cref{lem:shrinking_numer}]

Fix $\params\in\paramsbigball$ and for any value of $\eps > 0$ let $$\hesssmalltrueip(\eps) = \inf_{\re \in \ball{\retruep}{\universalradius}}\eigen_{\redim}(-\partial^{2}_{\re}\llhoodip(\re))/{\numpergroup_i^{1-\eps}}.$$
By the proof of Lemma~6 in \citet{bilodeau2021stochastic},
\*[
	\int_{\ballc{\condmodeip}{\shrinkingrad} \cap \ball{\retruep}{\universalradius} } \exp\{ \llhoodip(\re) - \llhoodip(\condmodeip) \}\dee\re
	\leq  
	\left(\frac{2\pi e}{\hesssmalltrueip} \right)^{\redim/2} \numpergroup_i^{-\shrinkingconst^2 \hesssmalltrueip/4 - \redim/2 + \eps}.
\]
By \cref{assn:consistency}, $\lim_{\numpergroup_i\to\infty} \trueprob{\numobs}\Big(\forall_{i = 1, \dots, \numgroups} \ \forall \params\in\paramsbigball \quad \condmodeip \in \reball \Big) = 1$, and by \cref{assn:hessian},
\*[
	\lim_{\minpergroup\to\infty} \trueprob{\numobs} \Big[\forall_{i = 1, \dots, \numgroups} \ \inf_{\params\in\paramsbigball} \hesssmalltrueip(\eps) \geq  \hesssmall\Big] = 1,
\]
giving the statement of the lemma.

\end{proof}

\subsection{Proof of \cref{fact:lowerbound_quad}}
\noindent Fix $\params\in\paramsbigball$.
Adapting Section~S.2.3 of \citet{bilodeau2021stochastic} it holds that if
$\condmodeip \in \ball{\retruep}{\universalradius/2}$, $\hesssmallmodeip \geq \hesssmall$, $\hessbigtrue \leq \hessbig$, and $\hessbigmodeip  \leq \hessbig$, 
\*[
	&\frac{\approxmargdensAQ_{i}(\obsall_{i};\params)}{\jointdens_{i}(\obsall_{i} \setdelim \condmodeip; \params)} \\
	&\geq 
	\priorsmall
	(\hessbig \, \numpergroup_i)^{-\redim/2 - \eps} \left[ (2\pi)^{\redim/2}  - 
	\frac{1}{6} \redim^3 (\hesssmall \, \numpergroup_i)^{-3/2 + \eps} \ \quadpointbound
	\
	\max_{\quadpointvec\in\quadpointset, (i_1, i_2, i_3 ) \in [\redim] } \abs{\partial^{i_1 i_2 i_3} \llhoodip(\requadmid)}
	\cdot \sum\limits_{\quadpointvec\in\quadpointset} 
	\abs{\weight(\quadpointvec)}  \right],
\]
where $\requadmid = \midweight_{\quadpointvec} (\cholip\,\quadpointvec+\condmodeip) + (1-\midweight_{\quadpointvec}) \condmodeip$ for some $\midweight_{\quadpointvec} \in [0,1]$ and for all $\eps > 0$.
Next, if $\hessbigmodeip  \leq \hessbig$,
\*[
	\abssmall{\cholip}
	= \sqrt{\absbig{[\hessip]^{-1}}} 
	= \absbig{\hessip}^{-1/2} 
	\geq \Big[\eigen_1(\hessip)\Big]^{-\redim/2 - \eps}
	= (\hessbigmodeip  \, \numpergroup_i)^{-\redim/2 -\eps}
	\geq (\hessbig \, \numpergroup_i)^{-\redim/2 -\eps}.
\]
Thus, by \cref{fact:constant_convergence} part i) and \cref{assn:kderiv},
\*[
	\lim_{\minpergroup \to \infty}
	\trueprob{\numobs}
	\Big[ \forall_{i = 1, \dots, \numgroups} \ 
	\sup_{\params\in\paramsbigball}
	\max_{\quadpointvec\in\quadpointset, (i_1, i_2, i_3 ) \in [\redim] } \abs{\partial^{i_1 i_2 i_3} \llhoodip(\requadmid)}
	\leq
	\numpergroup_i^{1 +\eps} \cdot \derivbound
	\Big]
	= 1.
\]
That is, for all $\eps^\prime>0$,
\*[
	\lim_{\minpergroup \to \infty}
	\trueprob{\numobs}
	\Big[
		&\forall_{i = 1, \dots, \numgroups}
	\sup_{\params\in\paramsbigball}
	\frac{1}{6} \redim^3 (\hesssmall \, \numpergroup_i)^{-3/2 + \eps} \quadpointbound
	\ \cdot
	\\
	&\max_{\quadpointvec\in\quadpointset, (i_1, i_2, i_3 ) \in [\redim] } \abs{\partial^{i_1 i_2 i_3} \llhoodip(\requadmid)}
	\cdot \sum\limits_{\quadpointvec\in\quadpointset} 
	\abs{\weight(\quadpointvec)}
	\leq
	\eps^\prime
	\Big]
	= 1.
\]
The statement of the lemma then follows from \cref{assn:consistency,assn:prior,assn:hessian}, as well as \cref{fact:constant_convergence}.
\manualendproof

\section{Supporting results for the proof of Theorem \ref{fact:consistency}}\label{sec:proofs-consistency}

\noindent The proof of \cref{fact:consistency} requires the following additional lemma:

\begin{lemma} \label{lemma:hessian-mle}
	For $0 < \hesssmall^\prime $, and for all $\eps > 0$ and for $\numgroups = \minpergroup^q$ for any $q> 0$
	\*[
		\lim_{\minpergroup \to \infty} \trueprob{\numobs}\left[\hesssmall^\prime \sum_{i=1}^\numgroups n_i^{1- \eps} \leq
		\inf_{\params\in\paramsbigball}
		\eigen_{\paramdim}(-\partial^{2}_{\params}\logmargdens(\obsall; \params)) \right] = 1.
	\]
\end{lemma}

\begin{proof}
	Recall that $\logmargdens(\obsall;\params) = \log\margdens(\obsall;\params) = \log\int\margdens(\obsall;\re, \params)\dee\re = \sum_{i=1}^{\numgroups}\log\int\margdens(\obsall_i;\re_i, \params)\dee\re_i$.
	Using Liebniz rule for exchanging integration with differentiation we have the following for all $\params\in\paramsbigball$:
	\*[
		\eigen_{\paramdim}(-\partial^{2}_{\params}\logmargdens(\obsall; \params)) &= \eigen_{\paramdim}\left(-\partial^{2}_{\params}\sum_{i=1}^{\numgroups}\log\int_{\re_i \in \Reals^\redim} \margdens(\obsall_i;\re_i, \params)\dee\re_i\right), \\
		&\geq \sum_{i=1}^{\numgroups} \eigen_{\paramdim}\left( -\partial^{2}_{\params}\log\int_{\re_i \in \Reals^\redim} \margdens(\obsall_i;\re_i, \params) \dee\re_i\right), \\
	\]
	by Weyl's inequality.
	The remainder of the proof is dedicated to showing that
	\*[
	\eigen_{\paramdim}\left( -\partial^{2}_{\params}\log\int_{\re_i \in \Reals^\redim} \margdens(\obsall_i;\re_i, \params) \dee\re_i\right) \geq C n_i^{1- \eps},
	\]
	for a constant $C>0$ independent of $i$, from which the statement of the Lemma follows.

	Note that we may write: 
	\[\label{eqn:hessian-split}
		&\log\int_{\re_i \in \Reals^\redim}\margdens(\obsall_i;\re_i, \params) \dee\re_i \\
		&= \underbrace{\log\left( \frac{\int_{\re_i \in \reball} \margdens (\obsall_i;\re_i, \params) \dee\re_i}{\int_{\paramsbigballfixed}\int_{\re_i \in \reball}\margdens (\obsall_i;\re_i, \params) \dee\re_i  \dee\params }\right)}_{A} + \underbrace{\log\left( \frac{\int_{\re_i \in \Reals^{\redim}} \margdens (\obsall_i;\re_i, \params) \dee\re_i}{\int_{\re_i \in \reball}\margdens (\obsall_i;\re_i, \params) \dee\re_i  }\right)}_B \\
		&+ \underbrace{\log\left(\int_{\paramsbigballfixed}\int_{\re_i \in \reball}\margdens (\obsall_i;\re_i, \params) \dee\re_i  \dee\params \right)}_C.
	\]
	The $C$ term in \cref*{eqn:hessian-split} is $0$ when differentiated with respect to $\params$, so it can be ignored. 
	The $A$ term in \cref*{eqn:hessian-split} is the marginal of a truncated posterior distribution of $\params$ with uniform prior on $\params$ on $\paramsbigballfixed$, truncated to $\paramsbigballfixed \times \reball$;
	by \cref{assn:hessian}, the smallest eigenvalue of the negative hessian of the log-density of this truncated posterior of $(\params, \re)$ is uniformly lower bounded by $\hesssmall\numpergroup_i^{1-\eps}$, implying it is a
	strongly log-concave distribution. As strong log-concavity is preserved by marginalization, it follows that the negative hessian of the first term is also lower bounded by $\hesssmall\numpergroup_i^{1- \eps}$, see \cite{saumard2014log} Proposition 2.24 c) and Theorems 3.8 for the statement and proof of these results. 

	We now show that the largest singular value of the matrix $B$  in \cref*{eqn:hessian-split} is asymptotically negligible. When differentiated with respect to $\params$, $B$ equals:
	\[\label{eqn:big-expansion-negligible}
		&-\partial^{2}_{\params} \log\left( \frac{\int_{\re_i \in \Reals^{\redim}} \margdens (\obsall_i;\re_i, \params) \dee\re_i}{\int_{\re_i \in \reball}\margdens (\obsall_i;\re_i, \params) \dee\re_i  }\right)\\
		&= -\partial^{2}_{\params} \log\left(\int_{\re_i \in \Reals^{\redim}} \exp \{ \log\margdens (\obsall_i;\re_i, \params) \}\dee\re_i \right) + \partial^{2}_{\params} \log\left(\int_{\re_i \in \reball} \exp \{ \log\margdens (\obsall_i;\re_i, \params) \}\dee\re_i \right)\\
		&= \underbrace{\frac{\int_{\re_i \in \Reals^{\redim}} -\partial^{2}_{\params} \log\margdens (\obsall_i;\re_i, \params) \margdens (\obsall_i;\re_i, \params) \dee\re_i  }{\int_{\re_i \in \Reals^{\redim}} \margdens (\obsall_i;\re_i, \params)\dee\re_i} - \frac{\int_{\re_i \in \reball} -\partial^{2}_{\params} \log\margdens (\obsall_i;\re_i, \params)\margdens (\obsall_i;\re_i, \params) \dee\re_i  }{\int_{\re_i \in \reball} \margdens (\obsall_i;\re_i, \params) \dee\re_i}}_{D}\\
		&+ \left( \frac{\int_{\re_i \in \Reals^{\redim}} \partial_{\params} \log\margdens (\obsall_i;\re_i, \params) \margdens (\obsall_i;\re_i, \params)\dee\re_i  }{\int_{\re_i \in \Reals^{\redim}} \margdens (\obsall_i;\re_i, \params) \dee\re_i} \right) \left( \frac{\int_{\re_i \in \Reals^{\redim}} \partial_{\params}^\top \log\margdens (\obsall_i;\re_i, \params) \margdens (\obsall_i;\re_i, \params) \dee\re_i  }{\int_{\re_i \in \Reals^{\redim}} \margdens (\obsall_i;\re_i, \params) \dee\re_i} \right) \\
		&- \left( \frac{\int_{\re_i \in \reball} \partial_{\params} \log\margdens (\obsall_i;\re_i, \params)\margdens (\obsall_i;\re_i, \params) \dee\re_i  }{\int_{\re_i \in \reball} \margdens (\obsall_i;\re_i, \params) \dee\re_i} \right) \left( \frac{\int_{\re_i \in \reball} \partial_{\params}^\top \log\margdens (\obsall_i;\re_i, \params) \margdens (\obsall_i;\re_i, \params) \dee\re_i  }{\int_{\re_i \in \reball} \margdens (\obsall_i;\re_i, \params) \dee\re_i} \right), 
	\]
	these terms can also be interpreted as the conditional expectation under the posterior measure, and of the truncated posterior measures of $\params | \re$. Consider $D$,  
	\[\label{eqn:hess-A-B-split}
		&\frac{\int_{\re_i \in \Reals^{\redim}} -\partial^{2}_{\params} \log\margdens (\obsall_i;\re_i, \params) \margdens (\obsall_i;\re_i, \params) \dee\re_i  }{\int_{\re_i \in \Reals^{\redim}} \margdens (\obsall_i;\re_i, \params)\dee\re_i} - \frac{\int_{\re_i \in \reball} -\partial^{2}_{\params} \log\margdens (\obsall_i;\re_i, \params)\margdens (\obsall_i;\re_i, \params) \dee\re_i  }{\int_{\re_i \in \reball} \margdens (\obsall_i;\re_i, \params) \dee\re_i}	\\
		&= \underbrace{\frac{\int_{\re_i \in \reball^C} -\partial^{2}_{\params} \log\margdens (\obsall_i;\re_i, \params) \margdens (\obsall_i;\re_i, \params) \dee\re_i  }{\int_{\re_i \in \Reals^{\redim}} \margdens (\obsall_i;\re_i, \params)\dee\re_i}}_{D.1} \\
		&+ \underbrace{\left(\frac{\int_{\re_i \in \Reals^\paramdim} \margdens (\obsall_i;\re_i, \params) \dee\re_i  }{\int_{\re_i \in \reball} \margdens (\obsall_i;\re_i, \params)\dee\re_i}  - 1\right) \frac{\int_{\re_i \in \reball} -\partial^{2}_{\params} \log\margdens (\obsall_i;\re_i, \params) \margdens (\obsall_i;\re_i, \params) \dee\re_i  }{\int_{\re_i \in \Reals^\paramdim} \margdens (\obsall_i;\re_i, \params)\dee\re_i}}_{D.2}.	
	\]
	Take $\delta^\prime > 0$ to be a constant to be specify later, multiply and divide $D.1$ by $\exp\{ - \log\margdens (\obsall_i;\retrue, \params)\}$ and take it's operator norm, resulting in
	\[\label{eqn:ratio-lemma-hess}
	&\frac{ \norm{\int_{\re_i \in \reball^C} -\partial^{2}_{\params} \log\margdens (\obsall_i;\re_i, \params) \exp\{ \log\margdens (\obsall_i;\re_i, \params) - \log\margdens (\obsall_i;\retrue, \params)\}\redens(\re_i) \dee\re_i}_{op} }{\int_{\re_i \in \Reals^{\redim}} \exp\{ \log\margdens (\obsall_i;\re_i, \params) - \log\margdens (\obsall_i;\retrue, \params)\}\redens(\re_i) \dee\re_i} \\
	&\leq \frac{\exp( - \numpergroup_i^{1 - \eps} \llhoodmargin ) \int_{\re_i \in \reball^C } \norm{ -\partial^{2}_{\params} \log\margdens (\obsall_i;\re_i, \params) \redens(\re_i)}_{op} \dee\re_i}{\int_{\re_i \in \ball{\retrue}{\delta_{\numpergroup_i}^\prime} } \exp\{ \log\margdens (\obsall_i;\re_i, \params) - \log\margdens (\obsall_i;\retrue, \params)\}\redens(\re_i) \dee\re_i}\\
	&\leq \frac{\exp( - \numpergroup_i^{1- \eps} \llhoodmargin ) \int_{\re_i \in \reball^C }\norm{ -\partial^{2}_{\params} \log\margdens (\obsall_i;\re_i, \params) \redens(\re_i)}_{op} \dee\re_i}{\int_{\re_i \in \ball{\retrue}{\delta_{\numpergroup_i}^\prime} } \exp\{ \log\margdens (\obsall_i;\re_i, \params) - \log\margdens (\obsall_i;\retrue, \params)\}\redens(\re_i) \dee\re_i},
	\]
	for every $\eps > 0$ and a constant $\llhoodmargin > 0$ independent of $i$ by \cref{assn:limsup} and $\delta_{\numpergroup_i} = \numpergroup_i^{\alpha^\prime}$ for some value of $0<\alpha^\prime$ to be specified later.
	We now upper bound the numerator and denominator of this fraction separately. The numerator of $D.1$ is upper bounded by
	\*[
		&\int_{\re_i \in \reball^C } \norm{-\partial^{2}_{\params} \log\margdens (\obsall_i;\re_i, \params)}_{op} \redens(\re_i) \dee\re_i \leq \int_{\re_i \in \Reals^\redim }\norm{ -\partial^{2}_{\params} \log\margdens (\obsall_i;\re_i, \params)}_{op} \redens(\re_i) \dee\re_i\\
		& = O(\numpergroup_i^k)
	\]
	by \cref{assn:prior}, where the $O(\cdot)$ term is uniform in $i$.
	We now bound the denominator of $D.1$ by writing it as:
	\*[
		 \exp\left\{ \underbrace{ \log\margdens (\obsall_i;\re_i, \params) - \log\margdens (\obsall_i;\condmodeip, \params)}_E + \underbrace{\log\margdens (\obsall_i;\condmodeip, \params) - \log\margdens (\obsall_i;\retrue, \params) }_F \right\}.
	\]
	The $F$ term:
	\*[
		\log\margdens (\obsall_i;\condmodeip, \params) - \log\margdens (\obsall_i;\retrue, \params)> 0.
	\]
	While, by a second order Taylor expansion, the $E$ term:
	\[ \label{eqn:max-eigen-lemma}
		\log\margdens (\obsall_i;\re_i, \params) - \log\margdens (\obsall_i;\condmodeip, \params) &= (\re_i - \condmodeip)^\top  \{\partial^{2}_{\params} \log\margdens (\obsall_i;\re_i, \params^\star)\} (\re_i - \condmodeip)\\
		&\geq - (\delta_{\numpergroup_i}^\prime)^2 \numpergroup_i^{1+\eps} \hessbig = - \numpergroup_i^{1+\eps -\alpha^\prime} \hessbig,
	\]
	for $\hessbig$ independent of $i$ and where
	\*[
		\left\{ \partial^{2}_{\params} \log\margdens (\obsall_i;\re_i, \params^\star) \right\}= \int_0^1 (1 - t) \partial^{2}_{\params} \log\margdens (\obsall_i;\condmodeip +t \re_i, \params)\dee t,
	\]
	by \cref{assn:hessian}, as
	for any positive definite matrix $A(t)$ indexed by a scalar random variable $t$, it is the case that by the variational representation of the maximal eigenvalue and the linearity of expectations that: 
	\[ \label{eq:eigen-lower}
		\eigen_{1} \left(\int A(t) \dee t \right) &= \max_{\norm{x} = 1} x^\top \left( \int A(t) \dee t\right)  \ x \\
		&= \max_{\norm{x} = 1}  \int x^\top A(t)  x \ \dee t\\
		&\leq    \int  \max_{\norm{x} = 1} x^\top A(t) x \ \dee t = \int \eigen_{1} (A(t)) \ \dee t.
	\]
	
	Combining the bounds on $E$ and $F$, we have that the denominator of $D.1$ is lower bounded by:
	\*[
		&\int_{\re_i \in \ball{\retrue}{\delta_{\numpergroup_i}^\prime} } \exp\{ \log\margdens (\obsall_i;\re_i, \params) - \log\margdens (\obsall_i;\retrue, \params)\}\redens(\re_i) \dee\re_i	\\
		&\geq  - 2(\delta^\prime)^2 \numpergroup_i^{1 + \eps - \alpha} \hessbig \int_{\re_i \in \ball{\retrue}{\delta_{\numpergroup_i}^\prime} }\redens(\re_i) \dee\re_i\\
		&= \exp\{- 2(\delta^\prime)^2 \numpergroup_i^{1 + \eps - \alpha^\prime} \hessbig\}  \PP[\re \in \ball{\retrue}{\delta_{\numpergroup_i}^\prime} ].
	\]
	We can lower bound the probability in the above line by:
	\*[
		\PP[\re \in \ball{\retrue}{\delta_{\numpergroup_i}^\prime} ] &= \int_{\ball{\retrue}{\delta_{\numpergroup_i}^\prime}} \frac{1}{ (2\pi)^{\redim/2} |\covmat(\resd)|^{\redim/2}}  \exp\left(-\frac{1}{2} \re\Tr \covmat(\resd) \re \right) \dee \re\\
		&\geq\int_{\ball{\retrue}{\delta_{\numpergroup_i}^\prime}} \frac{1}{ (2\pi)^{\redim/2} (\norm{\covmat(\resd)}_{op} )^{\redim/2}} \exp\left(-\frac{1}{2\norm{\covmat(\resd)}_{op}}\left\{ \norm{\re - \retruepi}_2^2 + \norm{\retruepi}_2^2 \right\} \right) \dee \re\\
		&\geq \frac{1}{ (2\pi)^{\redim/2} (\norm{\covmat(\resd)}_{op} )^{\redim/2}} \exp\left(-\frac{\left\{ \delta_{\numpergroup_i}^2 + \norm{\covmat(\resd)}_{op}\log(\numgroups) \right\}}{2\lambda_\redim(\covmat(\resd)) } \right) \int_{\ball{\retrue}{\delta_{\numpergroup_i}^\prime}} \dee \re\\
		&\geq \frac{ (\delta^\prime_{\numpergroup_i})^2 }{ (2\pi)^{\redim/2} (\norm{\covmat(\resd)}_{op} )^{\redim/2}} \exp\left(-\frac{\left\{ \delta_{\numpergroup_i}^2 + \norm{\covmat(\resd)}_{op}\log(\numgroups) \right\}}{2\lambda_\redim(\covmat(\resd)) } \right) \\
		&\geq \exp( C(\delta^\prime_{\numpergroup_i})^2)  = \exp(C\numpergroup_i^{-2\alpha^\prime})
	\]
	for some $C> 0$, as $\max_{i = 1,\dots, \numgroups} \norm{\retruepi}_2^2 = O(\log(\numgroups))$ with probability tending to $1$ by Theorem 1.14 in \cite{rigollet2023high}.
	Combining the bounds on the numerator and denominator of $D.1$: 
	\*[
		&\frac{ \norm{\int_{\re_i \in \reball^C} -\partial^{2}_{\params} \log\margdens (\obsall_i;\re_i, \params) \exp\{ \log\margdens (\obsall_i;\re_i, \params) - \log\margdens (\obsall_i;\retrue, \params)\}\redens(\re_i) \dee\re_i}_{op} }{\int_{\re_i \in \Reals^{\redim}} \exp\{ \log\margdens (\obsall_i;\re_i, \params) - \log\margdens (\obsall_i;\retrue, \params)\}\redens(\re_i) \dee\re_i} \\
		&\leq O(\numpergroup_i^k)	\exp\{ -\numpergroup_i^{1 - \eps} \llhoodmargin+ 2 \numpergroup_i^{1 + \eps - \alpha^\prime } \hessbig\}   \exp(C\numpergroup_i^{2\alpha^\prime}) =  o(1),
	\]
	by taking $\eps < \alpha^\prime < 1/2 - \eps/2$, uniformly in $i$.
	Now consider $D.2$ in \cref{eqn:hess-A-B-split}, note that for any fixed value of $\params \in \paramsbigball$
	\*[
		\left|\frac{\int_{ \re_i \in \Reals^{\redim}}  \margdens (\obsall_i;\re_i, \params) \dee\re_i  }{\int_{\re_i \in \reball} \margdens (\obsall_i;\re_i, \params)\dee\re_i} - 1 \right|\leq \frac{\int_{ \re_i \in \reball^C}  \margdens (\obsall_i;\re_i, \params) \dee\re_i  }{\int_{\re_i \in \reball} \margdens (\obsall_i;\re_i, \params)\dee\re_i} = O(\exp(-\numpergroup_i^{1 - \eps})), 
		\]
	by the same arguments used to bound D.1. Secondly, 
	\*[
		 \frac{\int_{\re_i \in \reball} \norm{ -\partial^{2}_{\params} \log\margdens (\obsall_i;\re_i, \params)}_{op} \margdens (\obsall_i;\re_i, \params) \dee\re_i  }{\int_{\re_i \in \Reals^\paramdim} \margdens (\obsall_i;\re_i, \params)\dee\re_i}  \leq \numpergroup_i^{k}  \hessbig, 
	\]
	thus the term $D.2 =  O(\exp(-\numpergroup_i^{1 - \eps}))$, which decays exponentially and is therefore asymptotically negligible, thereby showing that the entirely of $D$ is negligible.

	As for the other terms in \cref{eqn:big-expansion-negligible} which we did not yet consider, the same proof strategy applies, the only difference being that we use that prior expectation of the first derivative to be finite and we can use the Frobenius norm to upper-bound the largest eigenvalue of a matrix. Therefore the $B$ term in \cref*{eqn:hessian-split} is uniformly negligible, showing the desired result.

	
\end{proof}

\section{Proof of Proposition \ref{prop:glmmregularity}}\label{sec:proofs-exponentialfamily}

\subsection{Regularity conditions}\label{subsec:reg-expfam}

\noindent \cref{prop:glmmregularity} holds for all well-specified GLMMs for which the response
distribution is non-degenerate.
\cref{assn:non-degen,assn:well-spec} formalize what is meant by
these notions.
In the Assumptions which follow, assume that $\numgroups = \numpergroup_{\min}^q$ for some $q > 0$ and that $\numpergroup_i$ are increasing at the same rate.

\begin{assumption}[Non-Degenerate]\label{assn:non-degen}
The \glmm{} is \emph{non-degenerate} if
there exist $\universalradius>0$ and $\paramstrue\in\paramspace$
such that each of the following six statements are true with $\numgroups = \numpergroup_{\min}^q$ and $\numpergroup_{\min} \rightarrow \infty$.
\begin{itemize}
 		\item[(i)] The natural parameter space, $$\natparamspace = \left\{ \linpred\in\Reals: \int_{\samplespace}\exp\left\{ \frac{\obsi\linpred}{a(\obssd)} + c(y;\obssd)\right\}\dee\obsi<\infty\right\},$$is an open subset of $\Reals$ containing $\linpred=0$.
 		\item[(ii)] For every $C>0$,
 		$\inf_{\abs{\linpred}<C\log(\numpergroup_{\min})^{1-\eps}}\abs{b^{\prime\prime}(\linpred)}>\minpergroup^{-\eps}$, for all $\eps > 0$.
 		\item[(iii)] There exist $0 < \gausscovsmall \leq \gausscovbig < \infty$ such that 
		$$0<\gausscovsmall\leq\inf_{\params\in\paramsbigballfixed}\eigen_\redim\left\{\covmat(\resd)\right\}\leq\sup_{\params\in\paramsbigballfixed}\eigen_1\left\{\covmat(\resd)\right\}\leq\gausscovbig<\infty.$$
 		Further, there exists $\gausscovderivbig$ for all $\dumderivvec \subseteq \Nats^\resddim$ with $\abs{\dumderivvec} \in \{1,2\}$, $$\sup_{\params\in\paramsbigballfixed}\eigen_1\left\{\frac{\partial^{\abs{\dumderivvec}}}{\partial\resd^{\dumderivvec}}\covmat(\resd)\right\}\leq\gausscovderivbig<\infty.$$
		\item[(iv)] There exist $0 < \recovsmall \leq \recovbig < \infty$ such that for each $i\in[\numgroups]$, $$0<\recovsmall \numpergroup_i\leq\eigen_\redim\left\{\fulldesign_i\Tr\fulldesign_i\right\}\leq\eigen_1\left\{\fulldesign_i\Tr\fulldesign_i\right\}\leq\recovbig\numpergroup_i <\infty,$$ where $\fulldesign_i = [\design_i:\redesign_i]$ with $\design_i = (\cov_{i1}\Tr,\ldots,\cov_{i\numpergroup_i}\Tr)\Tr$ and $\redesign_i = (\recov_{i1}\Tr,\ldots,\recov_{i\numpergroup_i}\Tr)\Tr$. Furthermore for some $\epsilon > 0$, $ \max_{i,j} \lVert \cov_{ij} \rVert_2 = O(\log(\numpergroup_{\min})^{1 - \epsilon})$ and $ \max_{i ,j} \lVert \recov_i \rVert_2 = O(\log(\numpergroup_{\min})^{\frac{1 - \epsilon}{2}})$.
		\item[(v)] The derivatives $b^\prime (\eta)$ and $b^{\prime\prime} (\eta)$ are $o(\exp(\eta^{2- \eps}))$and for all $C>0$ 
		$$\sup_{|\eta| < C\log(n)^{1-\epsilon}} |b^{(k)}(\eta)| \leq n^\epsilon $$ 
		for all $\epsilon > 0$ and $k\in \Nats$.  
		\item[(vi)] For all $C > 0$ and all $\eps > 0$,  $\sup_{|\eta| < C\log(n )^{1-\epsilon}}\mathbb{E}[|c(Y;\obssd)|^{k}] = O(n^\epsilon)$, for all integers $k$ such that $k < q + 1$, where the expectation is taken with respect to the distribution of $Y$ with natural parameter $\eta$. 
 	\end{itemize}
\end{assumption}

\begin{assumption}[Well-Specified and Consistency]\label{assn:well-spec}
The \glmm{} is \emph{well-specified} if there exists $\paramstrue\in\paramspace$ and $\retrueptrue\in\respace$ for each $i\in[\numgroups]$ such that for each $j\in[\numpergroup_i]$, $\obsi_{ij} \setdelim \cov_{ij},\recov_{ij} \indsim\responsedist(\cov_{ij}\Tr\regparamtrue + \recov_{ij}\Tr\retrueptrue)$. Furthermore assume that the marginal maximum likelihood estimator for $\paramtrue$ is consistent at a rate of $\numgroups^{\eps}$ for some $\eps>0$.
\end{assumption}

\begin{remark}
In exponential models considered, the derivatives of the log-likelihood are functions of the natural parameter and the restriction of $|\eta| < C\log(n)^{1-\epsilon}$ in \cref{assn:non-degen} (ii), (V) and $(vi)$ is used to control the size of these derivatives.
\end{remark}

\begin{remark}
	\cref{assn:non-degen} i) and iv) are standard assumptions, while
	\cref{assn:non-degen} v) and vi) are satisfied for most commonly used exponential family such as the inverse normal, normal, gamma, Poisson and the binomial distribution. \cref{assn:non-degen} v) requires that growth rates of the mean and variance of the random variable as a function of the natural parameter $\eta$ are slower than $\exp(\eta^2)$ which is true of all models listed above. 
	\cref{assn:non-degen} vi) can be shown to hold for the Gamma distribution through Stirling approximation, for the inverse normal this condition is implied by the fact that the inverse $k$-th moments exists for all $k$ and for the other listed distributions this can be shown through direct calculation; example calculations for the Poisson is given below.
\end{remark}

\begin{example}
	We provide sample calculations for Poisson regression to show that \cref{assn:non-degen} v) and vi) are satisfied. This case is one of the more difficult to check as the mean function is increasing exponentially in the natural parameter, contrary to most other cases of GLMs used in practice.
	For \cref{assn:non-degen} v), note that for the Poisson $b(\eta) = \exp(\eta)$: 
	\*[
		\sup_{|\eta| < C\log(n)^{1-\epsilon}} b^{(k)}(\eta) = \sup_{|\eta| < C\log(n)^{1-\epsilon}} \exp(\eta) \leq \exp(C\log(n)^{1-\eps}) = o(n^\eps),
	\]
	for all $\eps > 0$. While for the Poisson distribution $c(Y) = \log(Y!) \leq Y^{1 + \eps^\prime}$ for any arbitrary $\eps^\prime > 0$ by Stirling's approximation, therefore for $k \in \Nats$:
	\*[
	\sup_{|\eta| < C\log(n)^{1-\epsilon}} \EE[ (Y^{1 + \eps^\prime})^k ] \leq \sup_{|\eta| < C\log(n)^{1-\epsilon}} \EE[ Y^{k+1} ] \leq k \exp( C(k+1)\log(n)^{1-\epsilon} ) = o(n^\eps),
	\]
	for any $\eps >0$ showing the desired result.
\end{example}

\begin{remark}
	The rate of consistency of the marginal maximum likelihood estimator is shown to be at the very least $\numgroups^{1/2}$ in \cite{jiang21glmm}, therefore \cref{assn:well-spec} can be easily satisfied.
\end{remark}

\subsection{Proof}\label{subsec:proof-expfam}

\noindent We now state a Lemma which will help us bound the natural parameter of the observations uniformly to control key quantities which will appear in the proof.

\begin{lemma}\label{lem:exponentialfamilyderivatives}
Under \cref{assn:non-degen} (iii) and (iv), for all $\params\in\paramsbigballfixed$ and $\re\in\reball$ there exist some constants $\linpredbound <\infty$ and $\epsilon > 0$: 
\[\lim_{n_{\min} \rightarrow \infty}\trueprob{\numobs} \left[
\max_{i, j} \abs{\linpredp_{ij}(\re)}  \leq \linpredbound \log(\numpergroup_{\min})^{1-\epsilon} \right] = 1. \label{eqn:eta-bound}
\]
\end{lemma}
\begin{proof}
For $\params\in\paramsbigballfixed$ and $\re\in\reball$, the following holds for some unit vector $z$ and $z^\prime$
\begin{align*}
	\abs{\linpredp_{ij}(\re)} &= \abs{\cov_{ij}^\top (\regparamtrue + \delta z)+ \recov_{ij}^\top (\retruepi + \delta z^\prime) }\\
	&\leq \lVert\cov_{ij} \rVert_2 \lVert\regparamtrue \rVert_2 + \delta(\lVert\cov_{ij}\rVert  +\lVert\recov_{ij}\rVert_2 )+ \lVert\recov_{ij} \rVert_2 \lVert\retruepi\rVert_2,
\end{align*}
where the first two terms are uniformly $O(\log(n)^{1 - \epsilon})$ and $O(\log(n)^{\frac{1-\epsilon}{2}} )$ by \cref{assn:non-degen} (iv), therefore it remains to bound the final term uniformly in probability. Note that $\retruepi$ follows a zero mean Gaussian whose covariance matrix has eigenvalues bounded by $\gausscovbig$ by \cref{assn:non-degen} iii), therefore  for each $i$, $\lVert\retruepi\rVert_2^2$ is stochastically dominated by a $\gausscovbig\chi^2_{\redim, i}$ Chi-squared random variable. By Lemma D.2 in \cite{tang2021laplace}, $\max_{i = 1,\dots ,\numgroups}(\chi^2_{\redim, i})^{1/2} = O(\log(\numgroups)^{1/2} ) = O(\log(\numpergroup_{\min})^{1/2} )$ with probability $O(1/\numgroups)$. Noting that $\numgroups \rightarrow \infty$ and that by Assumption $\max_{i ,j} \lVert \recov_i \rVert_2  = O(\log(n)^{\frac{1-\epsilon}{2}} )$ shows the desired result.

\end{proof}

We now control the probability that certain averages used in the subsequent proofs will be close to their expected value jointly as both $\numgroups \rightarrow \infty$ and $\numpergroup \rightarrow \infty$; recall that $m = n_{\min}^q$ for $q > 0$. The $\epsilon$ factor is needed here for the uniform control of the likelihood derivatives as the number of groups increases.

\begin{lemma}\label{lemma:uniform-markov}
	Under \cref{assn:non-degen}, there exists constants $B_1,B_2,B_3 > 0$ and a polynomial function $G(x)$ where $\lim_{x\rightarrow\infty} G(x) = \infty$ such that for all $\epsilon > 0$
	\begin{itemize}
		\item $\lim_{n_{\min} \rightarrow \infty}\trueprob{\numobs}\left(\forall_{i = i, \dots, m} \  \frac{1}{\numpergroup_i}\sum^{\numpergroup_i}_{j = 1} \left| \obsi_{ij} \right| > B_1 \numpergroup_{\min}^\epsilon \right) \rightarrow 0,$
		\item $\lim_{n_{\min} \rightarrow \infty}\trueprob{\numobs} \left(\forall_{i = i, \dots, m} \ \forall_{k = 1, \dots, \redim} \ \left| \frac{1}{\numpergroup_i}\sum^{\numpergroup_i}_{j = 1} \{\obsi_{ij}\recov_{jk} - \EE\obsi_{ij}\recov_{jk} \}\right| > \frac{B_2}{G(n_{\min}) } \right) \rightarrow 0.$
		\item $\lim_{n_{\min} \rightarrow \infty}\trueprob{\numobs} \left(\forall_{i = i, \dots, m} \ \forall_{k = 1, \dots, \redim} \ \left| \frac{1}{\sqrt{\numpergroup_i}}\sum^{\numpergroup_i\}}_{j = 1} \{\obsi_{ij}\recov_{jk} - \EE\obsi_{ij}\recov_{jk} \}\right| > B_2 \numpergroup_{\min}^\eps \right) \rightarrow 0.$
		\item $\lim_{n_{\min} \rightarrow \infty}\trueprob{\numobs} \left(\forall_{i = i, \dots, m} \  \frac{1}{\numpergroup_i}\sum^{\numpergroup_i}_{j = 1} |c(y_{ij};\phi)| > B_3 \numpergroup_{\min}^\eps \right) \rightarrow 0.$
	\end{itemize}
\end{lemma}

\begin{proof}
	These four inequalities will be proven with Markov's inequality and a union bound. 
	For each $i$ fixed and for any $t> 0$:
	\*[ \trueprob{\numobs}\left( \frac{1}{\numpergroup_i}\sum^{\numpergroup_i}_{j = 1} \left| \obsi_{ij} \right| > t \right) \leq \trueprob{\numobs}\left( \sqrt{\frac{1}{\numpergroup_i}\sum^{\numpergroup_i}_{j = 1}  \obsi_{ij}^2 } > t \right) = \trueprob{\numobs}\left( \frac{1}{\numpergroup_i}\sum^{\numpergroup_i}_{j = 1}  \obsi_{ij}^2  > t^2 \right),\]
	as $\sum^{\numpergroup_i}_{j = 1} \left| \obsi_{ij} \right| \leq \sqrt{\numpergroup_i} \sqrt{\sum^{\numpergroup_i}_{j = 1} \obsi_{ij}^2}$. 
	First consider the centred version of these sums, for some $t^\prime > 0$ and any even integer $a$:
	\begin{align*}
		\trueprob{\numobs}\left( \frac{1}{\numpergroup_i}\sum^{\numpergroup_i}_{j = 1} \left( \obsi_{ij}^2 - \EE\obsi_{ij}^2 \right) > t^\prime \right) &= 
		\trueprob{\numobs}\left( \left\{ \frac{1}{\numpergroup_i}\sum^{\numpergroup_i}_{j = 1} \left( \obsi_{ij}^2 - \EE\obsi_{ij}^2 \right) \right\}^a > (t^\prime)^a \right) \\
		&\leq \frac{\EE \left(\left\{\sum^{\numpergroup_i}_{j = 1} \left( \obsi_{ij}^2 - \EE\obsi_{ij}^2 \right) \right\}^a \right) }{\numpergroup_i^a (t^\prime)^a} \leq \frac{D_a \numpergroup_{\min}^\epsilon}{\numpergroup_i^{a/2}(t^\prime)^a},
	\end{align*}
for some constant which only depends on $a$ as only terms with even parity will be non-zero in above expectation by independence the observation for fixed $i$ and from \cref{lem:exponentialfamilyderivatives} and \cref{assn:non-degen} v) the moments of $\obsi_{ij}^2$ are uniformly bounded by any polynomial of $\numpergroup_{\min}$. Next, it is also the case that:
\begin{align*}
	\frac{1}{\numpergroup_i}\sum^{\numpergroup_i}_{j = 1} \EE \obsi_{ij}^2 = O(\numpergroup_{\min}^\eps) 
\end{align*}
uniformly in $i$ by \cref{lem:exponentialfamilyderivatives} and \cref{assn:non-degen} v). Combining these bounds and letting $t^\prime = 1$:
\*[
	\trueprob{\numobs}\left(\forall_{i = i, \dots, m} \  \frac{1}{\numpergroup_i}\sum^{\numpergroup_i}_{j = 1} \left| \obsi_{ij} \right| > B_1 \right) \leq \trueprob{\numobs}\left(\forall_{i = i, \dots, m} \ \frac{1}{\numpergroup_i}\sum^{\numpergroup_i}_{j = 1}  \obsi_{ij}^2  > 1 + E \right) \leq \frac{\numgroups D_a}{\numpergroup_{\min}^{a/2 -\epsilon}} \rightarrow 0
\]
as by the Assumption that $m =\minpergroup^q$ for some $q> 0$, $m = o(\numpergroup_{\min}^{a/2 - \epsilon})$ for some even value of $a$.

For the proof of the second statement, we use \cref{lem:exponentialfamilyderivatives} and \cref{assn:non-degen} v) to obtain
\*[
	\trueprob{\numobs}\left(\forall_{i = i, \dots, m} \ \forall_{k = 1, \dots, \redim} \ \left| \frac{1}{\numpergroup_i}\sum^{\numpergroup_i}_{j = 1} \{\obsi_{ij}\recov_{jk} - \EE\obsi_{ij}\recov_{jk} \}\right| > t \right) \leq \frac{\redim mD_a}{\numpergroup_{\min}^{a/2 - \epsilon}t^a},
\] 
from similar arguments as above. Letting $t = o(\numpergroup_{\min}^{-q^\prime})$ for $q^\prime < a/2 - q - \epsilon$ gives the desired result.
The third statement follows by simple substitution and noting that by \cref{assn:non-degen} v) $a$ in the above bound can be made arbitrarily large and $\eps$ can be made arbitrarily small. For any $\eps^\prime > 0$, specifically
\*[
	\trueprob{\numobs}\left(\forall_{i = i, \dots, m} \ \forall_{k = 1, \dots, \redim} \ \left| \frac{1}{\sqrt{\numpergroup_i}}\sum^{\numpergroup_i}_{j = 1} \{\obsi_{ij}\recov_{jk} - \EE\obsi_{ij}\recov_{jk} \}\right| > n^{\eps^\prime} \right) \leq \frac{\redim mD_a \numpergroup_{\min}^{\eps}}{\numpergroup_{\min}^{a\eps^\prime}},
\] 
take $a$ such that $a\eps^\prime > q + 1$ give that this probability tends to $0$ in the limit.
The final statement can be similarly shown with the same steps as in proof of the first statement and \cref{assn:non-degen} vi) and \cref{lem:exponentialfamilyderivatives}.
\end{proof}

\begin{lemma}\label{lem:exponentialfamilyboundeddensity}
For any \glmm{} satisfying \cref{assn:non-degen}, there exists $\derivbound>0$ such that for any $\eps >0$,
$$
\lim_{\numpergroup_{\min} \to \infty} \trueprob{\numobs}\Big[\forall_{i = 1,\dots, \numgroups} \ \sup_{\params\in\paramsbigballfixed} \sup_{\re\in\reball} \abs{\llhoodip(\re)} < \numpergroup_i^{1+\eps} \cdot \derivbound\Big]=1.
$$
\end{lemma}
\begin{proof}
Fix $i \in [\numgroups]$.
The joint log-likelihood is, up to constants,
\*[
\llhoodip(\re) &= \sum_{j=1}^{\numpergroup}\log\responsedens(\obsi_{ij} \setdelim \linpred_{ij})\redens(\re\setdelim\resd), \\
&= \frac{1}{a(\obssd)}\sum_{j=1}^{\numpergroup_i}\left\{ \obsi_{ij}\linpredp_{ij}(\re) - b(\linpredp_{ij}(\re))\right\} -\frac{1}{2}\log\abs{\covmat(\resd)} - \frac{1}{2}\re\Tr\covmat^{-1}(\resd)\re,
\]
where $\linpredp_{ij}(\re) = \cov_{ij}\Tr\regparam + \recov_{ij}\Tr\re$.
For all $\params\in\paramsbigballfixed$ and $\re\in\reball$,
\*[
\frac{1}{\numpergroup_i}\abs{\llhoodip(\re)} &\leq \frac{1}{2\numpergroup_i}\left\{ \log\abs{\covmat(\resd)} + \re\Tr\covmat^{-1}(\resd)\re\right\} \\
&\qquad+ \frac{1}{a(\obssd)\numpergroup_i}\sum_{j=1}^{\numpergroup_i}\left\{ \abs{\obsi_{ij}\linpredp_{ij}(\re)} + \abs{b(\linpredp_{ij}(\re))}\right\} \\
&\qquad+\frac{1}{\numpergroup_i}\sum_{j=1}^{\numpergroup_i} \abs{c(\obsi_{ij};\obssd)}.
\]
We bound each term individually. 
For the first term,
\*[
&\sup_{\params\in\paramsbigballfixed}
\sup_{\re\in\reball}\frac{1}{2}\left\{ \log\abs{\covmat(\resd)} + \re\Tr\covmat^{-1}(\resd)\re\right\}\\
&\leq \frac{1}{2}\left( \redim\log\gausscovbig + \gausscovsmall^{-1} \sup_{\re\in\reball} \norm{\re}_2^2 \right) \leq \frac{1}{2}\left( \redim\log\gausscovbig + \gausscovsmall^{-1} (\norm{\retrue}_2 + \delta)^2 \right),
\]
by \cref{assn:non-degen}(iv). Note that $\norm{\retrue, i}^2 = O_p(d)$ as $\retrue$ follows a multivariate Gaussian distribution with bounded eigenvalues. 
Therefore:
\*[
	\frac{1}{2\numpergroup_i}\left\{ \log\abs{\covmat(\resd)} + \re\Tr\covmat^{-1}(\resd)\re\right\}
	 = O_p\left(\frac{1}{\numpergroup_i} \right).
\]
For the second term, by \cref{lem:exponentialfamilyderivatives},
\*[
& \sup_{\params\in\paramsbigballfixed}\sup_{\re\in\reball}\frac{1}{a(\obssd)\numpergroup_i}\sum_{j=1}^{\numpergroup_i}\left\{ \abs{\obsi_{ij}\linpredp_{ij}(\re)} + \abs{b(\linpredp_{ij}(\re))}\right\} \\
&\qquad\leq
 \frac{1}{a(\obssd)}\left\{\linpredbound \numpergroup_{\min}^\eps \frac{1}{\numpergroup_i}\sum_{j=1}^{\numpergroup_i}\abs{\obsi_{ij}} + \frac{1}{\numpergroup_i}\sup_{\linpred\leq \linpredbound\log(\numpergroup_{\min})^{1- \eps} }\abs{b(\linpred)}\right\} \\
&\qquad\leq \frac{\linpredbound}{a(\obssd)}(B_1\numpergroup_{\min}^\eps + O(\numpergroup_{\min}^\eps))  = O(\numpergroup_{\min}^\eps)),
\]
with probability tending to 1 uniformly in the index $i$ by \cref{lemma:uniform-markov,lem:exponentialfamilyderivatives} and and \cref{assn:non-degen} v).

For the third term,  
$$
\frac{1}{\numpergroup_i}\sum_{j=1}^{\numpergroup_i} \abs{c(\obsi_{ij};\obssd)}
 \leq \numpergroup_{\min}^\eps B_3 
$$
with probability tending to $1$ uniformly in the index $i$ by \cref{lemma:uniform-markov}.

Combining these bounds, we have that for all $i$
$$\lim_{\numpergroup_i\to\infty} \frac{1}{\numpergroup_i}\abs{\llhoodip(\re)} = O(\numpergroup_{\min}^\eps)$$ 
with probability $1$ for arbitrarily small $\eps > 0 $, which implies the result.
\end{proof}

We now proceed with the main proof of \cref{prop:glmmregularity}.

Fix $i\in[\numgroups]$.
The derivatives of $\llhoodip(\re)$ are
\*[
\partial^{\dumderivvec}_{\re}\llhoodip(\re) = \begin{cases} \frac{1}{a(\obssd)}\sum_{j=1}^{\numpergroup_i}\left\{ \obsi_{ij} - b^{\prime}(\linpredp_{ij}(\re))\right\}\recov_{ij1}^{\dumderivvec_1}\cdots\recov_{ij\redim}^{\dumderivvec_\redim} + \sum_{t=1}^{\redim}\left(\covmat(\resd)\inv_{\dumderivvec_{t}\cdot}\right)^{\dumderivvec_t}\re & \abs{\dumderivvec} = 1, \\
-\frac{1}{a(\obssd)}\sum_{j=1}^{\numpergroup_i}b^{(\abs{\dumderivvec})}(\linpredp_{ij}(\re))\recov_{ij1}^{\dumderivvec_1}\cdots\recov_{ij\redim}^{\dumderivvec_\redim} + \sum_{t=1}^{\redim}\sum_{s=1}^{\redim}\left(\covmat(\resd)\inv_{\dumderivvec_{t}\dumderivvec_{s}}\right)^{\dumderivvec_{t}\dumderivvec_{s}} & \abs{\dumderivvec} = 2, \\
-\frac{1}{a(\obssd)}\sum_{j=1}^{\numpergroup_i}b^{(\abs{\dumderivvec})}(\linpredp_{ij}(\re))\recov_{ij1}^{\dumderivvec_1}\cdots\recov_{ij\redim}^{\dumderivvec_\redim} & \abs{\dumderivvec} \geq 3.
\end{cases}
\]

\textbf{\cref{assn:kderiv}}.

The $\abs{\dumderivvec}=0$ case follows directly from \cref{lem:exponentialfamilyboundeddensity}. Suppose $\abs{\dumderivvec} = 1$. Then, by \cref{assn:non-degen}(ii) and \cref{lem:exponentialfamilyderivatives},
\*[
\sup_{\params\in\paramsbigballfixed}
\sup_{\re\in\reball}\partial^{\dumderivvec}_{\re}\llhoodip(\re) &\leq \frac{\covbound}{a(\obssd)}\left\{\sum_{j=1}^{\numpergroup_i}\abs{\obsi_{ij}} +  \sup_{\abs{\linpred}<\linpredbound\log(\numpergroup_{\min})^{1- \eps} }\abs{b^{\prime}(\linpred)}\right\},
\]
from which it follows that for all $i$ 
\*[
\lim_{\numpergroup_i\to\infty}\frac{1}{\numpergroup_i}\sup_{\params\in\paramsbigballfixed}\sup_{\re\in\reball}\partial^{\dumderivvec}_{\re}\llhoodip(\re)
\leq \frac{\covbound}{a(\obssd)}\left\{B_1 \numpergroup_{\min}^\eps + \sup_{\abs{\linpred}<\linpredbound\log(\numpergroup_{\min})^{1- \eps}}\abs{b^{\prime}(\linpred)}\right\} = O(n^\eps),
\]
for all $\eps > 0$ with probability tending to 1 by
\cref{lemma:uniform-markov,lem:exponentialfamilyderivatives}.

If $\abs{\dumderivvec}=2$ then
\*[
&\frac{1}{\numpergroup_i}\sup_{\params\in\paramsbigballfixed}\sup_{\re\in\reball}\partial^{\dumderivvec}_{\re}\llhoodip(\re) \\
&\leq \frac{\covbound^{\abs{\dumderivvec}}}{a(\obssd)}\left\{\sup_{\abs{\linpred}<\linpredbound\log(\numpergroup_{\min})^{1- \eps}}\abs{b^{(\abs{\dumderivvec})}(\linpred)}\right\} + \sup_{\params\in\paramsbigballfixed}\sup_{\re\in\reball}\abs{\sum_{t=1}^{\redim}\left(\covmat(\resd)\inv_{\dumderivvec_t}\right)^{\dumderivvec_{t}\cdot}\re}
\]
and if $\abs{\dumderivvec}\geq3$ then
\*[
&\frac{1}{\numpergroup_i}\sup_{\params\in\paramsbigballfixed}\sup_{\re\in\reball}\partial^{\dumderivvec}_{\re}\llhoodip(\re) \\
&\leq \frac{\covbound^{\abs{\dumderivvec}}}{a(\obssd)}\left\{\sup_{\abs{\linpred}<\linpredbound \log(\numpergroup_{\min})^{1- \eps}}\abs{b^{(\abs{\dumderivvec})}(\linpred)}\right\} + \sup_{\params\in\paramsbigballfixed}\sup_{\re\in\reball}\abs{\sum_{t=1}^{\redim}\sum_{s=1}^{\redim}\left(\covmat(\resd)\inv_{\dumderivvec_{t}\dumderivvec_{s}}\right)^{\dumderivvec_{t}\dumderivvec_{s}}}\\
&= O(\numpergroup^\eps) + \sup_{\params\in\paramsbigballfixed}\sup_{\re\in\reball}\abs{\sum_{t=1}^{\redim}\sum_{s=1}^{\redim}\left(\covmat(\resd)\inv_{\dumderivvec_{t}\dumderivvec_{s}}\right)^{\dumderivvec_{t}\dumderivvec_{s}}}
\]
for all $\eps > 0$ by \cref{lem:exponentialfamilyderivatives}.
Furthermore we have that, by \cref{assn:non-degen} (iv), 
$$\sup_{\params\in\paramsbigballfixed}\sup_{\re\in\reball}\abs{\sum_{t=1}^{\redim}\left(\covmat(\resd)\inv_{\dumderivvec_t}\right)^{\dumderivvec_{t}\cdot}\re} < \gausscovsmall\inv\universalradius < \infty$$ 
and $\sum_{t=1}^{\redim}\sum_{s=1}^{\redim}\left(\covmat(\resd)\inv_{\dumderivvec_{t}\dumderivvec_{s}}\right)^{\dumderivvec_{t}\dumderivvec_{s}} < \gausscovsmall\inv < \infty$. \cref{assn:kderiv} now follows by noting that these bounds are independent of $i$.

\textbf{\cref{assn:hessian}}

The Hessian is
\*[
 -\partial^{2}_{(\params,\re)}\llhoodip(\re_i) &= \fulldesign_i\Tr\weightmatp_i(\re_i)\fulldesign_i + \partial^2_{(\params,\re)}\left\{\frac{1}{2}\log\abs{\covmat(\resd)} + \frac{1}{2}\re_i \Tr\covmat^{-1}(\resd)\re_i \right\}, \\
 &\equiv \hessip(\re_i) + \hessreip(\re_i),
\]
where $\fulldesign_i = \left[\design_i:\redesign_i\right]$ as defined in \cref{assn:non-degen} and $\weightmatp_i(\re_i) = \text{diag}\left\{ b^{\prime\prime}(\linpredp_{i1}(\re_i),\ldots,b^{\prime\prime}(\linpredp_{i\numpergroup_i}(\re_i))\right\}$. 
We use the elementary fact that if $A$ and $B$ are $p\times p$ real symmetric matrices then $\lambda_p(A) + \lambda_p(B) \leq \lambda_p(A + B) \leq \lambda_1(A+B) \leq \lambda_1(A) + \lambda_1(B)$; note that this also implies that the eigenvalues of a block matrix are upper and lower bounded by the eigenvalues of the blocks.

First, for all $\params\in\paramsbigballfixed$,
\*[
\sup_{\params\in\paramsbigballfixed}\sup_{\re_i\in\reball} \eigen_{1}(\hessip(\re_i))
&= \sup_{\params\in\paramsbigballfixed}\sup_{\re_i \in\reball}\norm{\weightmatp_i(\re_i)^{1/2}\fulldesign_i}_2^2, \\
&\leq \sup_{\params\in\paramsbigballfixed}\sup_{\re_i\in\reball}\norm{\weightmatp_i(\re_i)^{1/2}}_2^2 \norm{\fulldesign_i}_2^2 \\
&\leq \numpergroup_i\sup_{\abs{\linpred}\leq\linpredbound\log(\numpergroup_{\min})^{1-\eps}}\abs{b^{\prime\prime}(\linpred)}\covbound^2,
\]
where we have used \cref{assn:non-degen}(ii) and \cref{lem:exponentialfamilyderivatives}.
That is,
\*[
 \frac{1}{\numpergroup_i}\sup_{\params\in\paramsbigballfixed}\sup_{\re_i\in\reball} \eigen_{1}(\hessip(\re_i))\leq O( \numpergroup_{\min}^\eps),
\]
with probability tending to $1$ by \cref{lem:exponentialfamilyderivatives}.
Further, for all $\params\in\paramsbigballfixed$,
\*[
\inf_{\params\in\paramsbigballfixed}\inf_{\re_i\in\reball} \eigen_{\redim}(\hessip(\re_i))
&= \inf_{\params\in\paramsbigballfixed}\inf_{\re_i\in\reball}\inf_{\quadpointvec:\norm{\quadpointvec}_2=1}\quadpointvec\Tr\fulldesign_i\Tr\weightmatp_i(\re_i)\fulldesign_i\quadpointvec \\
&= \inf_{\params\in\paramsbigballfixed}\inf_{\re_i\in\reball}\inf_{\quadpointvec:\norm{\quadpointvec}_2=1} \norm{\weightmatp_i(\re_i)^{1/2}\fulldesign_i\quadpointvec}_2^2\\
&\geq \inf_{\abs{\linpred}\leq\linpredbound\log(\numpergroup_{\min})^{1-\eps} }\abs{b^{\prime\prime}(\linpred)}
\inf_{\quadpointvec:\norm{\quadpointvec}_2=1}
\norm{\fulldesign_i \quadpointvec}_2^2 \\
&= \inf_{\abs{\linpred}\leq\linpredbound\log(\numpergroup_{\min})^{1-\eps}}\abs{b^{\prime\prime}(\linpred)} \eigen_{1}(\fulldesign_i\Tr\fulldesign_i) \\
&\geq \recovsmall \numpergroup_i \inf_{\abs{\linpred}\leq \linpredbound\log(\numpergroup_{\min})^{1-\eps}}\abs{b^{\prime\prime}(\linpred)},
\]
where the last step follows from \cref{assn:non-degen}(v).
Since $\inf_{\abs{\linpred}\leq\linpredbound\log(\numpergroup_{\min})^{1-\eps}}\abs{b^{\prime\prime}(\linpred)} > \numpergroup_{\min}^{-\eps}$ for any $\eps > 0$ by \cref{assn:non-degen}(iii), we conclude that
\*[
\frac{1}{\numpergroup_i}\inf_{\params\in\paramsbigballfixed} \inf_{\re_i\in\reball} \eigen_{\redim}(\hessip(\re_i))> \numpergroup_{\min}^{-\eps}.
\]

For $\hessreip(\re_i)$, its second order derivative is make of the following three pieces which we now evaluate:
\*[
-\partial^2_{\re_i\re_i\Tr}\left\{\frac{1}{2}\log\abs{\covmat(\resd)} + \frac{1}{2}\re_i\Tr\covmat^{-1}(\resd)\re_i\right\} &= \covmat(\resd)\inv, \\
-\partial^2_{(\param_j)\re_i\Tr}\left\{\frac{1}{2}\log\abs{\covmat(\resd)} + \frac{1}{2}\re_i\Tr\covmat^{-1}(\resd)\re_i\Tr\right\} &= -\covmat(\resd)\inv\left\{ \partial_{\resd_{(j-\regparamdim + \dispparamdim)}}\covmat(\resd)\right\}\covmat(\resd)\inv\re_i \\ 
-\partial^2_{\param_{j}\param_{l}}\left\{\frac{1}{2}\log\abs{\covmat(\resd)} + \frac{1}{2}\re_i\Tr\covmat^{-1}(\resd)\re_i \right\} & \\
= \frac{1}{2}\tr\left\{\covmat(\resd)\inv\partial^2_{\resd_{(j-\regparamdim + \dispparamdim)}\resd_{(l-\regparamdim + \dispparamdim)}}\covmat(\resd)\right\} &- \frac{1}{2}\re_i\Tr\left[ \covmat(\resd)\inv\left\{\partial^2_{\resd_{(j-\regparamdim + \dispparamdim)}\resd_{(l-\regparamdim + \dispparamdim)}}\covmat(\resd)\right\}\covmat(\resd)\inv\right]\re_i, 
\]
for $j,l = \regparamdim + \dispparamdim + 1,\ldots, \regparamdim + \dispparamdim + \resddim$; the respective terms are $0$ otherwise.
This matrix does not depend on $\numpergroup_i$, and \cref{assn:non-degen} (iv) is sufficient to upper and lower bound its spectrum.
Note that for the first term, 
$$
0<\gausscovbig\inv\leq\inf_{\params\in\paramsbigballfixed}\lambda_{\redim}\left\{\covmat(\resd)\inv\right\}\leq\sup_{\params\in\paramsbigballfixed}\lambda_{1}\left\{\covmat(\resd)\inv\right\}\leq\gausscovsmall\inv<\infty.
$$ 
We can bound the second term by
\*[
\norm{\covmat(\resd)\inv\left\{ \partial_{\resd_{(j-\regparamdim + \dispparamdim)}}\covmat(\resd)\right\}\covmat(\resd)\inv\re_i}_{2} &\leq \norm{\covmat(\resd)\inv}_{2}^2\norm{\re_i}_2 \norm{\partial_{\resd_{(j-\regparamdim + \dispparamdim)}}\covmat(\resd)}_{2},
\]
where $\norm{A}_2\equiv\norm{A}_{op} = \lambda_1(A)^{1/2}$.
It follows that
$$
\sup_{\params\in\paramsbigballfixed}\sup_{\re_i\in\reball} \norm{-\covmat(\resd)\inv\left\{ \partial_{\resd_{(j-\regparamdim + \dispparamdim)}}\covmat(\resd)\right\}\covmat(\resd)\inv\re_i}_{2} = O_p(\sqrt{\log(\numgroups)}),
$$
uniformly in $i$ as the maximum of $\norm{\re_i}$ is of order $O_p(\sqrt{\log(\numgroups)})$ by Theorem 1.14 in \cite{rigollet2023high} and \cref{assn:non-degen} iii) .
Further, note that for $A\in\Reals^{d\times d}$, $\norm{A}_{2} \leq d\norm{A}_{+}$ where $\norm{A}_{+} = \text{max}_{1\leq i,j \leq d} \abs{A_{ij}}$.
Since the dimension $\redim$ is fixed, it suffices to bound the third block element-wise.
For the trace term, we apply Holder's inequality and the fact that $\tr(A) = \eigen_1(A)+\cdots+\eigen_\redim(A)$ to write
\*[
\tr\left\{\covmat(\resd)\inv\partial^2_{\resd_{(j-\regparamdim + \dispparamdim)}\resd_{(l-\regparamdim + \dispparamdim)}}\covmat(\resd)\right\} &\leq \norm{\covmat(\resd)\inv}_{op}\tr\left\{\partial^2_{\resd_{(j-\regparamdim + \dispparamdim)}\resd_{(l-\regparamdim + \dispparamdim)}}\covmat(\resd)\right\}, \\
&\leq \redim\eigen_1\left\{\covmat(\resd)\inv\right\}^{1/2}\eigen_1\left\{\partial^2_{\resd_{(j-\regparamdim + \dispparamdim)}\resd_{(l-\regparamdim + \dispparamdim)}}\covmat(\resd)\right\}, \\
&= \redim\eigen_p\left\{\covmat(\resd)\inv\right\}^{-1/2}\eigen_1\left\{\partial^2_{\resd_{(j-\regparamdim + \dispparamdim)}\resd_{(l-\regparamdim + \dispparamdim)}}\covmat(\resd)\right\}, \\
&\leq \redim\gausscovsmall^{-1/2}\gausscovderivbig. \\
\]
We conclude that
$$
\sup_{\params\in\paramsbigballfixed}\sup_{\re\in\reball} \tr\left\{\covmat(\resd)\inv\partial^2_{\resd_{(j-\regparamdim + \dispparamdim)}\resd_{(l-\regparamdim + \dispparamdim)}}\covmat(\resd)\right\} \leq \redim\gausscovsmall^{-1/2}\gausscovderivbig < \infty
$$
by \cref{assn:non-degen} (iv).

Using $\eigen_1(AB) \leq \eigen_1(A)\eigen_1(B)$ for any real, symmetric, positive definite $A$ and $B$, we also have
\*[
\re_i\Tr\left[ \covmat(\resd)\inv\left\{\partial^2_{\resd_{(j-\regparamdim + \dispparamdim)}\resd_{(l-\regparamdim + \dispparamdim)}}\covmat(\resd)\right\}\covmat(\resd)\inv\right]\re_i &\leq \eigen_1\left\{\covmat(\resd)\inv\left\{\partial^2_{\resd_{(j-\regparamdim + \dispparamdim)}\resd_{(l-\regparamdim + \dispparamdim)}}\covmat(\resd)\right\}\covmat(\resd)\inv\right\}, \\
&\leq \eigen_1\left\{ \covmat(\resd)\inv\right\}^2\eigen_1\left\{\partial^2_{\resd_{(j-\regparamdim + \dispparamdim)}\resd_{(l-\regparamdim + \dispparamdim)}}\covmat(\resd)\right\}, \\
&\leq \gausscovsmall^{-2}\gausscovderivbig\norm{\re_i}_2^2,
\]
and so by \cref{assn:non-degen} (iv) we conclude that
\*[
&\sup_{\params\in\paramsbigballfixed}\sup_{\re_i\in\reball} \re_i\Tr\left[ \covmat(\resd)\inv\left\{\partial^2_{\resd_{(j-\regparamdim + \dispparamdim)}\resd_{(l-\regparamdim + \dispparamdim)}}\covmat(\resd)\right\}\covmat(\resd)\inv\right]\re_i\\
 &\leq \gausscovsmall^{-2}\gausscovderivbig \norm{\re_i + \delta Z}_2^2 = O_p(\log(\numgroups) ),
\]
uniformly in $i$, where $z$ is some unit vector and the last statement follows again from the fact that $\re_i$ follows a zero mean Gaussian distribution .
We have therefore shown that
$$
\lambda_1\left\{\hessreip(\re_i)\right\} = O_p(\log(\numgroups)).
$$
Combined with the fact that $\hessreip(\re)$ does not depend on $\numpergroup_i$, we conclude that
$$
\lambda_1\left\{\numpergroup_i\inv\hessreip(\re_i)\right\} = O_p(\numpergroup_i\inv\log(\numgroups)),
$$
and hence
$$
\lambda_{\redim}\left\{\numpergroup_i\inv\hessreip(\re_i)\right\} = O_p(\numpergroup_i\inv \log(\numgroups))
$$
as well for all $\eps > 0$.
Noting that these bounds are either independent of or uniform in $i$, therefore combining all these bounds \cref{assn:hessian} now follows.

\textbf{\cref{assn:limsup}}
We first prove the following two lemmas:
\begin{lemma}\label{lem:convexlikelihood}
	For all $i = 1, \dots, \numgroups$,
$-\log\jointdens_{i}(\obsall_{i} \setdelim \re;\params)$ are almost surely globally convex as a function of $\re\in\respace\subseteq\Reals^{\redim}$ for each $\params\in\paramsbigballfixed$.
\end{lemma}
\begin{proof}
We have 
$$
\jointdens(\obsall_{i}, \re_i;\params) = \prod_{j=1}^{\numpergroup_{i}}\responsedens(\obsi_{ij} \setdelim \re_{i};\regparam)\redens(\re_{i};\resd).
$$
Observe that $-\sum_{j=1}^{\numpergroup_i}\log\responsedens(\obsi_{ij} \setdelim \linpred)$ is almost surely globally convex in $\linpred\in\natparamspace$ (see \cref{assn:non-degen} (i)), 
which follows immediately from the identity $\hessip(\re) = \fulldesign_i\Tr\weightmatp_i(\re)\fulldesign_i$ shown in the verification of \cref{assn:hessian}
and from \cref{assn:non-degen} (iii).
Note as well that $\eta$ is linear, and hence convex, in $\re$.
Further, note that $\redens(\re\setdelim\resd)$ is a Gaussian density and hence $-\log\redens(\re\setdelim\resd)$ is convex in $\re$.
The result follows because sums and compositions of convex functions are convex.
\end{proof}
\begin{lemma}\label{lem:convexboundary}
	Let $\complementmodeip = \argmax_{\re\in[\reball]^c}\log\jointdens_{i}(\obsall_{i} \setdelim \re;\params)$ for arbitrary $\params\in\paramsbigballfixed$.
	Then
	$$
	\lim_{\numpergroup_{\min} \to \infty} \trueprob{\numobs}\left( \max_{i = 1, \dots, \numgroups} \ \norm{\complementmodeip - \retruepi}_2 = \universalradius\right) = 1.
	$$
\end{lemma}
\begin{proof}
	By \cref{assn:consistency},
	$$
	\lim_{\numpergroup_{\min} \to \infty} \trueprob{\numobs}\left[\forall_{i = 1,\dots, \numgroups} \ \condmodeip\in\reball\right] = 1,
	$$
	which shows that $\complementmodeip\neq\condmodeip$ for large enough $\numobs$.
	Define
	$$
	\mb{v}_\universalradius = \retruepi + \frac{\complementmodeip - \retruepi}{\lVert\complementmodeip - \retruepi\rVert}_2\universalradius,
	$$
	and note that $\norm{\mb{v}_\universalradius - \retruepi}_2 = \universalradius$.
	But it follows from \cref{lem:convexlikelihood} that
	\*[
	\log\jointdens_{i}(\obsall_{i} \setdelim \retruepi;\params) - \log\jointdens_{i}(\obsall_{i} \setdelim \mb{v}_\universalradius;\params) \leq \log\jointdens_{i}(\obsall_{i} \setdelim \retruepi;\params) -  \log\jointdens_{i}(\obsall_{i} \setdelim \complementmodeip;\params),
	\]
	almost surely, and hence $\log\jointdens_{i}(\obsall_{i} \setdelim \mb{v}_\universalradius;\params) \geq \log\jointdens_{i}(\obsall_{i} \setdelim \complementmodeip;\params)$,
	which in turn implies that $\mb{v}_\universalradius = \complementmodeip$ by definition of $\complementmodeip$.
	The proof is completed by noting that this argument holds for every $i$ uniformly.
\end{proof}
We proceed with the main verification of \cref{assn:limsup}.
Fix a value of $\params \in \paramsbigball$. 
Taylor expansion yields:
\begin{align*}
	&\frac{1}{\numpergroup_i}\{\log\jointdens_{i}(\obsall_{i} \setdelim \re_i;\params) - \log\jointdens_{i}(\obsall_{i} \setdelim \retruep;\params)\} \\
	&= \frac{1}{\numpergroup_i}\left\{ (\re_i - \retruepi)^\top \partial_{\re}\log\jointdens_{i}(\obsall_{i} \setdelim \retruepi;\params) + (\re_i - \retruepi)^\top \partial_{\re}\partial^\top_{\re}\log\jointdens_{i}(\obsall_{i} \setdelim \re^\circ;\params)(\re_i - \retruepi) \right\}\\
	&= \frac{1}{\numpergroup_i}\left\{(\re_i - \retruepi)^\top \partial_{\re}\log\jointdens_{i}(\obsall_{i} \setdelim \retruepi;\paramstrue) + (\re_i - \retruepi)^\top \partial_{\re}\partial^\top_{\params}\log\jointdens_{i}(\obsall_{i} \setdelim \retruepi;\params^\circ)(\params - \paramstrue)\right\} \\
	&+\frac{1}{\numpergroup_i}(\re_i - \retruepi)^\top \partial_{\re}\partial^\top_{\re}\log\jointdens_{i}(\obsall_{i} \setdelim \re^\circ;\params)(\re_i - \retruepi),
\end{align*}
where $\params^\circ = t_1\params + (1-t_1)\paramstrue$ and
$\re^\circ = t_2\re_i + (1-t_2)\retruepi$ for some $t_1, t_2 \in [0,1]$.
Our approach is to use this expansion to argue that
$$
\lim_{\numpergroup_i\to\infty}\frac{1}{\numpergroup_i}\{\log\jointdens_{i}(\obsall_{i} \setdelim \complementmodeip;\params) - \log\jointdens_{i}(\obsall_{i} \setdelim \retruepi;\params)\} < 0,
$$
which implies the result.

To bound the first term, observe that
\*[
	 &\max_{i = 1, \dots, \numgroups} \left|(\complementmodeip - \retruepi)\Tr \left\{\frac{\partial_{\re}\log\jointdens_{i}(\obsall_{i} \setdelim \retruepi;\paramstrue)}{\numpergroup_i}\right\}\right| \\
	 &= \max_{i = 1, \dots, \numgroups} \norm{\complementmodeip - \retruepi} \left\{  \max_{k = 1, \dots, \redim}  \frac{ \sqrt{\redim}
	 \left| \frac{1}{\numpergroup_i}\sum^{\numpergroup_i}_{j = 1} \{\obsi_{ij}\recov_{jk} - \EE\obsi_{ij}\recov_{jk} \}\right|}{\numpergroup_i} + \frac{\norm{\covmat\inv(\paramstrue)\retruepi}}{\numpergroup_i} \right\}\\
	 &\leq \delta\left\{O_p\left(\frac{1}{G(\numpergroup_{\min})} \right) + O_p\left( \frac{\sqrt{\log(\numgroups)} }{\numpergroup_{\min}} \right)\right\},
\]
by \cref{lemma:uniform-markov,lem:convexboundary} and the fact that the maximum of the norm of $\numgroups$ gaussian vectors is $O_p(\sqrt{\log(\numgroups)})$ by Theorem 1.14 in \cite{rigollet2023high}.
The third term satisfies
\*[
	(\complementmodeip - \retruepi)^\top \frac{\partial_{\re}\partial^\top_{\re}\log\jointdens_{i}(\obsall_{i} \setdelim \re^\circ;\params)}{\numpergroup_i} (\complementmodeip - \retruepi) 
	\leq -\norm{\complementmodeip - \retruepi}_2^2 \hesssmall \numpergroup_{\min}^{-\eps}
\]
by \cref*{assn:hessian} for every $\eps > 0$.
We therefore conclude
\*[
\lim_{\numpergroup_i\to\infty}\frac{1}{\numpergroup_i}(\complementmodeip - \retruepi)^\top \partial_{\re}\partial^\top_{\re}\log\jointdens_{i}(\obsall_{i} \setdelim \re^\circ;\params) (\complementmodeip - \retruepi) \leq - \universalradius^2 \hesssmall \numpergroup_{\min}^{-\eps} ,
\]
by \cref{lem:convexboundary}.

As for the second term: note that $\numpergroup_i\inv\hessip(\complementmodeip) \geq C\minpergroup^\eps$ for any $\eps > 0$ and for some $C > 0$, and therefore any of its sub-matrix must have singular values growing at most at the same order. 
Since $ \params \in \paramsbigball$ and $\paramradius \downarrow 0$ at a polynomial rate, we have
\begin{align*}
	&\frac{1}{\numpergroup_i}|(\re_i - \retruepi)^\top \partial_{\re}\partial^\top_{\params}\log\jointdens_{i}(\obsall_{i} \setdelim \retruepi;\params^\circ)(\params - \paramstrue)| \\
	&\leq \universalradius\paramradius \norm{\frac{1}{\numpergroup_i} \partial_{\re}\partial^\top_{\params}\log\jointdens_{i}(\obsall_{i} \setdelim \retruepi;\params^\circ)}_{op} = O(n^{\eps - \alpha}),   
\end{align*}
for some $\alpha > 0$ by \cref{assn:consistency}, therefore as $\eps$ is arbitrary, this terms decreases at polynomial rate.
\cref{assn:limsup} follows by combining these three bounds.

\textbf{\cref{assn:limsup-out}}

There exists a $\universalradius^\prime < \universalradius$ such that $ [\ball{(\paramstrue, \retruepi)}{\universalradius^\prime}]^{c} \subset[\paramsbigballfixed]^{c} \times [\reball]^{c}$, where $\times$ is the direct product operator. We show the stronger assumption that there exists $\llhoodmargin >0$ such that
for all $i\in[\numgroups]$,  
\*[
	\lim_{\numpergroup_{\min} \to \infty} \trueprob{\numobs}\Big[ \forall_{i = 1,\dots, \numgroups} \ \sup_{(\params, \re) \in [\ball{(\paramstrue, \retruepi)}{\delta^\prime}]^{c}} \log\jointdens_{i}(\obsall_{i} \setdelim \re;\params) - \log\jointdens_{i}(\obsall_{i} \setdelim \retruepi;\paramstrue) \leq - \numpergroup_i \cdot \llhoodmargin\Big] = 1.
\]
By \cref{assn:consistency} and \cref{assn:well-spec},
we have that
\begin{equation}\label{eqn:a5modeinball}
\lim_{\numpergroup_{\min} \to \infty} \trueprob{\numobs}\left\{ \forall_{i = 1,\dots, \numgroups} \ (\paramsmle, \condmodeipmle) \in \bigball{(\paramstrue, \retruepi)}{\universalradius^\prime}\right\} = 1,
\end{equation}
where $\hat\theta$ is the marginal maximum likelihood estimator.

By arguments analogous to the proof of \cref{lem:convexboundary}
(see the verification of \cref{assn:limsup}), it can be shown that
\begin{equation}\label{eqn:a5modeonboundary}
\lim_{\numpergroup_{\min} \to \infty} \trueprob{\numobs}\left(\forall_{i = 1,\dots, \numgroups} \ \norm{(\paramscomplementmode, \complementmodeipmode) - (\paramstrue, \retruepi)} = \universalradius^\prime\right) = 1,
\end{equation}
where
$$
(\paramscomplementmode, \complementmodeipmode) = \argmax_{(\params,\re)\in\bigballc{(\paramstrue, \retruepi)}{\universalradius^\prime}}\log\jointdens_{i}(\obsall_{i} \setdelim \re;\params).
$$
It follows that:
\*[
	&\sup_{(\paramstrue, \retrue) \in [\ball{(\paramstrue, \retrue)}{\delta^\prime}]^{c}} \frac{1}{\numpergroup_i} \left\{ \log\jointdens_{i}(\obsall_{i} \setdelim \re;\params) - \log\jointdens_{i}(\obsall_{i} \setdelim \retruepi;\paramstrue)\right\} \\
	&=\frac{1}{\numpergroup_i}\left\{ \log\jointdens_{i}(\obsall_{i} \setdelim \complementmodeipmode; \paramscomplementmode) - \log\jointdens_{i}(\obsall_{i} \setdelim \retruepi;\paramstrue)\right\} \\
	&= \numpergroup_i^{-1} \{\log\jointdens_{i}(\obsall_{i} \setdelim \complementmodeipmode; \paramscomplementmode) - \log\jointdens_{i}(\obsall_{i} \setdelim \condmodeipmle; \hat\params)\} + \numpergroup_i^{-1}\{\log\jointdens_{i}(\obsall_{i} \setdelim \condmodeipmle; \paramsmle) -\log\jointdens_{i}(\obsall_{i} \setdelim \retruepi;\paramstrue)\}.
\]
By \cref{assn:consistency},
\*[
	\numpergroup_i^{-1}\{\log\jointdens_{i}(\obsall_{i} \setdelim \condmodeipmle; \hat\params) -\log\jointdens_{i}(\obsall_{i} \setdelim \retruepi;\paramstrue)\} = O(\numpergroup_{\min}^\eps),	
\]
almost surely at a uniform rate.
By \cref{eqn:a5modeinball,eqn:a5modeonboundary}, there exists
a $\universalradius^{\prime\prime} > 0$ such that
$$
\lim_{\numpergroup_i \to \infty} \trueprob{\numobs}\left(\forall_{i = 1,\dots, \numgroups} \ \norm{ (\paramsmle, \condmodeipmle) - (\paramscomplementmode,\complementmodeipmode)} > \universalradius^{\prime\prime}\right) = 1.
$$
By \cref{assn:hessian} and second-order Taylor expansion around $(\condmodeipmle, \hat\params)$ followed by a first order Taylor expansion on the score around $(\retruepi, \paramstrue)$ we have:
\[\label{eq:A5-1}
	&\numpergroup_i^{-1} \{\log\jointdens_{i}(\obsall_{i} \setdelim \complementmodeipmode; \paramscomplementmode) - \log\jointdens_{i}(\obsall_{i} \setdelim \condmodeipmle; \hat\params)\} \\
	&= \numpergroup_i^{-1}(\complementmodeipmode  - \condmodeipmle, \paramscomplementmode -  \hat\params)^\top \{\partial_{\params} \log\jointdens_{i}(\obsall_{i} \setdelim \condmodeipmle; \hat\params)\} \\
	&+ \numpergroup_i^{-1}(\complementmodeipmode  - \condmodeipmle, \paramscomplementmode -  \hat\params)^\top \{\partial_{\params}^2 \log\jointdens_{i}(\obsall_{i} \setdelim \re^\star; \params^\star)\}(\complementmodeipmode  - \condmodeipmle, \paramscomplementmode -  \hat\params)/2\\
	&= \numpergroup_i^{-1}(\complementmodeipmode  - \condmodeipmle, \paramscomplementmode -  \hat\params)^\top \{\partial_{\params} \log\jointdens_{i}(\obsall_{i} \setdelim \retruepi; \paramstrue)\} \\
	&+ \numpergroup_i^{-1}(\complementmodeipmode  - \condmodeipmle, \paramscomplementmode -  \hat\params)^\top \{\partial_{\params} \log\jointdens_{i}(\obsall_{i} \setdelim \re^\prime; \params^\prime)\} (\condmodeipmle -\retruepi, \hat\params - \paramstrue)
	\\
	&+ \numpergroup_i^{-1}(\complementmodeipmode  - \condmodeipmle, \paramscomplementmode -  \hat\params)^\top \{\partial_{\params}^2 \log\jointdens_{i}(\obsall_{i} \setdelim \re^\star; \params^\star)\}(\complementmodeipmode  - \condmodeipmle, \paramscomplementmode -  \hat\params)/2
\]
Where in the above, the score evaluated at the data generating value tends to $0$ uniformly by \cref{lem:exponentialfamilyderivatives} and
\*[
	&\partial_{\params}^2 \log\jointdens_{i}(\obsall_{i} \setdelim \re^\prime; \params^\prime) = \int_0^1 (1 - t) \partial^{2}_{\params} \log\jointdens_{i} (\obsall_i;\condmodeipmle +t \retrue, \hat\params + t \paramstrue)\dee t\\
	&\partial_{\params}^2 \log\jointdens_{i}(\obsall_{i} \setdelim \re^\star; \params^\star) = \int_0^1 (1 - t) \partial^{2}_{\params} \log\jointdens_{i} (\obsall_i;\condmodeipmle +t \complementmodeipmode, \hat\params + t \paramscomplementmode)\dee t.
\]
by the same argument as in \cref*{eq:eigen-lower} the maximal eigenvalue of these matrices are lower bounded by $\numpergroup_i^{1 - \eps} \hesssmall $ for any $\eps > 0$.
Noting that $(\condmodeipmle -\retruepi, \hat\params - \paramstrue) = o_p(1)$ uniformly in the index $i$ by \cref{assn:consistency,assn:well-spec}, only the final term in \cref{eq:A5-1} is non-negligible. Therefore: 
\*[
	\cref{eq:A5-1}
	\leq o_p(1) + \norm{\complementmodeipmode  - \condmodeipmle, \paramscomplementmode -  \hat\params}_2 \eigen_{1}\{\numpergroup_i^{-1} \partial_{\params}^2 \log\jointdens_{i}(\obsall_{i} \setdelim \re^\star; \params^\star)\} \leq -\delta^2\hesssmall \numpergroup_{\min}^{-\eps},
\]
with probability uniformly tending to 1.
\cref{assn:limsup-out} holds by taking $\llhoodmargin = \delta^2 \hesssmall/2 $ and noting that the upper bound does not depend on $i$.

\textbf{\cref{assn:consistency}}

Fix $\params\in\paramsbigball$.
For any $\params$ and $i\in[\numgroups]$, the conditional mode $\condmodeip$ satisfies 
$$
\partial_{\re}\log\margdens(\obsall_i;\condmodeip,\params) + \partial_{\re}g(\condmodeip;\params) = 
	\{\obsall_i - \mb{b}^{\prime}_i(\params,\condmodeip)\}\mb{V}_i + \covmat\inv(\params)\condmodeip =  
	\zero,
$$
where $\mb{b}_i^\prime(\params,\re) = (b^\prime(\linpred_{i1}(\params,\re)),\ldots,b^\prime(\linpred_{i\numpergroup_i}(\params,\re))\Tr$.
A first-order Taylor expansion of the first term of this equation about $(\paramstrue,\retruepi)$ evaluated at $(\params, \condmodeip)$ gives
\*[
&\{\obsall_i - \mb{b}^{\prime}_i(\paramstrue,\retruepi)\}\redesign_i - \design_i\Tr\weightmat^{\params^\circ}_i(\re^\circ)\redesign_i\left(\regparam - \regparamtrue\right)\\
&\qquad  - \left\{\redesign_i\Tr\weightmat^{\params^\circ}(\re^\circ)\redesign_i + \covmat\inv(\params)\right\}\left(\condmodeip - \retruepi\right) - \covmat\inv(\params)\retruepi = \zero,
\]
where $\weightmatp_i(\re) = \text{diag}\left\{ b^{\prime\prime}(\eta^{\tilde\params_1}_{i1}(\tilde\re_1),\ldots,b^{\prime\prime}(\eta^{\tilde{\params}_{\numpergroup_i}}_{i\numpergroup_i}(\tilde\re_{\numpergroup_i}))\right\}$, $\tilde{\params}_i = (1 - t_i)\params + \paramstrue$, $\tilde{\re}_j = (1 - t_j)\re + \retrue$ for $t_i, t_j \in [0,1]$ and $\re^\circ = (\tilde\re_1, \dots, \tilde\re_{\numpergroup_i})$, $\params^\circ = (\tilde\params_1, \dots, \tilde\params_{\numpergroup_i})$.
Rearranging, we obtain
\*[
\left(\condmodeip - \retruepi\right) &= \left\{\redesign_i\Tr\weightmat_i^{\params^\circ}(\re^\circ)\redesign_i + \covmat\inv(\params)\right\}\inv\{\obsall_i - \mb{b}^{\prime}_i(\paramstrue,\retruepi)\}\redesign_i \\
&- \left\{\redesign_i\Tr\weightmat_i^{\params^\circ}(\re^\circ)\redesign_i + \covmat\inv(\params)\right\}\inv\design_i\Tr\weightmatp_i(\re^\circ)\redesign_i\left(\regparam - \regparamtrue\right)\\
&+ \left\{\redesign_i\Tr\weightmat_i^{\params^\circ}(\re^\circ)\redesign_i + \covmat\inv(\params)\right\}\inv \covmat\inv(\params)\retruepi
\]
and hence,
\begin{align}\label{eqn:condmodebound}
&\paramradius\inv\norm{\condmodeip - \retruepi}_2 \leq  \norm{\left\{\redesign_i\Tr\weightmat^{\params^\circ}(\re^\circ)\redesign_i + \covmat\inv(\params)\right\}\inv}_2 \\
&\times \left\{\paramradius\inv\norm{\{\obsall_i - \mb{b}^{\prime}_i(\paramstrue,\retruepi)\}\redesign_i}_2  
 + \paramradius^{-1}\norm{\covmat\inv(\params)\retruepi}_2 + \paramradius\inv\norm{\regparam - \regparamtrue}_2\norm{\design_i\Tr\weightmatp_i(\re^\circ)\redesign_i}_2  \right\},\nonumber
\end{align}
where we recall that $\paramradius = (\min_{i=1,\ldots,\numgroups}\numpergroup_i)^{-\alpha}$ with $0 < \alpha < 1/4$.
\cref{eqn:condmodebound} will be used to show both statements in the Assumption.

For the first statement, we show that the two terms converges in probability to zero and the second term is bounded in probability uniformly by any polynomial function in both the index $i$ and for any $\params\in\paramsbigball$.
Recall from the proof of \cref{assn:hessian} that we have
\*[
\frac{1}{\numpergroup_i}\sup_{\params\in\paramsbigballfixed}\sup_{\re\in\reball} \eigen_{1}(\fulldesign_i\Tr\weightmatip(\re)\fulldesign_i)< \hessbig\minpergroup^{\eps}.
\]
and
\*[
\frac{1}{\numpergroup_i}\inf_{\params\in\paramsbigballfixed} \inf_{\re\in\reball} \eigen_{\redim}(\fulldesign_i\Tr\weightmatip(\re)\fulldesign_i)> \hesssmall \minpergroup^{-\eps}.
\]
for some any $\eps > 0$, where $\fulldesign_i = [\design_i:\redesign_i]$.
We therefore have
$$
\sup_{\params\in\paramsbigball}\norm{\design_i\Tr\weightmatp_i(\re^\circ)\redesign_i}_2 < \hessbig\minpergroup^{\eps}
$$
and
$$
\sup_{\params\in\paramsbigball}\norm{\left\{\redesign_i\Tr\weightmat^{\params^\circ}(\re^\circ)\redesign_i + \covmat\inv(\params)\right\}\inv}_2 < \left(\hesssmall \numpergroup^{1-\eps} + \gausscovsmall\right)\inv < \hesssmall\inv \minpergroup^{-(1-\eps)} .
$$
By definition we have $\sup_{\params\in\paramssmallball}\norm{\regparam-\regparamtrue}_2 \leq \paramradius$. 
Combining these, we have that the third term:
\*[
&\sup_{\params\in\paramsbigball}\paramradius\inv\norm{\regparam - \regparamtrue}_2\norm{\design_i\Tr\weightmatp_i(\re^\circ)\redesign_i}_2\norm{\left\{\redesign_i\Tr\weightmat^{\params^\circ}(\re^\circ)\redesign_i + \covmat\inv(\params)\right\}\inv}_2\\
 &\leq \paramradius\paramradius\inv \hesssmall\inv \hessbig\minpergroup^{ 1 +\eps} \minpergroup^{-(1-\eps)} C_2\inv = O(\numpergroup_{\min}^\eps),
\]
for any $\eps > 0$.
By \cref{lemma:uniform-markov},
$$
\max_{i = 1, \dots, \numgroups} \PP \left( \numpergroup_i^{-1}\norm{\{\obsall_i - \mb{b}^{\prime}_i(\paramstrue,\retruepi) \}\redesign_i }_2 \leq \frac{\sqrt{\redim} B_2}{G(\numpergroup_{\min})} \right) \geq 1 - o(1).
$$
We then have:
\*[
&\paramradius\inv\norm{\{\obsall_i - \mb{b}^{\prime}_i(\paramstrue,\retruepi)\}\redesign_i}_2\norm{\left\{\redesign_i\Tr\weightmat^{\params^\circ}(\re^\circ)\redesign_i + \covmat\inv(\params)\right\}\inv}_2, \\
&\leq \left[\numpergroup_i^{-1}\norm{\{\obsall_i - \mb{b}^{\prime}_i(\paramstrue,\retruepi)\}\redesign_i}_2\right]\cdot\left[ \paramradius\inv\numpergroup_i\numpergroup_i^{-(1 + \eps)} C_2\inv\right], \\
&\leq  \frac{\sqrt{\redim} \numpergroup_{\min}^{\eps - \alpha} B_2 }{G(\numpergroup_{\min})} \to 0,
\]
with probability tending to 1 uniformly in $i$.
For the second term, by a similar argument: 
\[
	\norm{\left\{\redesign_i\Tr\weightmat^{\params^\circ}(\re^\circ)\redesign_i + \covmat\inv(\params)\right\}\inv}_2 \norm{\paramradius^{-1}\covmat\inv(\params)\retruepi}_2\leq \numpergroup_i^{-1 + \eps} C_2^{-1} \paramradius^{-1} \bar\xi_1 \norm{\retruepi}_2 \rightarrow 0,
\]
by \cref{assn:non-degen} iii) and the fact that $\max_{i = 1,\dots, \numgroups}\norm{\retruepi}_2 = O_p(\sqrt{\log(\numgroups)})$ by Theorem 1.14 in \cite{rigollet2023high}.
This proves the first statement in \cref{assn:consistency}.

For the second statement, we evaluate the same Taylor expansion at $(\paramstrue, \condmodeiptrue)$ to obtain
\*[
\frac{\numpergroup_i^{1/2}}{G^\prime(\numpergroup_i)}\norm{\condmodeiptrue - \retruepi}_2 \leq &\frac{\numpergroup_i^{1/2}}{G^\prime(\numpergroup_i)}\norm{\{\obsall_i - \mb{b}^{\prime}_i(\paramstrue,\retruepi)\}\redesign_i}_2\norm{\left\{\redesign_i\Tr\weightmat^{\paramstrue}(\re^\circ)\redesign_i + \covmat\inv(\params)\right\}\inv}_2,
\]
where now $\re^\circ = t\condmodeiptrue + (1-t)\retruepi$ for $t\in[0,1]$ and we recall that $G^\prime$ is any nonzero function with $\lim_{n\to\infty}G(n)=\infty$.
We have
\*[
&\frac{\numpergroup_i^{1/2}}{G^\prime(\numpergroup_i)}\norm{\{\obsall_i - \mb{b}^{\prime}_i(\paramstrue,\retruepi)\}\redesign_i}_2\norm{\left\{\redesign_i\Tr\weightmat^{\paramstrue}(\re^\circ)\redesign_i + \covmat\inv(\params)\right\}\inv}_2, \\
&\leq \frac{1}{G^\prime(\numpergroup_i)}\left[\numpergroup_i^{-1/2}\norm{\{\obsall_i - \mb{b}^{\prime}_i(\paramstrue,\retruepi)\}\redesign_i}_2\right]\left[\numpergroup_i\norm{\left\{\redesign_i\Tr\weightmat^{\paramstrue}(\re^\circ)\redesign_i + \covmat\inv(\params)\right\}\inv}_2\right], \\
&\leq \frac{1}{G^\prime(\numpergroup_i)}\left[\numpergroup_i^{-1/2}\norm{\{\obsall_i - \mb{b}^{\prime}_i(\paramstrue,\retruepi)\}\redesign_i}_2\right]C_2\inv \numpergroup_i^\eps= O(\numpergroup_i^\eps), 
\]
by \cref{lemma:uniform-markov}.
This proves the second statement in \cref{assn:consistency}.

\textbf{\cref{assn:prior}}

The first statement of \cref{assn:prior} follows directly from the fact that our prior is Gaussian with full support.

For the second and third statement for all $\params \in \paramsbigball$, in GLMMs the score function is: 
\*[
	&\int_{\re \in \Reals^\redim} \norm{\partial_{\params}\log\margdens(\obsall_i;\re,\params)}_2 \bm{\phi}(\re; \zero, \bm\Sigma(\bm\sigma) ) \dee \re\\
	&\leq \numpergroup_i^{1/2} \norm{\design_i}_{op} \int_{\re \in \Reals^\redim} \left\{\frac{\norm{\obsall_i - \mb{b}^{\prime}_i(\paramstrue,\retruepi)}_2}{\numpergroup_i^{1/2}} + \frac{\norm{\mb{b}^{\prime}_i(\paramstrue,\retruepi) - \mb{b}^{\prime}_i(\params,\re)}_2}{\numpergroup_i^{1/2}}\right\}  \bm{\phi}(\re; \zero, \bm\Sigma(\bm\sigma)) \dee \re\\
	&\leq \recovbig \numpergroup_i^{1 + \eps} \left\{o_p(\numpergroup_{\min}^\eps) + \int_{\re \in \Reals^\redim} \frac{\norm{\mb{b}^{\prime}_i(\paramstrue,\retruepi) - \mb{b}^{\prime}_i(\params,\re)}_2}{\numpergroup_i^{1/2}}  \bm{\phi}(\re; \zero, \bm\Sigma(\bm\sigma)) \dee \re \right\} ,
\]
uniformly in $i$ by \cref{lemma:uniform-markov} and by  \cref{assn:non-degen} iv) $\norm{\design_i}_{op} \leq \recovbig \numpergroup_i^{1/2 +\eps} $. The term within the integrand for every $\re$
\*[
	\frac{\norm{\mb{b}^{\prime}_i(\paramstrue,\retruepi) - \mb{b}^{\prime}_i(\params,\re)}_2}{\numpergroup_i^{1/2}} &= \norm{(\paramstrue - \params,\retruepi - \re) \circ \mb{b}_i^{\prime\prime} (\tilde\params,\tilde\re)}_2 \\
	&\leq \norm{(\paramstrue - \params,\retruepi - \re)}_2 \norm{\mb{b}_i^{\prime\prime} (\tilde\params,\tilde\re)}_2\\
	&\leq \left\{ \paramradius + \norm{\retruepi}_2 + \norm{\re}_2 \right\}\\ 
	&\times o\left(\exp\left(-\log(\numpergroup_{\min})^{1 - \eps} \left\{\norm{\re}_2^{2 - \eps} +  2\norm{\retruepi}_2^{2 - \eps} \right\} \right)\right) \leq C <\infty
\]
for some constant $C>0$, where $\circ$ denotes the element wise product and $\tilde\params,\tilde\re$ are vectors whose components satisfies $\tilde\params_j = t_j \params + (1-t_j)\paramstrue$ and $\tilde\re_j = t^\prime_j \re + (1-t_j^\prime)\retruepi$ for $t_j, t^\prime_j \in [0,1]$ and $j = 1,\dots, \numpergroup_i$. 
The last inequality holds as the natural parameter is linear in $\re$ and the largest $\norm{\redesign_i}_2$ is $O(\log(\numpergroup_{\min})^{\frac{1-\eps}{2}} )$ and by \cref{assn:non-degen} v).
The expectation of this term is uniformly bounded as $\max_{i =1,\dots, \numgroups} \norm{\retruepi}_2 = O_p(\sqrt{\log(\numgroups)} )$, and therefore this integrand is uniformly $o(\exp(\norm{\re}_2^{2 - \eps}))$ and therefore integrable. 
This holds as the natural parameter is linear in $\re$ and by \cref*{assn:non-degen} v). 

As for the final statement, for $\weightmatp_i(\re)$ as defined in the proof of \cref{assn:consistency}
\*[
	&\int_{\re \in \Reals^\redim} \norm{\partial^2_{\params}\log\margdens(\obsall_i;\re,\params)}_{op} \bm{\phi}(\re; \zero, \bm\Sigma(\bm\sigma) ) \dee \re\\
	&\leq \int_{\re \in \Reals^\redim} \norm{\design_i\Tr \weightmatp_i(\re)\design_i }_{op}\bm{\phi}(\re; \zero, \bm\Sigma(\bm\sigma)) \dee \re\\
	&\leq \int_{\re \in \Reals^\redim} \max \{{\weightmatp_i(\re)}\} \norm{\design_i\Tr \design_i }_{op}\bm{\phi}(\re; \zero, \bm\Sigma(\bm\sigma)) \dee \re,
\]
where $\max \{{\weightmatp_i(\re)}\}$ is the maximum entry of the matrix and $\norm{\design_i\Tr \design_i }_{op} \leq \recovbig \numpergroup_i^{1+\eps} $ for all $\eps > 0$. The maximum variance of an observation can then be show to grow slower than $\exp(-\norm{\re}_2^{2-\eps})$ by \cref*{assn:non-degen} v) through similar steps as before, showing the desired result.
\manualendproof

\section{Additional Simulation Results}\label{supp:additionalsims}

\noindent This section gives additional results for the simulation study of \cref{subsec:empirical}.
We report bias, coverage, and root-mean square error (RMSE) for $\beta_0$ and $\beta_1$.

\begin{figure}[ht]
\centering
\includegraphics[width=5in]{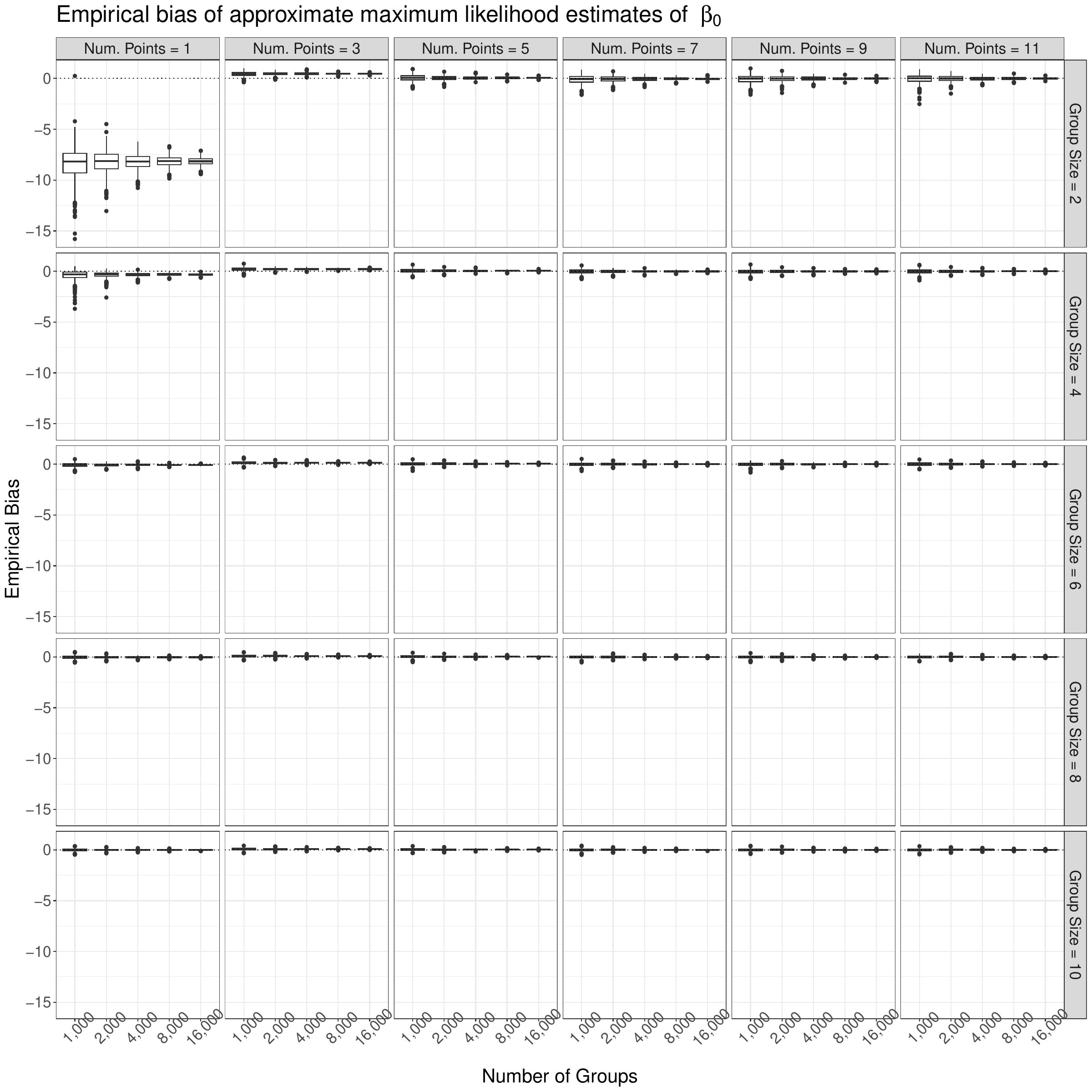}
\caption{Empirical bias of $\beta_0$ for the simulation of \cref{subsec:empirical}. A Bernoulli random intercept model (\cref{eqn:bernoullimodel})
was fit $1000$ times to each combination of numbers of groups ($\numgroups$, x-axis), group size ($\numpergroup$, rows) and numbers of quadrature points ($\quadnum$, columns). The y-axis shows the empirical bias of the approximate
MLE, $\approxparamsmle$ as an estimate of $\params$. 
Shown are empirical bias across $1000$ simulations ($\bullet$) along with Monte Carlo confidence intervals ($- - -$).}
\label{fig:moreresultsbeta0bias}
\end{figure}

\begin{figure}[ht]
\centering
\includegraphics[width=5in]{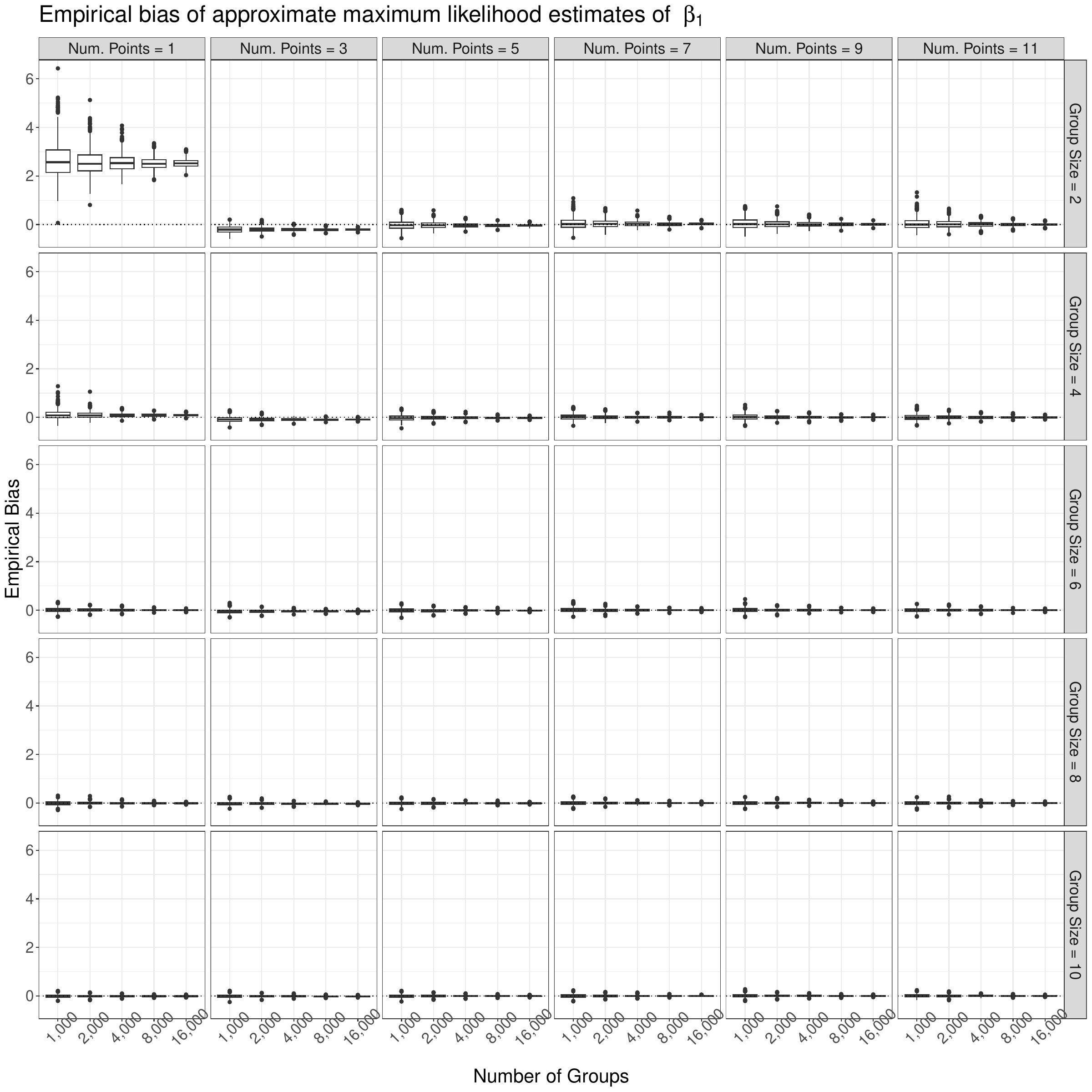}
\caption{Empirical bias of $\beta_1$ for the simulation of \cref{subsec:empirical}. A Bernoulli random intercept model (\cref{eqn:bernoullimodel})
was fit $1000$ times to each combination of numbers of groups ($\numgroups$, x-axis), group size ($\numpergroup$, rows) and numbers of quadrature points ($\quadnum$, columns). The y-axis shows the empirical bias of the approximate
MLE, $\approxparamsmle$ as an estimate of $\params$. 
Shown are empirical bias across $1000$ simulations ($\bullet$) along with Monte Carlo confidence intervals ($- - -$).}
\label{fig:moreresultsbeta1bias}
\end{figure}

\begin{figure}[ht]
\centering
\includegraphics[width=5in]{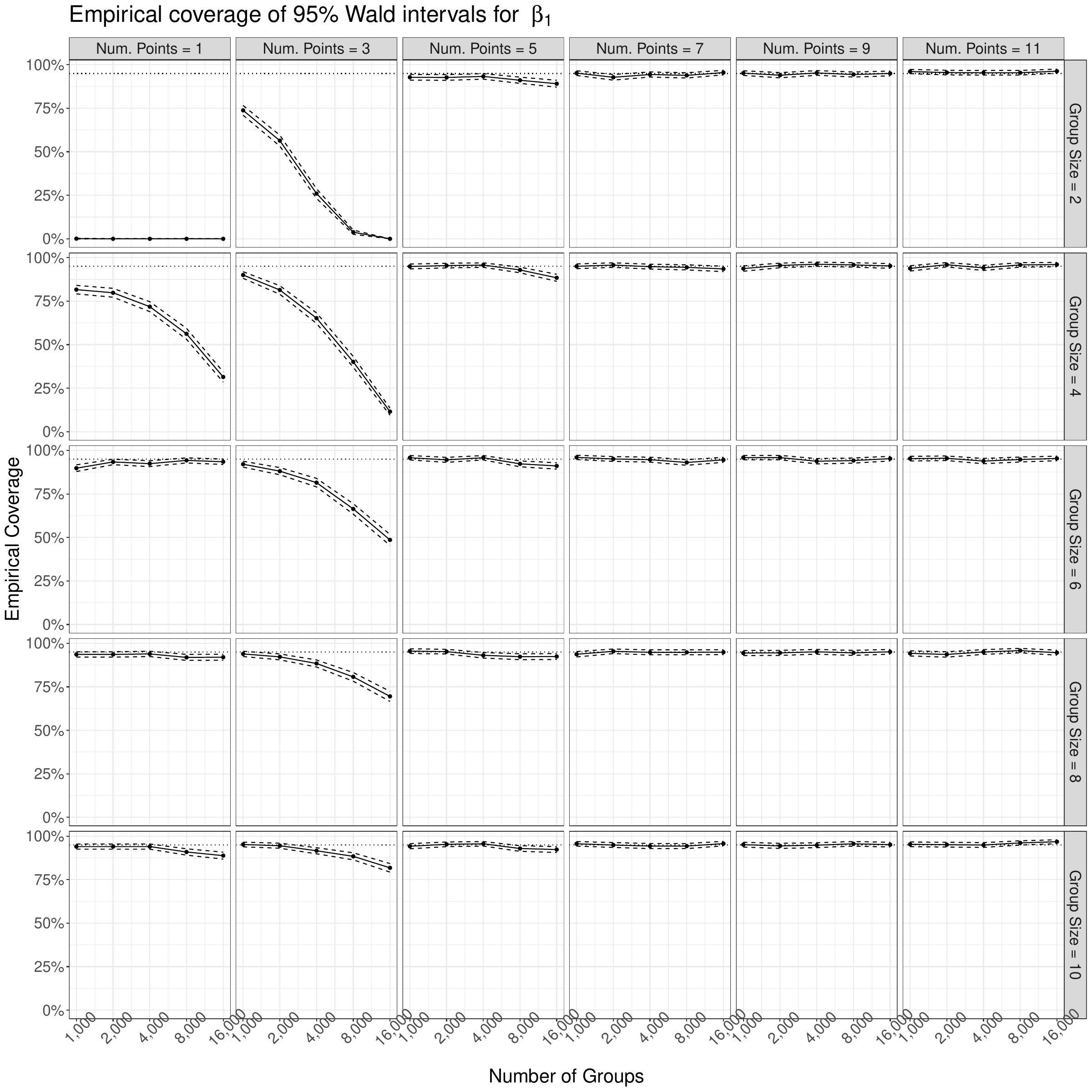}
\caption{Empirical coverage of $\beta_1$ for the simulation of \cref{subsec:empirical}. A Bernoulli random intercept model (\cref{eqn:bernoullimodel})
was fit $1000$ times to each combination of numbers of groups ($\numgroups$, x-axis), group size ($\numpergroup$, rows) and numbers of quadrature points ($\quadnum$, columns). The y-axis shows empirical coverages of $95\%$ Wald confidence intervals centred at the approximate
MLE, $\approxparamsmle$, with standard errors computed using the diagonal of the inverse Hessian of the approximate marginal likelihood.
Shown are empirical coverage proportions across $1000$ simulations ($\bullet$) along with Monte Carlo confidence intervals ($- - -$).}
\label{fig:moreresultsbeta1covr}
\end{figure}

\begin{figure}[ht]
\centering
\includegraphics[width=5in]{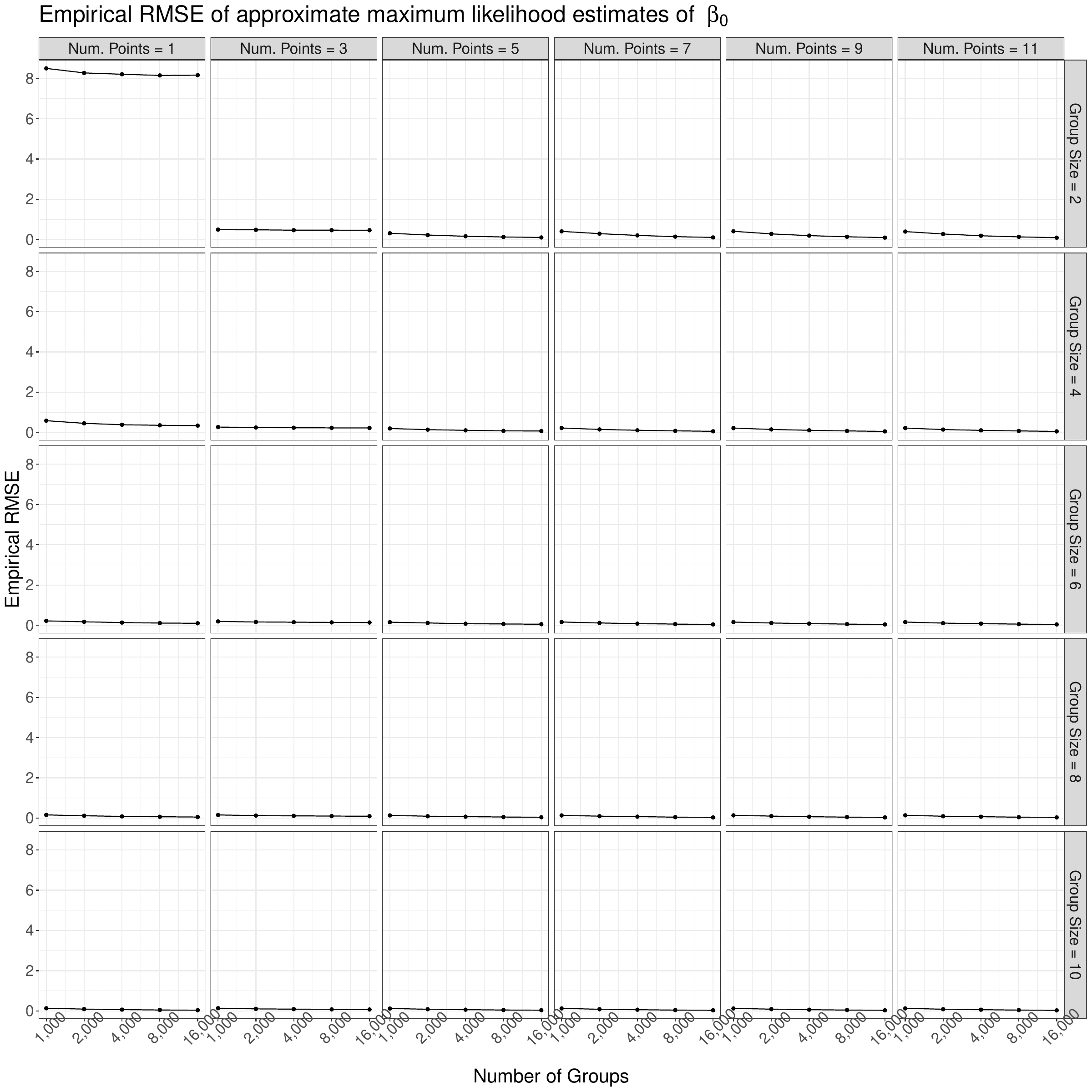}
\caption{Empirical RMSE of $\beta_0$ for the simulation of \cref{subsec:empirical}. A Bernoulli random intercept model (\cref{eqn:bernoullimodel})
was fit $1000$ times to each combination of numbers of groups ($\numgroups$, x-axis), group size ($\numpergroup$, rows) and numbers of quadrature points ($\quadnum$, columns). The y-axis shows the empirical RMSE of the approximate
MLE, $\approxparamsmle$ as an estimate of $\params$. 
Shown are empirical RMSE across $1000$ simulations ($\bullet$) along with Monte Carlo confidence intervals ($- - -$).}
\label{fig:moreresultsbeta0rmse}
\end{figure}

\begin{figure}[ht]
\centering
\includegraphics[width=5in]{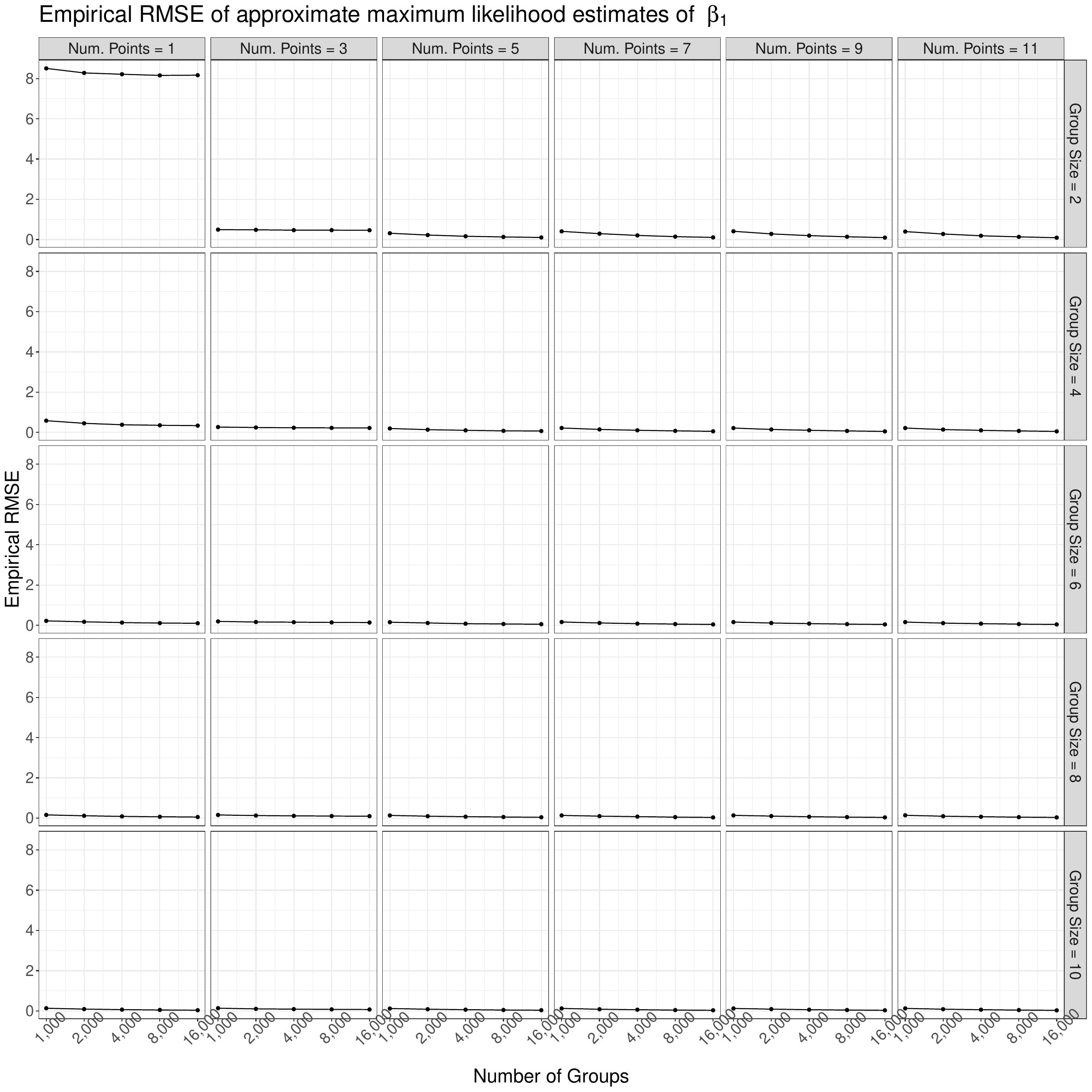}
\caption{Empirical RMSE of $\beta_1$ for the simulation of \cref{subsec:empirical}. A Bernoulli random intercept model (\cref{eqn:bernoullimodel})
was fit $1000$ times to each combination of numbers of groups ($\numgroups$, x-axis), group size ($\numpergroup$, rows) and numbers of quadrature points ($\quadnum$, columns). The y-axis shows the empirical RMSE of the approximate
MLE, $\approxparamsmle$ as an estimate of $\params$. 
Shown are empirical RMSE across $1000$ simulations ($\bullet$) along with Monte Carlo confidence intervals ($- - -$).}
\label{fig:moreresultsbeta1rmse}
\end{figure}

\end{appendix}

\end{document}